\documentclass[longauth]{aa}

\usepackage{graphicx}
\usepackage{txfonts}
\usepackage[]{hyperref}
\hypersetup{
  colorlinks   = true, 
  urlcolor     = blue,
  linkcolor    = blue, 
  citecolor   = blue 
} 

\usepackage[percent]{overpic} 
\usepackage{pdflscape}
\usepackage{amsmath}
\usepackage{comment}
\usepackage{enumitem}
\usepackage{booktabs} 
\usepackage{multirow} 
\usepackage{siunitx}
\usepackage{makecell} 
\usepackage[export]{adjustbox} 
\usepackage{caption} 

\usepackage{amssymb}
\usepackage[dvipsnames]{xcolor}
\usepackage{cancel}

\newcommand{\Mj}{\rm M_{Jup}}

\newcommand{\discminer}{\textsc{discminer}}
\newcommand{\twCOfull}{$^{12}$CO\,$J=2-1$}
\newcommand{\twCO}{$^{12}$CO}
\newcommand{\thCO}{$^{13}$CO}
\newcommand{\thCOfull}{$^{13}$CO\,$J=2-1$}
\newcommand{\eiCO}{C$^{18}$O}
\newcommand{\eiCOfull}{C$^{18}$O\,$J=2-1$}

\newcommand{\mwc}{MWC\,480}
\newcommand{\hd}{HD\,163296}
\newcommand{\as}{AS\,209}
\newcommand{\im}{IM\,Lup}
\newcommand{\gm}{GM\,Aur}

\newcommand{\redarrow}{\textcolor{red}{$\uparrow$}}
\newcommand{\Redarrow}{\textcolor{red}{$\Uparrow$}}

\newcommand{\redplus}{\textcolor{red}{$+$}}
\newcommand{\Redplus}{\textcolor{red}{$\oplus$}}

\newcommand{\bluearrow}{\textcolor{blue}{$\downarrow$}}
\newcommand{\Bluearrow}{\textcolor{blue}{$\Downarrow$}}

\newcommand{\blueminus}{\textcolor{blue}{$-$}}
\newcommand{\Blueminus}{\textcolor{blue}{$\ominus$}}

\defcitealias{izquierdo+2021}{Paper I}

\begin{document}

   \title{The Disc Miner\\ \Large{II. Revealing gas substructures and kinematic signatures from planet-disc interaction through line profile analysis} }
   
    \titlerunning{The Disc Miner II}

   \subtitle{}

   \author{A. F. Izquierdo\inst{1}\fnmsep\inst{2}
            \and
             L. Testi\inst{3}\fnmsep\inst{4}
          \and
          S. Facchini\inst{5}
          \and
          G. P. Rosotti\inst{2}\fnmsep\inst{5}\fnmsep\inst{6}
          \and
          E. F. van Dishoeck\inst{2}\fnmsep\inst{7}
           \and
          L. W\"olfer\inst{2}
          \and
          T. Paneque-Carre\~no\inst{1}\fnmsep\inst{2}
          }

   \institute{European Southern Observatory, Karl-Schwarzschild-Str. 2, 85748 Garching bei München, Germany\\
              \email{andres.izquierdo.c@gmail.com}
         \and
             Leiden Observatory, Leiden University, P.O. Box 9513, 2300 RA Leiden, The Netherlands
         \and 
             Dipartimento di Fisica e Astronomia ``Augusto Righi'' - Alma Mater Studiorum Universit\`a di Bologna, Via Gobetti 93/2, I-40129 Bologna, Italy
         \and
             INAF – Osservatorio Astrofisico di Arcetri, Largo E. Fermi 5, 50125 Firenze, Italy
         \and
             Dipartimento di Fisica, Universit\`a degli Studi di Milano, Via Celoria, 16, Milano, I-20133, Italy
         \and
             School of Physics and Astronomy, University of Leicester, University Road, Leicester, LE1 7RH, UK
         \and 
             Max-Planck-Institut für extraterrestrische Physik, Gießenbachstr. 1 , 85748 Garching bei München, Germany
             }

   \date{Accepted April 2023}


\abstract
{Detecting planets in the early stages of formation is key to reconstructing the history and diversity of fully developed planetary systems. The aim of this work is to identify potential signatures from planet-disc interaction in the circumstellar discs around \mwc{}, \hd{}, \as{}, \im{}, and \gm{}, through the study of molecular lines observed as part of the ALMA large program MAPS. Extended and localised perturbations in velocity, line width, and intensity have been analysed jointly using the \discminer{} modelling framework, in three bright CO isotopologues, \twCO{}, \thCO{}, and \eiCO{} $J=2-1$, to provide a comprehensive summary of the kinematic and column density substructures that planets might be actively sculpting in these discs. We find convincing evidence for the presence of four giant planets located at wide orbits in three of the discs in the sample: two around \hd{}, one in \mwc{}, and one in \as{}. One of the planet candidates in \hd{}, P94, 
previously associated with velocity signatures detected in lower velocity resolution \twCO{} data, is confirmed and linked to localised velocity and line width perturbations in \thCO{} and \eiCO{} too. We highlight that line widths are also powerful tracers of planet-forming sites as they are sensitive to turbulent motions triggered by planet-disc interactions. In \mwc{}, we identified non-axisymmetric line width enhancements around the radial separation of candidate planet-driven buoyancy spirals,
which we used to narrow the location of the possible planet to an orbital radius of $R=245$\,au and $\rm{PA}=193^\circ$, at a projected distance of $1.33"$ from the star. In the disc of \as{}, we found excess \twCO{} line widths centred at $R=210$\,au, $\rm{PA}=151^\circ$, at a projected distance of $1.44"$, spanning around the immediate vicinity of a circumplanetary disc candidate proposed previously, 
which further supports its presence. We report no clear localised or extended kinematic signatures in the discs of \im{} or \gm{} that could be associated with the presence of planets or gravitational instabilities. On the other hand, we demonstrate that pressure minima exhibit line width minima counterparts in optically thick emission, making them robust tracers of gaps in the gas surface density when analysed together with azimuthal velocity flows. Finally, we show that nine out of eleven millimetre dust continuum rings in the sample are co-located with pressure bumps traced by kinematic modulations, indicating that aerodynamic confinement via pressure traps is a common mechanism for the formation of dust substructures in these discs. Overall, our analysis reveals that all discs in the sample present a remarkable level of substructure in all the line profile observables considered, regardless of the CO isotopologue. However, the magnitude and morphology of the substructures vary between discs and tracers, indicating that the kinematics and thermodynamic properties are likely shaped by different physical mechanisms in each object. We propose that the main kinematic signatures identified in the discs of \mwc{}, \hd{}, and \as{} have a planetary origin, although they do not always manifest as highly localised perturbations, while the discs of \im{} and \gm{} do not yield clear signatures pointing to the presence of massive planets.
Our simultaneous analysis of multiple tracers and observables aims to lay the groundwork for robust studies of molecular line properties focused on the search for young planets in discs.}


   \maketitle

\section{Introduction}

The formidable precision and sensitivity provided by the Atacama Large Millimeter/submillimeter Array (ALMA) have revolutionised the way we study and understand protoplanetary discs. High angular resolution observations of these planet-forming environments have revealed a myriad of dust substructures possibly shaped by the hydrodynamic interaction of embedded planets with the disc itself, but also due to instabilities in the gas to which the dust grains respond according to their composition and morphology \citep[see][and references therein]{testi+2014, andrews+2020, bae+2022_ppvii}.

ALMA has also opened a unique window into the bulk of the gas disc through deep and high-resolution observations of molecular lines, which provide a vast amount of information that has only recently started to be exploited in the context of planetary formation \citep[see reviews by][]{dutrey+2014, miotello+2022, pinte+2022}. For instance, a simultaneous analysis of multiple molecular lines gives access to the three-dimensional structure of discs as different chemical species are arranged differently across the radial and vertical extent of the disc \citep[][]{vanzadelhoff+2001,walsh+2010, vandishoeck+2020, oberg_bergin+2021, paneque+2022}. Along with thermochemical modelling, this can be used as input information for detailed studies of the formation and destruction sites of molecular species \citep{booth+2021, leemker+2021, zhang+2021}, which in addition allows one to gain knowledge on internal and external thermodynamic processes that are key to predict the evolution of discs and planets within \citep[][]{benz+2014}. In parallel, by characterising the morphology and Doppler shift of these lines, it is possible to directly probe the kinematics and temperature of the disc, again in a three-dimensional fashion, but also to perform indirect estimates of other intrinsic properties such as the gas surface density and pressure \citep[as in e.g.][]{rosenfeld+2013, teague+2018a, teague+2018b, dullemond+2020, yu+2021}, quantify non-thermal motions \citep[as in e.g.][]{teague+2016, flaherty+2017, flaherty+2020, liu+2018}, or even derive stellar \citep{czekala+2015, czekala+2017} and disc masses \citep{veronesi+2021, lodato+2022} through analyses of the disc background velocity field. 

Delving into smaller scales, the study of molecular lines is presently the workhorse of kinematic studies focused on the detection of velocity perturbations to the Keplerian rotation of the gas in planet-forming discs, and with good reason. As several studies have theorised, these kinematic features represent a window into a wide variety of physical processes, including the response of the disc to the presence of embedded planets \citep[][]{perez+2015, perez+2018, dong+2019, bae+2021, bollati+2021, izquierdo+2021, rabago+2021}, stellar companions \citep{facchini+2018, rosotti_hd100453+2020} and flybys \citep[see][for a review]{cuello+2022}, but also to several (magneto--)hydrodynamic instabilities \citep[see][for a summary]{pinte+2022, bae+2022_ppvii}. In the planetary scenario, these studies predict that an embedded planet can simultaneously trigger localised and extended perturbations, not only in velocity, but also in density and temperature, which translate into characteristic features observable in molecular line emission. Indeed, a select number of works have claimed the discovery of planet-driven signatures based on the detection of `kinks' in intensity channel maps of molecular lines \citep[][]{pinte+2018b, pinte+2019, pinte+2020, calcino+2022, verrios+2022}, but also on the analysis of small- and large-scale velocity perturbations to the Keplerian rotation of the gas disc \citep[as in][]{casassus+2019, teague+2018a,  teague+2021, teague+2022, izquierdo+2022}.

Recently, the Molecules with ALMA at Planet-forming Scales (MAPS) 
survey mapped the spatial and spectral distribution of more than 20 chemical species in five protoplanetary discs around the young stars \mwc{}, \hd{}, \as{}, \im{}, and \gm{} \citep{oberg+2021}. 
Several studies of these data have revealed that the molecular line emission from simple and complex species in these discs is rich in substructure, both radially and vertically \citep[][]{law+2021_maps3,law+2021_maps4,zhang+2021,guzman+2021,ilee+2021,legal+2021, paneque+2022}, providing unparalleled clues about the chemical composition of planets that will eventually configure a planetary system such as ours. 
More specifically, \citet[][]{teague+2021} present detailed kinematic analyses of the discs around \mwc{} and \hd{}, identifying numerous velocity signatures in \twCO{}, and tentatively also in \thCO{} and \eiCO{}, possibly connected to the hydrodynamic and gravitational interaction of these discs with embedded planets. Recently, the disc of \as{} has gained additional interest in the context of the detection of planet-driven features as \citet[][]{bae+2022} propose the presence of a circumplanetary disc (CPD) after observing a localised signal in the intensity of \thCO{} channels, within a gap on the outskirts of the circumstellar disc, at a radial distance of $\sim\!200$\,au from the star. Nevertheless, given the amount and wealth of the data provided by the MAPS survey, much remains to be investigated, especially in light of new modelling techniques that allow not only the kinematics of discs to be mined, but also other properties encoded in molecular line profiles with greater precision than previous methods.

In this article, we employ the line profile analysis techniques developed in \citet[][or Paper I hereafter]{izquierdo+2021}, implemented in the \discminer{} package, to search for gas substructures and kinematic signatures that may point to, or refute, the presence of embedded planets in the discs observed by the ALMA large program MAPS. The paper is organised as follows. Section \ref{sec:data} gives an overview of the molecular lines and sources studied. Section \ref{sec:discminer} presents a summary of the modelling strategy, as well as the line profile observables considered in Sections \ref{sec:kinematics} and \ref{sec:gas_substructure} to characterise kinematic and gas substructures in the discs. In Section \ref{sec:discussion}, the interpretation of these features and their possible relationship with embedded planets is discussed, and we conclude with the main findings of the paper in Section \ref{sec:conclusions}.


\section{Observations and models} \label{sec:data}

   \begin{figure*}
   \centering
    \includegraphics[width=1.0\textwidth]{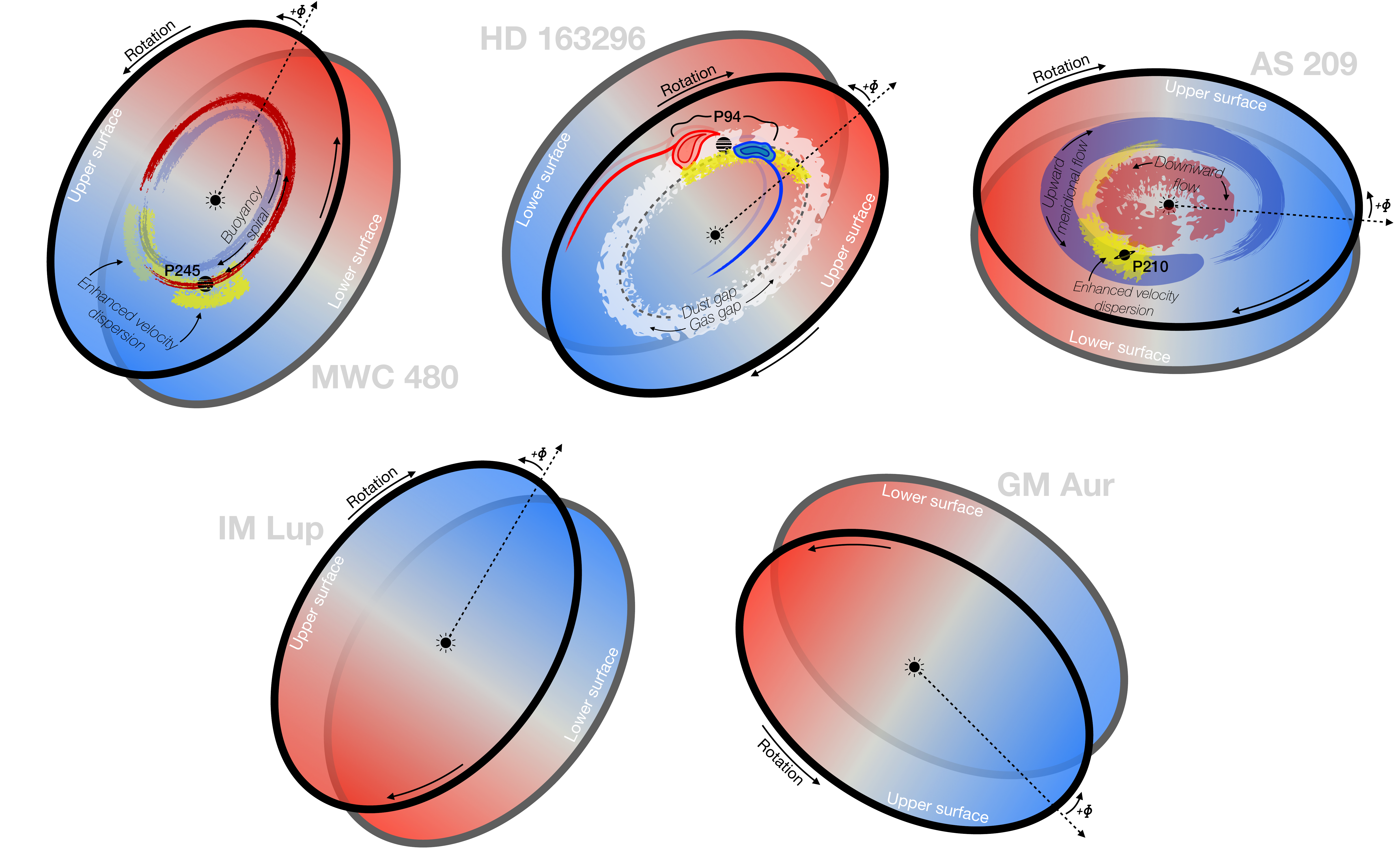}
      \caption{Illustrating the orientation and rotation direction of the discs observed by the ALMA large program MAPS and analysed here. Dashed arrows mark the origin of each disc coordinate system. For convenience, we assume this reference axis to be parallel to the projected semi-major axis that is closest to the north celestial axis in counterclockwise rotation. Also shown are the signatures reported in this work as potential planet-disc interaction features in the form of both coherent and turbulent velocity fluctuations.
              }
         \label{fig:disc_sketches}
   \end{figure*}

   \begin{figure*}
   \centering
    \includegraphics[width=0.8\textwidth]{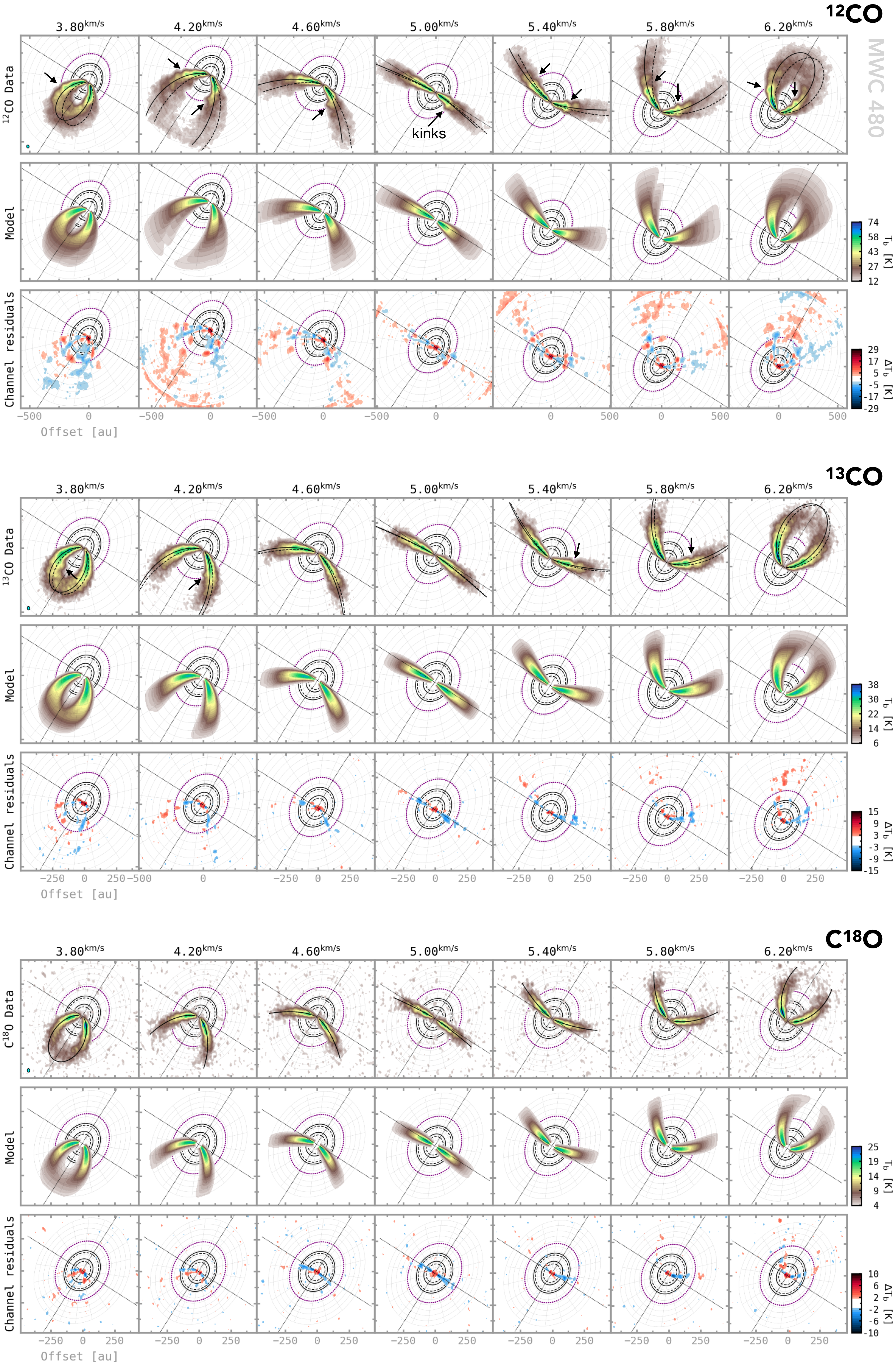}
      \caption{Selected CO isotopologue intensity channel maps from the disc around \mwc{}, along with best-fit channels obtained with \discminer{} and their corresponding residuals. For comparison, isovelocity contours from the model upper and lower surfaces are overlaid on the data channels as solid and dashed lines, respectively. The ellipses in the background represent millimetre dust rings (solid) and gaps (dashed) as reported by \citet{law+2021_maps3}, projected on the upper emission surface of the corresponding model. Kink-like features apparent to the eye are marked with arrows. The radial location of the most prominent kink in \twCO{} is highlighted by the dotted purple ellipse. The beam size is shown in the bottom left corner of the top left panel for each isotopologue. 
              }
         \label{fig:channel_maps_mwc480}
   \end{figure*}

In this work we make use of publicly available data from molecular line observations of five protoplanetary discs around the young stars \mwc{}, \hd{}, \as{}, \im{}, and \gm{}, as part of the ALMA large program MAPS \citep{oberg+2021}. We use the JvM-corrected images with a robust parameter of 0.5 \citep[][]{czekala+2021}\footnote{Publicly available on \url{https://alma-maps.info/data.html}}. Although there are multiple molecules and transitions at hand within the MAPS survey, our analysis focuses on the three brightest CO isotopologues, \twCO{}, \thCO{} and \eiCO{}, in the $J=2-1$ rotational transition, observed at an angular resolution of $0.15''$, and a velocity channel spacing of $0.2$\,km\,s$^{-1}$. 
The unprecedented angular and spectral resolution together with the high sensitivity of these observations allow us to apply dedicated state-of-the-art techniques to study the kinematics and morphology of these objects with exquisite precision. Cartoons illustrating the orientation and rotation direction of the discs analysed in this work are presented in Figure \ref{fig:disc_sketches}.

\subsection{Discminer models} \label{sec:discminer}

The majority of the analysis presented in this paper is performed with the \discminer{} channel-map modelling tool, whose main capabilities were introduced in \citetalias{izquierdo+2021} and first applied to actual data in \citet{izquierdo+2022} with the aim of looking for planetary signatures in the kinematics of the disc around \hd{} using \twCOfull{} line observations. 
In this chapter, we describe the parameters and attributes that make up the channel-map models of the five discs of interest, and present a comparison between selected model and data channels. 

\subsubsection{Model parameters and attributes}

\discminer{}\footnote{Publicly available on \url{https://github.com/andizq/discminer}} aims to reproduce the channel-by-channel intensity of molecular line emission from discs by modelling various types of parameters which can be classified in four groups:

\begin{enumerate}
    \item Orientation parameters that control the projected appearance of the disc in the sky, 
    such as inclination, $i$, position angle, PA, and offset, $x_c$ and $y_c$.
    \item Velocity parameters that shape the rotation velocity of the gas disc, such as the stellar mass, $M_\star$, and systemic velocity, $\upsilon_{\rm sys}$. Other parameters such as the gas mass or the steepness of the radial pressure profile can also play an important role in setting the disc background velocity \citep[see e.g.][]{rosenfeld+2013}. Although accountable by \discminer{}, modelling these parameters is computationally expensive and out of the scope of this paper. 
    Thus, the model rotation velocity is assumed Keplerian for all discs, with vertical differential rotation sustained by balance of forces in the $z-$direction.
    \item Surface parameters that determine the elevation of upper and lower emission surfaces above and below the disc midplane.
    \item Line profile parameters that account for the shape of the line profiles, controlling the variation of line peak intensity, line width and line slope across the disc radial and vertical extent.
\end{enumerate}

These parameters are then assembled together in a two-component line profile kernel (one per emission surface), or in a single-component profile (if the lower surface is not resolved), to compute the model intensity, $I_m$, as a function of the disc cylindrical coordinates, $(R,z)$, and the velocity channel, $\upsilon_{\rm ch}$. To combine the contribution of the upper and lower emission surfaces to the total line profile, we take the highest intensity between the two components on each pixel and velocity channel. 
For each component, we adopt the same kernel of \citetalias{izquierdo+2021}, a generalised bell function of the form, 
\begin{equation}
    I_m(R, z; \upsilon_{\rm ch}) = I_p \left(1+\left|\frac{\upsilon_{\rm ch} - \upsilon_{\rm k^{l.o.s}}}{L_w} \right|^{2L_s} \right)^{-1},
\end{equation}
where $I_p$ is the line peak intensity, and $\upsilon_{\rm k^{l.o.s}}$ is the Keplerian velocity projected along the line of sight. The line width, $L_w$, is the half width of the profile at half power. The dimensionless attribute, the line slope $L_s$, controls how steep the signal drops at the wings and in turn it also determines the spectral extent of the plateau at the top of the profile. As explained in \citetalias{izquierdo+2021}, this kernel performs better than a Gaussian at reproducing optically thick lines which are flat at the top and decay rapidly towards the wings, as is usually the case for \twCO{} and \thCO{} lines \citep[][]{beckwith+1993}, but at the same time it is flexible enough to account for optically thinner lines if needed. 
A summary of the assumed functional forms and parameters involved in each of the model attributes is presented in Table \ref{table:attributes_parameters}.

Finally, to find best-fit model channel maps to the data we couple \discminer{} with a Markov chain Monte Carlo (MCMC) random sampler, \textsc{emcee} \citep{foreman+2013}, which walks over a vast range of parameters to determine a subset of them that best reproduces the projected intensity of the input datacube on each velocity channel. The best-fit model parameters obtained for each disc and tracer are summarised in Tables \ref{table:pars_mwc480}--\ref{table:pars_gmaur}. To explore the influence of disc inclination and different tracers on the retrieved dynamical mass, we also made runs assuming disc inclination fixed at the continuum value for all sources and tracers, and reported the obtained parameters in the same tables. We note that differences in $M_\star$ range between $\sim\!5-10$\% for variations in inclination within $\sim\!1-4^\circ$, which is a consequence of the strong correlation between these two parameters \citep[see e.g. Fig. 11 of][]{izquierdo+2022}.

\setlength{\tabcolsep}{10.5pt} 

\begin{table}
\centering
{\renewcommand{\arraystretch}{1.5}
\caption{List of attributes adopted in the \discminer{} models for this work.}  \label{table:attributes_parameters}

\begin{tabular}{ ll } 

\toprule
\toprule
Attribute & Prescription \\
\midrule


Orientation & $i$, PA, $x_c$, $y_c$ \\ \midrule

Velocity & $\upsilon_{\rm rot} = \sqrt{\frac{GM_\star}{r^3}}R$, $\upsilon_{\rm sys}$ \\ \midrule

Upper surface & $z_u = z_{0} (R/D_0)^p \exp{[-(R/R_t)^q]}$ \vspace{0.25cm} \\

Lower surface & $z_l = -z_{0} (R/D_0)^p \exp{[-(R/R_t)^q]}$ \\

\midrule
Peak intensity & $I_p$ = $I_0 (R/D_0)^p (z/D_0)^q$ \vspace{0.25cm} \\

Line width & $L_w$ = $L_{w0} (R/D_0)^p (z/D_0)^q$ \vspace{0.25cm} \\ 

Line slope & $L_s$ = $L_{s0} (R/D_0)^p$ \\

\bottomrule

\end{tabular}
\caption*{\textbf{Note.} $G$ is the gravitational constant; $D_0=100$\,au is a normalisation factor. In disc coordinates, $z$ is the height above the disc midplane, $R$ is the cylindrical radius, and $r$ is the spherical radius. The remaining variables are free parameters. All free parameters involved in each attribute are modelled independently, even if they are named alike.}

  }
\end{table}

   \begin{figure*}
   \centering
   \includegraphics[width=1.0\textwidth]{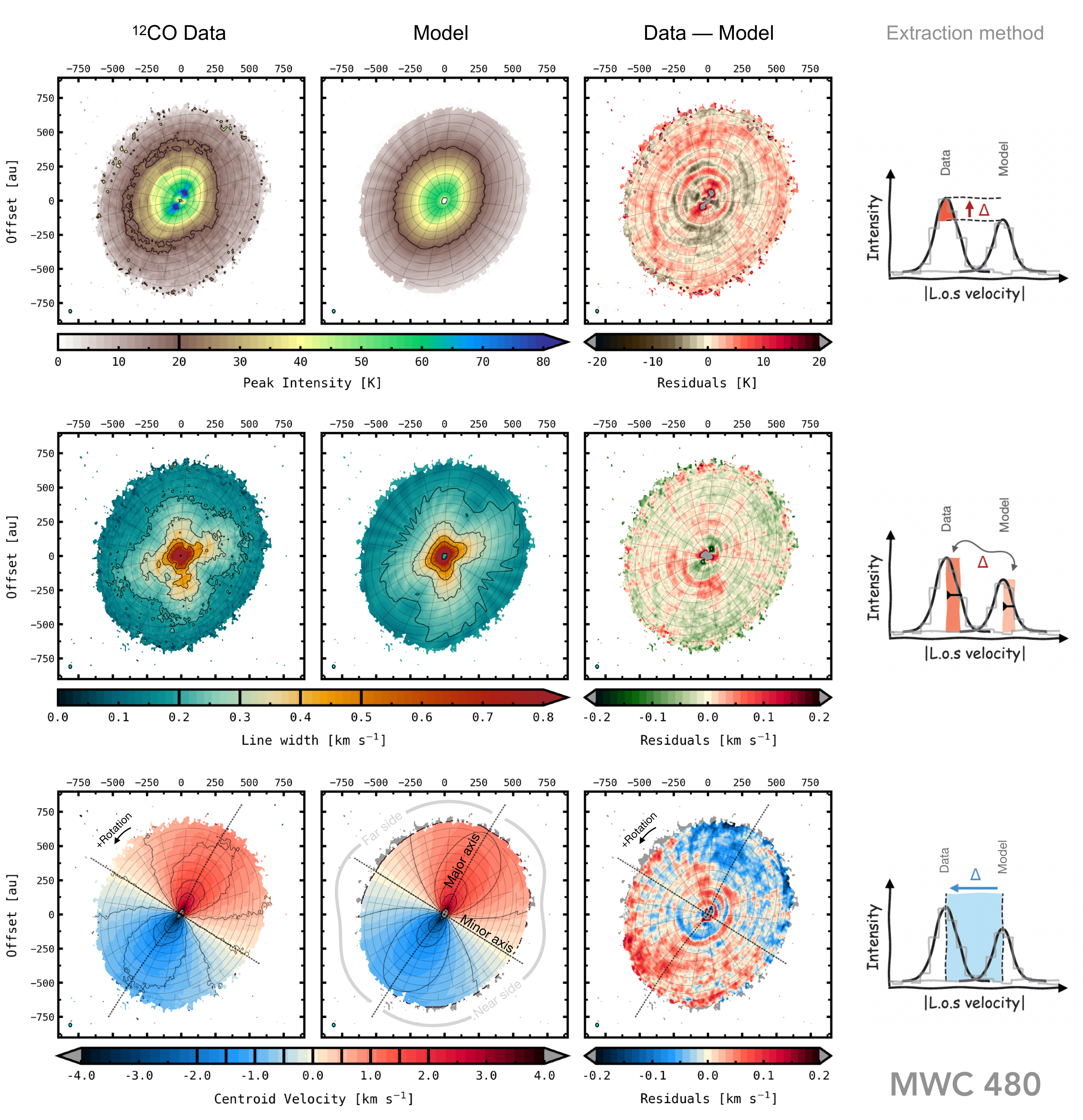}
      \caption{Extraction of the line profile observables considered in this work, using \mwc{} \twCO{} data as example. Peak intensities (top), line widths (middle), and centroid velocities (bottom) are displayed for both data and best-fit model, as well as residual maps showing the difference between them. Sketches on the rightmost column illustrate how each line profile observable is computed on the data cube and subsequently compared to those derived on the smooth, Keplerian model cube obtained by \discminer{}. The black solid lines are reference contours to ease comparison. Their levels are indicated in the colour bars. 
      Residuals on each of the observables are driven by temperature, density and velocity fluctuations in the gas disc.
              }
         \label{fig:observables}
   \end{figure*}

\subsubsection{Data versus model channel maps} \label{sec:channels}

In Figure \ref{fig:channel_maps_mwc480}, we present a comparison of selected intensity channel maps of the three CO isotopologues studied in this work, for the disc of \mwc{}, their corresponding model channels computed with \discminer{}, and residuals obtained by subtracting the intensity of each model channel to that of the data. Data and model channel maps for the discs of \hd{}, \as{}, \im{} and \gm{} are presented in Figures \ref{fig:channel_maps_hd163296}--\ref{fig:channel_maps_gmaur}.

We note that there are multiple features that can be easily grasped after quick visual inspection of the data channels. First, it is clear that the five discs yield different radial and vertical extents. 
Second, non-negligible residuals in channel intensities 
suggest that a smooth Keplerian model is not the perfect match for these data. However, it is precisely through these deviations from Keplerian that the underlying physics acting on the discs can be uncovered. A massive planet, for example, is expected to trigger perturbations to the Keplerian rotation of the gas in its vicinity, and at the same time launch spiral density waves, with characteristic kinematic and intensity structures, that propagate over large azimuthal and radial portions of the disc \citep[][]{perez+2018, bae+2021, rabago+2021}. 
Indeed, many of the intensity channels of all discs in this sample display wiggles, or kinks, which are usually linked, but not limited, to the presence of velocity perturbations in the gas disc. In Sections \ref{sec:kinematics} and \ref{sec:gas_substructure}, we characterise variations in the line profile observables to study the origin of these fluctuations at first sight evident in channel maps, but also of those that remain hidden to the eye.

We assume axisymmetric intensity fields for all model channels except for the \twCO{} disc of \as{}. The reason is that the \twCOfull{} line intensity of this disc is on average twice as weak on (almost all of) the blueshifted side than on the redshifted side due to foreground cloud absorption \citep{oberg+2011}. To take this effect into account, we used a different intensity normalisation factor for each side of the disc: $0.65I_0$ for $-80^\circ<\phi<80^\circ$, and $1.25I_0$ otherwise, where $I_0$ is the intensity normalisation reported in table \ref{table:pars_as209} for \twCO{}, and $\phi$ is the azimuthal coordinate of the disc reference frame.

\subsection{Line profile observables to compare data and models} \label{sec:observables}

   \begin{figure}
   \centering
   \includegraphics[width=0.69\columnwidth]{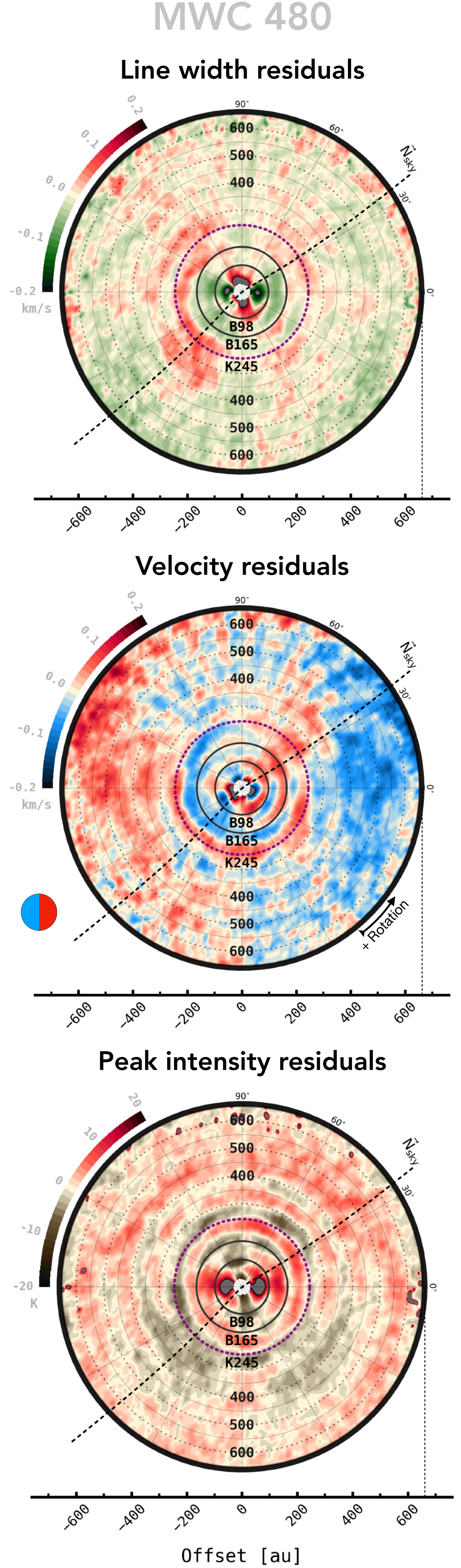}
    \caption{Line width (top), velocity (middle), and peak intensity (bottom) residuals computed for the \twCO{} disc of \mwc{}, as in Fig. \ref{fig:observables}, but deprojected on the disc reference frame. The vertical axis in this frame is parallel to the projected minor axis on the sky. The solid lines mark the radial location of millimetre dust rings, and the purple dotted line that of the most prominent kink identified by eye in \twCO{} channels. Also shown is the north-sky (or $\rm{PA}=0^\circ$) axis for reference. The oblique shape of this axis is due to the deprojected geometry of the disc emission surface. 
              }
         \label{fig:residuals_12co_mwc480}
   \end{figure}

We employ three observables from line intensity profiles, (i) peak intensity, (ii) line width, and (iii) centroid velocity, to search for gas surface density, temperature and velocity substructures in all discs and tracers considered in this work. The three observables are extracted from Gaussian kernels fitted to both the data and model line profiles, and compared against each other to produce residual maps, as illustrated in Figure \ref{fig:observables} for the \twCO{} disc around \mwc{}. Using information of the disc vertical structure and orientation retrieved by \discminer{}, these line profile observables and their residuals can then be deprojected onto the disc reference frame for quantitative studies of the physical location, extent, and magnitude of the retrieved signatures in the gas disc. Deprojected views of the observable residuals for the same disc and tracer are presented in Figure \ref{fig:residuals_12co_mwc480}. 
Radial profiles of each observable extracted from the data alone, as function of disc and isotopologue, including rotation curves and elevation of model emission surfaces, are summarised in Figure \ref{fig:attributes_all_co}.

In practice, each line profile observable carries information about different thermodynamic and kinematic properties of the disc. (i) The line peak intensity can be used for studies of the local temperature and density of a molecular species \citep[as in e.g.][]{facchini+2021,zhang+2021}. Likewise, (ii) the line width depends on the gas temperature and density, but it too responds to local turbulent motions \citep[see e.g.][]{hacar+2016, flaherty+2020}. Finally, (iii) the shift in velocity of the line profile centroid with respect to the systemic velocity, or simply centroid velocity, accounts for organised gas motions driven by different, usually coupled, physical mechanisms such as the gravitational and hydrodynamic interaction between the disc, star(s) and planet(s), as well as a range of (magneto--)hydrodynamic instabilities that may develop in the gas disc under appropriate conditions \citep[see][and references therein for a summary]{pinte+2022}. 
In Figure \ref{fig:discminer_workflow}, we outline the general recipe for computing channel-map models and extracting observables from a given data cube with \discminer{}, and present a summary of the physical properties of discs and planets that can be learned based on the analysis of such observables.

\subsubsection{Upper and lower emission surfaces}

One of the strengths of \discminer{} is its ability to model line profile properties on the upper and lower emission surfaces of the disc, simultaneously. 
As discussed extensively in \citetalias{izquierdo+2021}, the impact of the lower emission surface can be critical when it comes to kinematical analyses of high resolution observations of circumstellar discs \citep[see also][and references therein]{pinte+2022}. The lower surface can systematically shift the centroid of the observed intensity profile in an uneven fashion as a function of the disc coordinates, affecting the velocities derived via first moment maps or via parametric fits to the line profile \citepalias[see e.g. Figure A2 of][]{izquierdo+2021}. In practice, this unphysical shift is source-specific as it depends on the aspect ratio and intensity contrast between the two emitting surfaces.

To overcome this, at the cost of velocity precision, some methods measure velocities on, or around, the peak of the line profile to account primarily for the centroid velocity of the disc upper surface \citep[see e.g.][]{teague+2018_bettermoments, teague+2021}. However, when the emission is optically thin, or just marginally optically thick, these methods struggle at distinguishing between upper and lower surfaces because the intensity contrast between both is generally small in these scenarios. 
The inclusion of the lower surface contribution to the line profiles of our model aims to partly alleviate this effect and enables us to consider the bulk of velocity channels to measure the line centroid shift in velocity with little loss of precision.

\subsubsection{Decomposition method to extract azimuthal and meridional velocities } \label{sec:azimuthal_meridional_velocities}

In this Section, we present a new pipeline to quantify the vertical and azimuthal components of the disc velocity field, $\upsilon_z$ and $\upsilon_{\phi}$, by decomposing two types of deprojected velocity residual maps; the standard and the absolute residual map. 
While the former consists of a direct subtraction of the model centroid velocities to those retrieved from the data, $\upsilon_{0, d} - \upsilon_{0, m}$ (as in e.g. Fig. \ref{fig:residuals_kinematics_12co}), the latter is the difference between the absolute values referred to the model systemic velocity, $|\upsilon_{0, d} - \upsilon_{{\rm sys}, m}| - |\upsilon_{0, m} - \upsilon_{{\rm sys}, m}|$. Taking one or another allows to highlight variations in the three-dimensional components of the velocity in different ways. In a standard velocity residual map, an axisymmetric azimuthal perturbation, $\delta\upsilon_{\phi}$, would appear as two joint semi-rings with the same magnitude but different signs split by the disc minor axis \citep[see e.g. Fig. 5 of][]{teague+2019}. In the absolute velocity residuals, the same perturbation would look as a full ring, namely one where there is no sign flip involved around the disc minor axis. The reciprocal occurs for an axisymmetric vertical perturbation; a ring-like pattern would be observed in the standard residuals, and two semi-rings with different sign in the absolute residuals.

Now, assuming that the three-dimensional components of the velocity field are axisymmetric, one can obtain the azimuthal and vertical components on each radius by computing azimuthal averages of these absolute and standard velocity residuals, respectively. In the first case, the vertical and radial velocity components cancel out over $2\pi$ averages, as illustrated below. In the second scenario, it is the azimuthal and radial components that vanish. 
Radial profiles of the azimuthal and vertical velocity components computed with this pipeline for all discs and tracers are presented in Sect. \ref{sec:gas_substructure}. \\

\noindent\textit{Computing $\upsilon_z$}. A standard velocity residual map is the direct difference between the observed and modelled line-of-sight velocities, $\upsilon_{0, d} - \upsilon_{0, m}$, which can be decomposed into three orthogonal components of the velocity field as follows, 
\begin{multline} \label{eq:vdata}
    \upsilon_{0, d} - \upsilon_{0, m} = \bcancel{\upsilon_{{\rm sys}, d}} - \bcancel{\upsilon_{{\rm sys}, m}} + \left[\upsilon_{\phi, d} - \upsilon_{\phi, m}\right]\cos{\phi}\sin{i} \\ - \upsilon_{r, d}\sin{\phi}\sin{i} - \upsilon_{z, d}\cos{i},
\end{multline}
where the subscripts $d$ and $m$ stand for data and model, respectively. 
Therefore, an azimuthally averaged profile of this map, which is just a sum of integrals applied to each of the above terms, computed on a given annulus of the disc can be written as,
\begin{align} \label{eq:vz}
 \left<\upsilon_{0, d} - \upsilon_{0, m}\right>_{2\pi}
     =& \quad \frac{1}{2\pi} \left[ \cancelto{0}{\int_{0}^{2\pi} \left[\upsilon_{\phi, d} - \upsilon_{\phi, m}\right]\cos{\phi}\sin{i} \,d\phi} \right. \nonumber \\ 
     & \quad - \cancelto{0}{\int_{0}^{2\pi} \upsilon_{r, d}\sin{\phi}\sin{i} \,d\phi} \nonumber \\ 
     & \quad - \left. \int_{0}^{2\pi} \upsilon_{z, d}\cos{i} \,d\phi \right] \nonumber \\
    \therefore \left<\upsilon_{0, d} - \upsilon_{0, m}\right>_{2\pi} =&  \quad - \upsilon_{z,d}\cos{i} \nonumber \\ \nonumber \\
    \iff \upsilon_{z,d} =&  \quad - \frac{1}{\cos{i}} \left<\upsilon_{0, d} - \upsilon_{0, m}\right>_{2\pi},
\end{align}
which naturally results in an expression to compute the vertical component of the velocity field, $\upsilon_{z,d}$, thanks to the azimuthal symmetry of the other two components. We note that since the model velocity field is fully azimuthal it is actually irrelevant for this derivation, and thus an azimuthal average of $\upsilon_{0, d} - \upsilon_{{\rm sys}, m}$ alone would result in the same outcome. Either way, the model-dependent parameters involved in this calculation include systemic velocity, disc inclination, and of course the disc vertical structure which in practice determines the topology of the projected annular paths where azimuthal averages are measured.

Finally, it is straightforward to show that if the azimuthal section $\psi$, over which azimuthal averages are taken, is symmetric around the disc major and minor axes the same equality holds,
\begin{equation}
    \upsilon_{z,d} = - \frac{1}{\cos{i}} \left<\upsilon_{0, d} - \upsilon_{0, m}\right>_{\psi}
\end{equation} \\
\noindent\textit{Computing $\upsilon_{\phi}$}. If the same procedure is applied on absolute velocity residuals instead, following Appendix \ref{sec:appendix_absolute_residuals}, we obtain that the azimuthal component of the velocity perturbation field, $\Delta \upsilon_{\phi}$, can be written as,
\begin{align} \label{eq:vphi}
\Delta\upsilon_\phi = \frac{\pi}{2\sin{i}} \left<\left|\upsilon_{0, d} - \upsilon_{{\rm sys}, m}\right| - \left|\upsilon_{0, m} - \upsilon_{{\rm sys}, m}\right|\right>_{2\pi},
\end{align}
whose generalisation does depend this time on the total extent of the azimuthal section $\psi$ considered,
\begin{equation} \label{eq:vphi}
    \Delta\upsilon_\phi = \frac{\psi}{4\sin{\frac{\psi}{4}}\sin{|i|}} \left<\left|\upsilon_{0, d} - \upsilon_{{\rm sys}, m}\right| - \left|\upsilon_{0, m} - \upsilon_{{\rm sys}, m}\right|\right>_{\psi}.
\end{equation}

These general forms of $\upsilon_z$ and $\upsilon_\phi$ are useful in cases where either physical or unphysical contaminating velocities present over wide azimuthal sections of the disc need to be masked out of the analysis in order to remove, or at least minimise, their influence on the computation of velocity profiles, as will be clear later in Sect. \ref{sec:extended_perturbations} for the \twCO{} discs of \im{} and \gm{}.

As a final note, we highlight that velocity perturbations to the Keplerian rotation are not necessarily axisymmetric as originally imposed by this type of decomposition methods \citep[see also e.g.][]{teague+2019nat}. For instance, the presence of spiral features is an immediate proof that this assumption can break. Using a representative sample of (transition) discs, \citet[][]{woelfer+2022} showed that non-axisymmetric substructures may be relatively common features in both intensity and velocity fields. However, despite the fact that such features would impact the amplitude of the velocity components retrieved by this method, other less-affected observable quantities such as the radial location of velocity peaks, troughs, and gradients, can still provide invaluable clues about the presence and direction of azimuthal and vertical flows of material, generally interpreted as due to variations in the disc pressure structure, as explained and illustrated in Sect. \ref{sec:gas_substructure}. 

\section{Kinematic signatures} \label{sec:kinematics}

   \begin{figure*}
   \centering
   \includegraphics[width=1.0\textwidth]{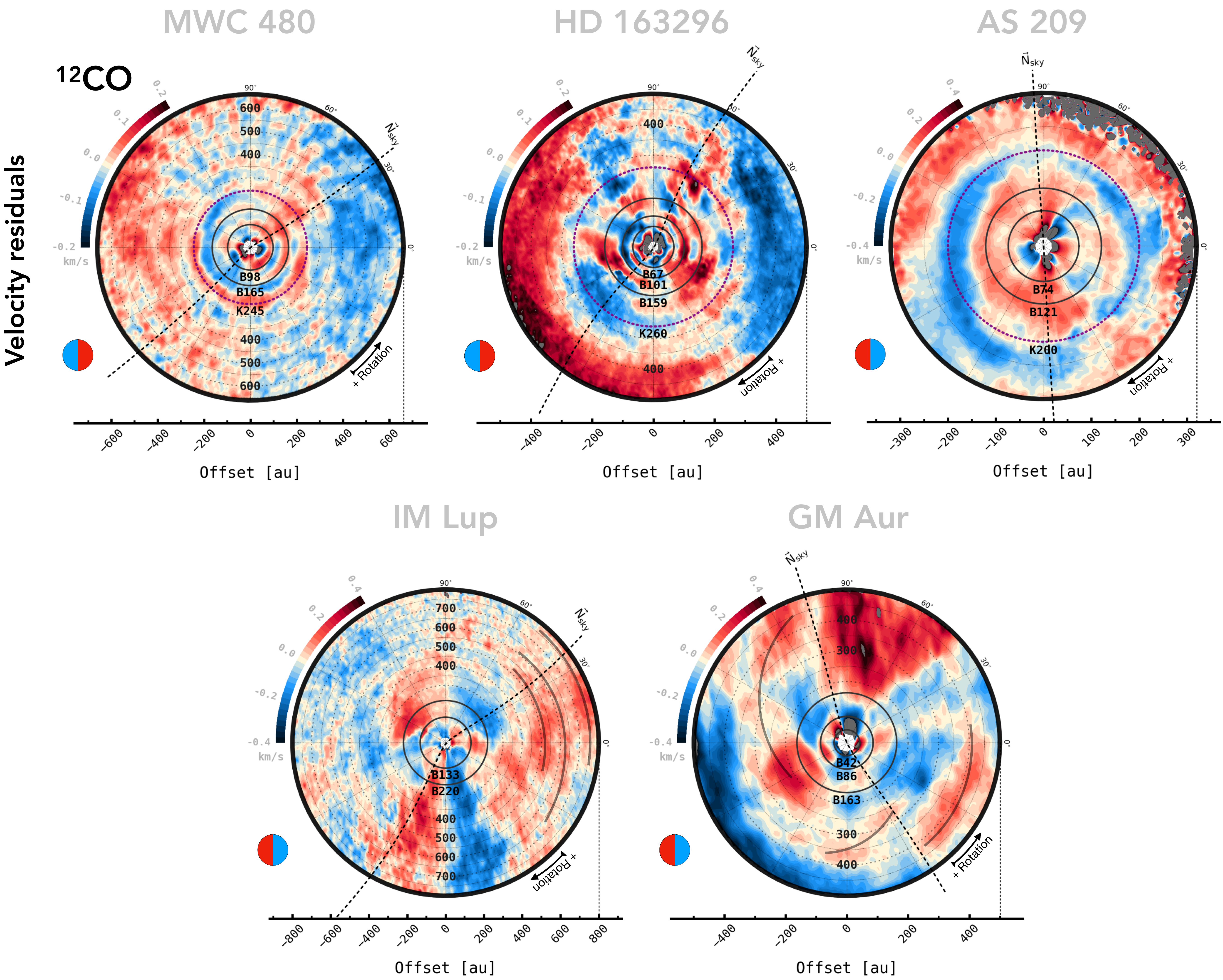}
    \caption{Centroid velocity residuals traced by \twCO{} emission for all discs, referred to each disc reference frame. The annotated solid lines mark the radial location of millimetre dust rings (denoted as Bxxx), and the purple dotted line that of the most prominent kink identified by eye in \twCO{} intensity channels (denoted as Kxxx). Also shown is the north-sky (or $\rm{PA}=0^\circ$) axis for reference. The oblique shape of this axis is due to the deprojected geometry of each disc vertical structure retrieved by the best-fit models. The 
    circle in the bottom left corner of each panel indicates the location of the blue-- and redshifted side for each disc. These maps illustrate the deviations from Keplerian rotation of the gas component located at high elevations ($0.2<z/r<0.4$) over the midplane.
              }
         \label{fig:residuals_kinematics_12co}
   \end{figure*}

In this Chapter, we present a gallery of the kinematic substructures identified in the discs of \mwc{}, \hd{}, \as{}, \im{}, and \gm{}, traced by \twCO{}, \thCO{}, and \eiCOfull{} line emission. 
More specifically, we report the location and magnitude of significantly localised perturbations, possibly driven by unseen massive planets, and comment on the plausible link between the presence of large-scale kinematic signatures and planet candidates proposed in recent literature. In Table \ref{table:planet_locations}, we summarise the relative location of the signatures attributed to the presence of embedded planets reported here and in other works. 

As was already manifest in the channel maps presented in Figs. \ref{fig:channel_maps_mwc480} and \ref{fig:channel_maps_hd163296}--\ref{fig:channel_maps_gmaur}, the five discs in the sample display intensity fluctuations at several places, especially in \twCO{} emission. Qualitatively, it is easy to note that in the disc of \mwc{}, for example, there are intensity kinks that span over multiple channels at different radial and azimuthal locations (see arrows in Fig. \ref{fig:channel_maps_mwc480}). In the disc of \hd{}, there seems to be fewer, yet very prominent kinks as originally reported by \citet[][]{pinte+2018b}. Likewise, the disc of \as{} appears highly perturbed and void around an annular region at a seemingly constant radial distance of $\sim\!200$\,au. At the same radius, a rather striking kink-like feature in the south of the sky is evident in channels near the systemic velocity, but it is apparent too that such a signature extends over almost the entire azimuth of the disc\footnote{Due to cloud absorption of the \twCOfull{} emission \citep{oberg+2011}, the blueshifted half of the disc of \as{}, to the west of the sky plane, is dimmer than the redshifted side.}. Nevertheless, despite being illustrative and guiding, the nature of these apparent signatures can only be revealed by quantitative analyses that provide a precise estimate of the location, magnitude, and extent of the underlying perturbations in the disc velocity field.

To this end, in the following sections we focus on the study of centroid velocity and line width residuals computed according to Sect. \ref{sec:observables}, and illustrated in Figures \ref{fig:residuals_kinematics_12co}, \ref{fig:residuals_kinematics_13co}, and \ref{fig:residuals_kinematics_c18o}, for 
\twCO{}, \thCO{}, and \eiCO{} velocities, for all discs, and in Figures 
\ref{fig:residuals_linewidth_12co}, \ref{fig:residuals_linewidth_13co}, and \ref{fig:residuals_linewidth_c18o} for line widths. We note that some of the prominent signatures identified by eye in intensity channels are now explicit and quantifiable in these residual maps. For instance, the seemingly extended kinks observed in \twCO{} channels of the disc around \mwc{} render as ring-like structures in centroid velocity residuals, revealing the presence of radially localised deviations from Keplerian rotation in this system. Also, it is now clear that the strong kink observed in the south of the \as{} disc appears to be the consequence of a large-scale, filament-like kinematic signature that spans over the entire azimuth of the disc, rather than due to a localised feature. In the following sections we assess how localised these velocity perturbations actually are across all discs and tracers in a quantitative manner.

\subsection{Localised velocity and line width perturbations} \label{sec:localised_perturbations}

Strong and localised velocity perturbations are expected in the vicinity of massive planets, that is, around and along spiral wakes triggered by the gravity of the planet and its hydrodynamic interaction with the host disc \citep[see e.g.][]{perez+2018, rabago+2021}. 
In \citetalias{izquierdo+2021}, we demonstrated that this fact can be readily exploited to pinpoint the azimuthal and radial location of embedded planets with remarkable precision. 
Our detection technique uses a clustering algorithm to spot those areas of the disc where the variance of the velocity (or line width) perturbations is significantly greater than that of the background velocity (or line width) fluctuations. To do so, we first split the disc radial extent into annuli with a width of a quarter of the beam size. Next, in each annulus, we perform an azimuthal scan to extract the peak residual and take record of its magnitude and azimuthal location. Finally, using a K-means algorithm, we group these peak residuals into clusters along both the radial and azimuthal coordinates of the disc. A cluster is classified as \textit{significant} when the spectral variance of the peak residuals within the cluster is above a threshold of 3$\sigma$ set by the variances of the other clusters. The region of the disc where the significant clusters intersect in azimuth and radius is attributed to a strong, localised, and coherent perturbation to the gas Keplerian velocity, possibly related to the presence of an unseen planet. In a quest for new planet candidates, we apply this same methodology to the velocity and line width residual maps of the five discs and CO isotopologues studied in this work.

\subsubsection{Detections in the disc of \hd{}: Velocity}

\begin{figure*}
   \centering
   \includegraphics[width=1.0\textwidth]{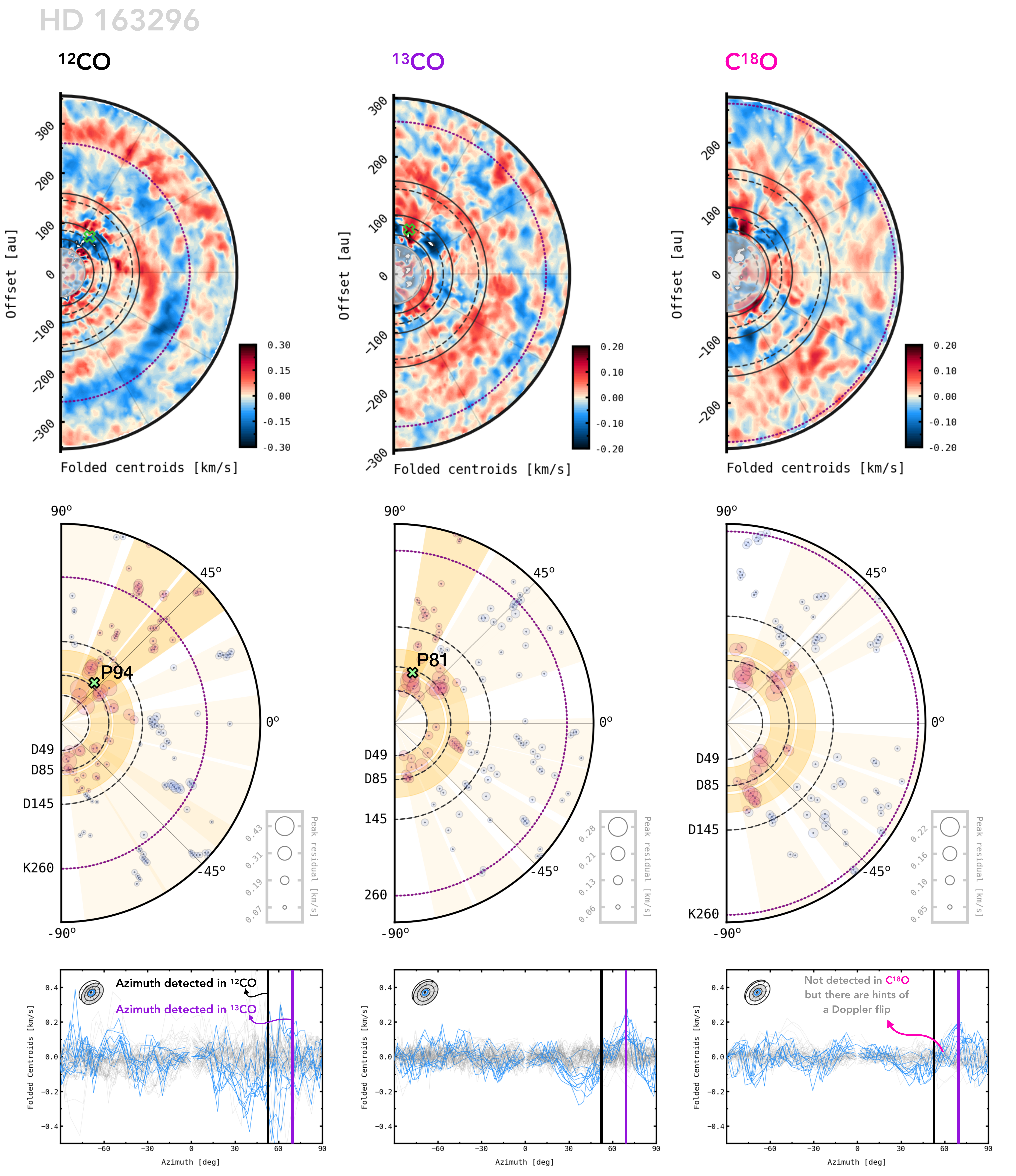}
     \caption{Analysis of localised velocity perturbations in the gas disc of \hd{}. \textit{Top row}: Folded velocity residuals computed for \twCO{} \thCO{}, and \eiCOfull{} lines. \textit{Middle row}: Azimuthal and radial extent of clusters identified by our detection algorithm. Regions in bold yellow highlight the accepted clusters of peak residuals whose variances are significantly greater than those in the background clusters. The weighted average of the 2D location of peak residuals within azimuthal \textit{and} radial clusters determines the centre of the localised velocity perturbations reported in the text. These are marked as green crosses for \twCO{} and \thCO{}. In \eiCO{}, only radial clusters were identified as localised by the technique and, therefore, no perturbation was tagged as detected in this tracer.
     \textit{Bottom row}: Azimuthal profiles of folded velocities (i.e. those in the top row) to highlight the blue-- redshifted, or Doppler flip, morphology of the perturbations detected in \twCO{} and \thCO{}, and of the tentative perturbation in \eiCO{}.
          }
        \label{fig:localised_perturbations_hd163296}
\end{figure*} 

In \citet{izquierdo+2022}, we applied this technique on DSHARP data \citep[][]{andrews+2018} of the \twCO{} disc around \hd{}, finding two significantly localised perturbations possibly driven by massive planets embedded in the disc at wide orbits: one at 94\,au, referred to as P94, and another at 261\,au, or P261. The latter is potentially related to the kink-like feature reported by \citet{pinte+2018b}, attributed in the same work to a planet twice the mass of Jupiter. Here, very similar results are obtained if the same analysis is carried out on the MAPS observations of this disc in the three CO isotopologues of interest.

\begin{figure*}
   \centering
   \includegraphics[width=0.75\textwidth]{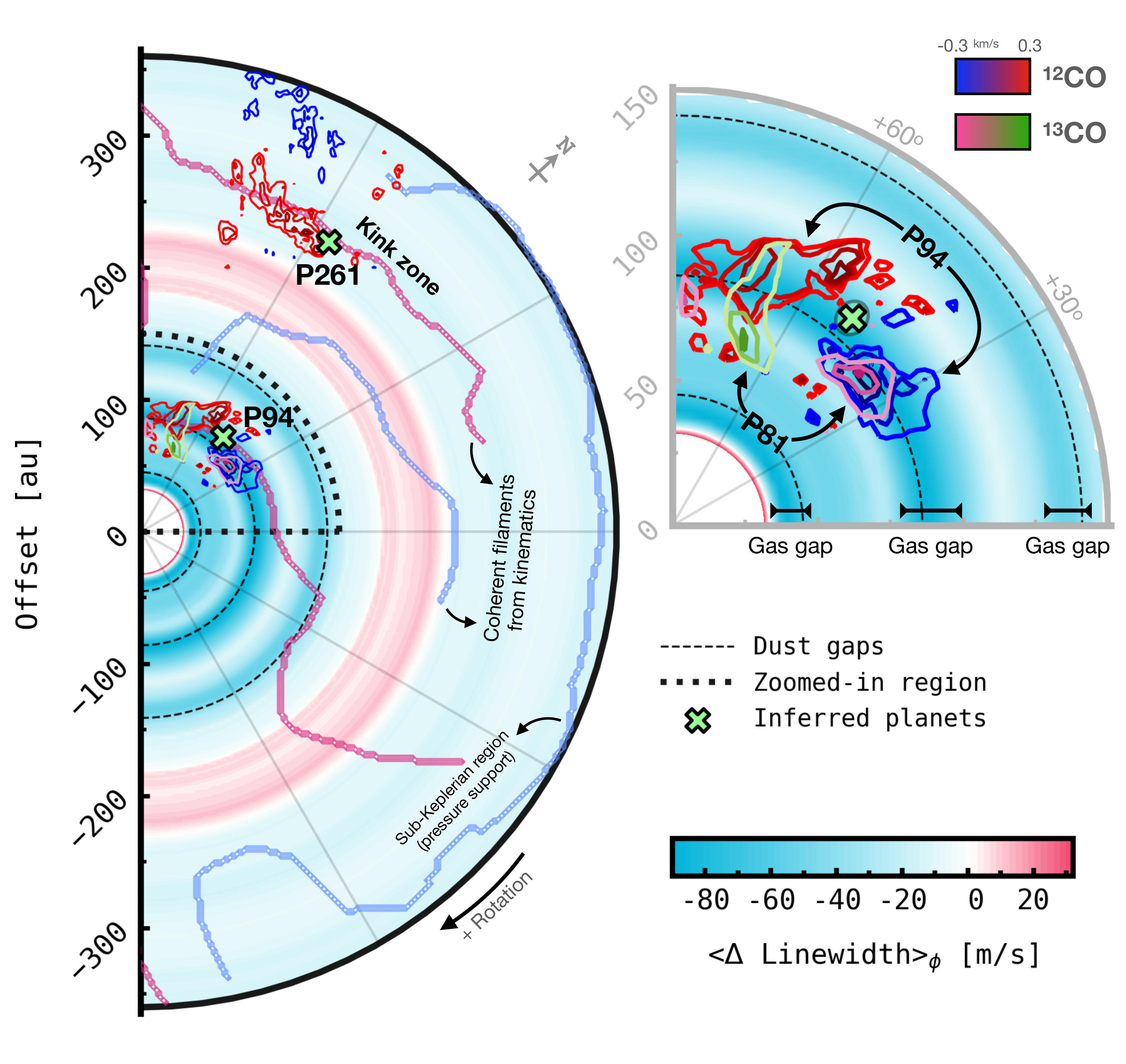}
     \caption{Summary of the localised velocity perturbations detected by our clustering algorithm in the disc of \hd{} in \twCO{} and \thCO{} velocities, using DSHARP and MAPS data. Localised red and blue contours highlight the morphology of the perturbations detected in \twCO{}, P94 and P261, whereas green and purple contours outline the localised signature detected in \thCO, P81, closely overlapping with the P94 signal. The green crosses mark the inferred location of the planet candidates linked to the aforementioned perturbations. Deep blue colours in the background highlight the location of \twCO{} line width minima which are co-spatial with positive gradients of azimuthal velocity flows probing gas surface density gaps.
     }
        \label{fig:summary_hd163296}
\end{figure*}

Figure \ref{fig:localised_perturbations_hd163296} illustrates the folded velocity maps computed from
\twCO{}, \thCO{}, and \eiCO{} centroid velocities. In the same Figure, we present the results of applying the aforementioned clustering algorithm to these maps, as well as azimuthal deprojections of the magnitude of the folded velocity residuals. Our algorithm detects two significantly localised velocity perturbations in different isotopologues, one in \twCO{} and another in \thCO{}. 
The \twCO{} perturbation, detected at an orbital radius of $R=93$\,au and an azimuth of $\phi=51^\circ$ in the disc reference frame, is in excellent agreement with the P94 perturbation found by \citet{izquierdo+2022}. The P94 convention will thus be kept throughout the work. 
The \thCO{} perturbation, or P81 hereafter, is detected at $R=81$\,au and $\phi=70^\circ$, namely interior to the D86 dust gap, and centred at an azimuth closer to the disc minor axis than that of the P94 perturbation. This is illustrated in Figure \ref{fig:summary_hd163296}, where we overlay contours of the velocity perturbations detected in \twCO{} and \thCO{}, along with more extended kinematic residuals for reference. Azimuthally averaged line widths are coloured in the background to highlight the location of gas gaps traced by line width minima (see Sect. \ref{sec:gas_substructure}). Even though the centres of the \twCO{} and \thCO{} perturbations appear distant from each other, we note that both the blue-- and the redshifted parts of these 
signatures overlap in their azimuthal extent, suggesting that they are likely connected to a common origin. 

While the \twCO{} perturbation yields a folded magnitude of $\sim\!0.4$\,km\,s$^{-1}$, the \thCO{} perturbation peaks at $\sim\!0.3$\,km\,s$^{-1}$ velocities. In \eiCO{}, the clustering algorithm does not detect any significantly localised signature. However, it is not less interesting as it also displays a Doppler flip pattern similar to that of \thCO{}, at the same location. The folded magnitude of this perturbation is weaker, around $\sim\!0.2$\,km\,s$^{-1}$. Taken together, it is possible that we are witnessing for the first time the structure of a planet-driven perturbation traced at different scale heights in a disc. The varying magnitude of the tentative and the two detected perturbations could be indicating that different velocity components are being probed as one goes from layers in the disc atmosphere to regions closer to the midplane. The morphology of the Doppler shift pattern suggests that the radial component of the perturbation is dominant on layers closer to the disc midplane.

\subsubsection{Detections in the disc of \hd{}: Line width}
\label{sec:detected_linewidths}

\begin{figure*}
   \centering
   \includegraphics[width=1.0\textwidth]{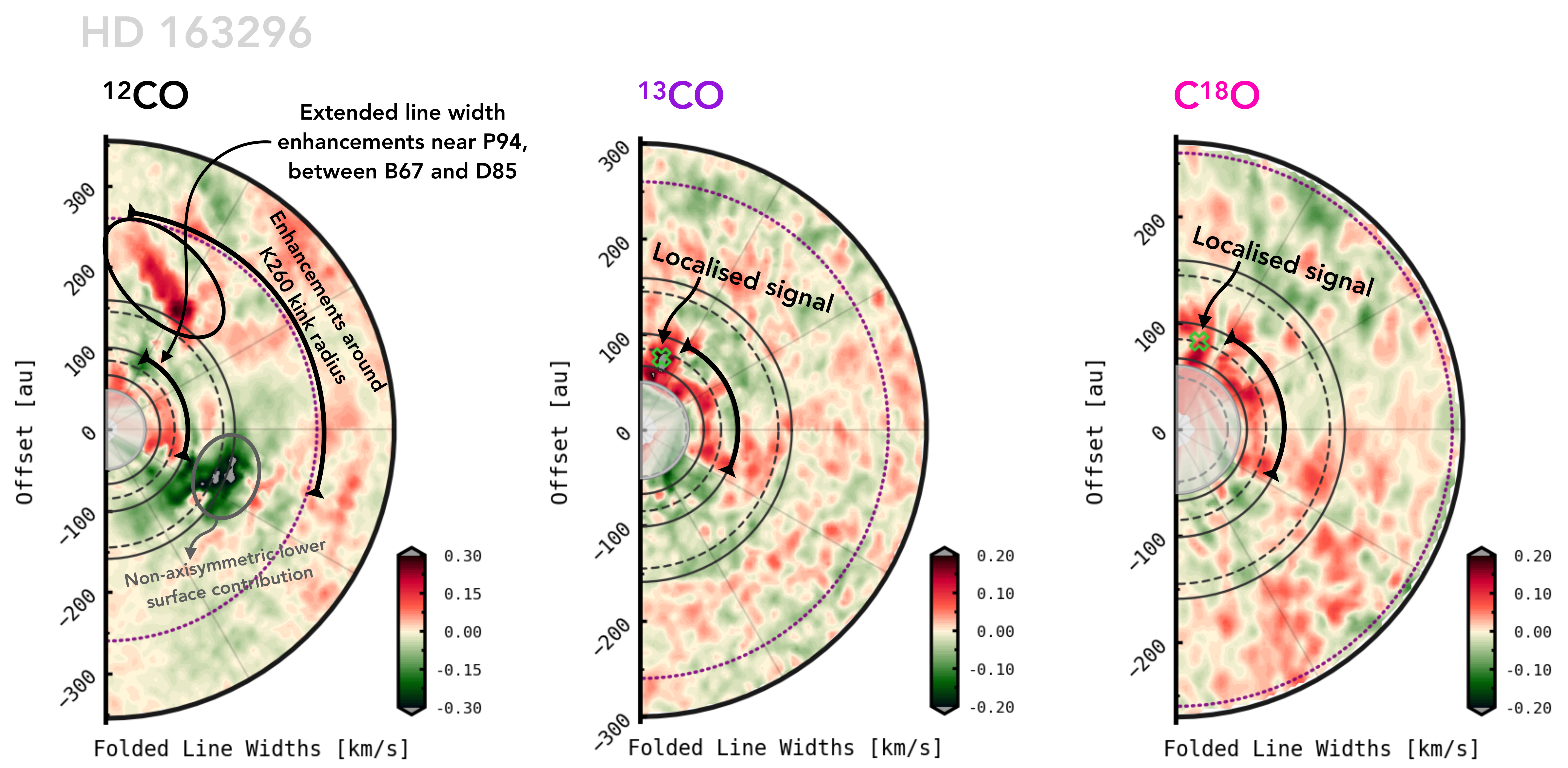}
     \caption{Folded line width residuals to highlight asymmetric line width enhancements near the orbit of the planet candidate P94, as well as the location of localised line width perturbations detected by our clustering algorithm in \thCO{} and \eiCO{}.
     }
        \label{fig:folded_linewidths_hd163296}
\end{figure*} 

In addition to producing organised velocity perturbations to the Keplerian rotation of the gas disc, massive embedded planets can also trigger turbulent motions in the neighbouring fluid \citep[][]{diskdynamics+2020, pinte+2022}. These non-thermal motions translate into broadening of the line intensity profiles on top of the local thermal broadening, and are expected to be observable if the driving planet is massive enough \citep[][]{dong+2019}. In this Section, we report the detection of localised line width enhancements in the disc of \hd{}. Line width perturbations identified as elongated substructures in the other discs of the sample are discussed in Sect. \ref{sec:extended_perturbations}.

In Figure \ref{fig:folded_linewidths_hd163296}, we present folded line width residuals for the disc of \hd{} traced by \twCO{}, \thCO{} and \eiCO{}. In all cases, we identify positive non-axisymmetric line width fluctuations of the order of $0.1$\,km\,s$^{-1}$, near the orbit of the planet candidate associated with the P94 velocity perturbation found in \twCO{}. These line width signatures extend for about $120^\circ$ between the radial location of the dust ring, B67, and the dust gap, D85. Moreover, our clustering algorithm detects two significantly localised perturbations in the line widths of \thCO{} and \eiCO{}, both associated with strong positive residuals. The centre of the \thCO{} perturbation is detected at $R=79$\,au, $\phi=74^\circ$, and that of the \eiCO{} perturbation is detected at $R=86$\,au, $\phi=75^\circ$. We note that the location of these perturbations broadly coincides with the centre of the localised velocity perturbation, P81, detected in the kinematics of \thCO{} as reported in Sect. \ref{sec:localised_perturbations}. These localised line widths further reinforce the hypothesis of the presence of a giant planet associated with the P94 velocity perturbation, at the radial and azimuthal location proposed by \citet[][]{izquierdo+2022}, and represent the first robust detection of localised line broadening features potentially driven by a massive planet embedded in a disc. 
Although not significantly localised, we also identify \twCO{} line width enhancements at larger radial distances, near the planet candidate P261 associated with the kink-like feature, K260, which include a strong $\sim\!0.2$\,km\,s$^{-1}$ signature between $R=160-260$\,au and $45^\circ<\phi<90^\circ$, and weaker $\sim\!0.05$\,km\,s$^{-1}$ fluctuations around the proposed planet orbit around $R=260$\,au, spanning over a much larger azimuthal extent.

\subsubsection{Non-detections}

\begin{figure*}
   \centering
   \includegraphics[width=0.9\textwidth]{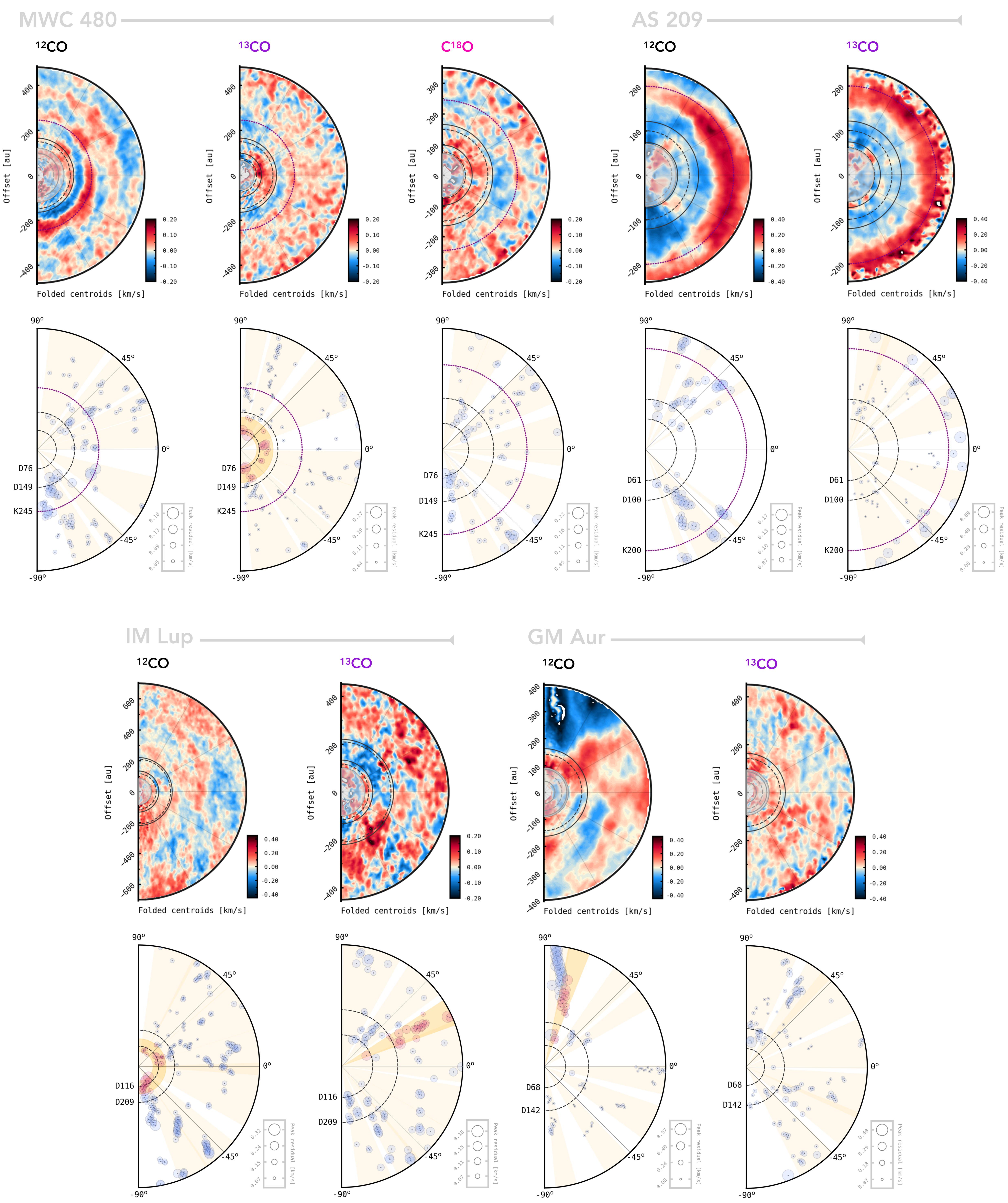}
     \caption{Folded velocity residuals for the discs where no localised velocity perturbations are identified by our clustering algorithm. 
     }
        \label{fig:folded_velocities_nondetections}
\end{figure*}

Figure \ref{fig:folded_velocities_nondetections} presents folded velocity maps for the other discs where our clustering algorithm does not detect any localised perturbation simultaneously in the radial and azimuthal coordinate, in any of the CO isotopologues. We remind the reader that these folded velocity residuals highlight asymmetric substructures in the azimuthal component of the velocity perturbations on one side of the disc with respect to the other, split by the disc minor axis as projected on the sky. These folded residual maps also highlight the presence of radial and/or vertical velocity fields regardless of their symmetry around the disc minor axis. 

As illustrated in the second and fourth rows of Fig. \ref{fig:folded_velocities_nondetections}, the asymmetric, or non-azimuthal, kinematic features in these discs appear to span across large spatial extents. In the discs of \mwc{} and \as{}, these velocity perturbations display as coherent filamentary structures, localised in the radial direction, and with azimuthal extents that comprise almost the entire angular coordinate. Similarly, for the disc of \im{} we report asymmetries in the form of extended filamentary structures, although weaker in magnitude than in the previous cases. No significantly localised perturbations associated with the presence of massive planets are found in this disc. Finally, in the disc of \gm{}, velocity perturbations with an elongated morphology are also found. In \twCO{}, a strong asymmetry can be seen in the northern part of the system due to the interaction of the uppermost layers of the disc with material possibly infalling from a remnant envelope. These signatures, classified as extended kinematic substructures, will be discussed further in Sect. \ref{sec:extended_perturbations} for each disc.

\setlength{\tabcolsep}{0.5pt} 

\begin{table*}
\centering
{\renewcommand{\arraystretch}{1.9}
\caption{Location of potential planet-driven signatures in discs reported in this and in previous works based on gas velocity perturbations and substructures, assuming that the planet candidate is in the disc midplane.} \label{table:planet_locations}

\begin{tabular}{ ccccccc } 

\toprule
\toprule
\multirow{2}{*}{Planet} & \multicolumn{4}{c}{Location} & \multirow{2}{*}{Description of Signature} & \multirow{2}{*}{Reference / Data from $^{\rm id}$} \\
& R\,[au] & $\phi$\,[$^\circ$] & R\,[''] & PA\,[$^\circ$] & &  \\

\midrule


\hd{}\,c & 94 & 50 & 0.77 & 352.1 & Localised velocity perturbation in \twCO{} & \citet{izquierdo+2022} / DSHARP $^{\rm a}$ \\
& 93 & 51 & 0.76 & 354.7 & Localised velocity perturbation in \twCO{} & This work (Fig. \ref{fig:localised_perturbations_hd163296}) / MAPS $^{\rm b}$ \\
& 83 & -- & -- & -- & Surface density gap traced by \eiCO{}  & \citet{teague+2018a} / \citet{isella+2016} $^{\rm c}$ \\
& 81 & 70 & 0.60 & 16.8 & Localised velocity perturbation in \thCO{}. & This work (Fig. \ref{fig:localised_perturbations_hd163296}) / MAPS \\ 
& 79 & 74 & 0.58 & 22.0 & Localised line width enhancement in \thCO{} & This work (Fig. \ref{fig:folded_linewidths_hd163296}) / MAPS \\
& 86 & 75 & 0.61 & 22.7 & Localised line width enhancement in \eiCO{}  & This work (Fig. \ref{fig:folded_linewidths_hd163296}) / MAPS \\
& 67--85 & -- & -- & -- & Extended line width enhancements in all COs  & This work (Fig. \ref{fig:folded_linewidths_hd163296}) / MAPS \\

\hd{}\,b & 260 & 54* & 2.20 & 357.0 & Kink(s) in \twCO{} intensity channels & \makecell[c]{\citet{pinte+2018a,pinte+2020} / \citet{isella+2016} \\ \citet{calcino+2022} / MAPS } \\
 & 261 & 57 & 2.06 & 359.4 & Localised velocity perturbation in \twCO{} & \citet{izquierdo+2022} / DSHARP \\
& 260--300 & -- & -- & -- & Extended velocity perturbation in \twCO{} & \makecell[c]{\citet{teague+2021} / MAPS \\
\citet{izquierdo+2022} / DSHARP} \\
& 250--320 & -- & -- & -- & Extended line width perturbations in \twCO{} & This work (Fig. \ref{fig:folded_linewidths_hd163296}) / MAPS \\

\midrule

\mwc{}\,b & 245 & -- & -- & -- & Extended vertical flow, buoyancy spiral & \makecell[c]{\citet{teague+2021} / MAPS \\ This work (Figs. \ref{fig:polar_deproj_mwc480}, \ref{fig:averaged_residuals_mwc480}) / MAPS} \\
& 245 & $-128$ & 1.33 & 192.6 & Peak line width enhancement in \twCO{} & This work (Fig. \ref{fig:polar_deproj_mwc480}) / MAPS \\ \midrule

\as{}\,b & 203* & $-102$* & 1.40 & 161.0 & Localised intensity feature in \thCO{} & \citet{bae+2022} / MAPS \\
 & 210 & $-110$ & 1.44 & 151.3 & Peak line width enhancement in \twCO{} & This work (Fig. \ref{fig:polar_deproj_as209}) / MAPS \\
 & 160--200 & -- & -- & -- & Extended high line widths in \twCO{} and \thCO{} & This work (Fig. \ref{fig:polar_deproj_as209}) / MAPS \\ 
 & 200 & $-150$, $-60$ & -- & -- & Bending of iso-velocities in \twCO{} & This work (Figs. \ref{fig:polar_deproj_as209}, \ref{fig:skeleton_as209}) / MAPS \\

\bottomrule

\end{tabular}
\caption*{\textbf{Note.} Radial locations and azimuths given in au and degrees, respectively, are referred to the disc reference frame. Radial separations and position angles given in arc-seconds and degrees are referred to the sky plane. The origin of sky coordinates is offset from the disc centre according to the model ($x_c$, $y_c$) values reported in tables \ref{table:pars_mwc480}--\ref{table:pars_gmaur} for each disc. The disc orientation parameters used to convert coordinates from one reference system to the other are also summarised in said tables. $^*$Value converted from sky coordinates reported in the referenced work to disc coordinates using orientation parameters retrieved in this work. \textit{Project IDs} -- $^{\rm a}$(2016.1.00484.L; PI: S. Andrews). $^{\rm b}$(2018.1.01055.L; PI: K. \"Oberg). $^{\rm c}$(2013.1.00601.S; PI: A. Isella).  
}

  }
\end{table*}

\subsection{Extended velocity and line width perturbations} \label{sec:extended_perturbations}

Embedded planets can also trigger extended velocity perturbations in the gas disc in the form of axisymmetric and non-axisymmetric substructures. Axisymmetric signatures driven by planets are primarily related to pressure modulations induced by these objects in the host disc. If massive enough, the gravitational interaction of a planet exerts a torque onto the disc fluid around its vicinity, which would in turn carve a density gap whose width and depth depend on the planet mass and on the local viscosity of the fluid. These pressure variations translate into ring-like deviations from Keplerian rotation in the azimuthal component of the disc velocity field \citep[][]{kanagawa+2015, teague+2018a}. It has also been demonstrated that gas gaps can lead to meridional circulation of material from the atmosphere into the disc midplane and vice-versa, which is observable through inspection of the vertical component of the velocity field with respect to the disc reference frame \citep{morbidelli+2014, teague+2019nat, yu+2021}. The location and magnitude of these axisymmetric velocity perturbations linked to the presence of gas substructures are summarised in Sect. \ref{sec:gas_substructure}.

Non-axisymmetric planetary signatures in the kinematics are mainly associated with spiral-like perturbations of two types. Spiral waves triggered by the gravitational potential of the planet at Lindblad resonances, or simply Lindblad spirals, and spirals driven at buoyancy resonances, or buoyancy spirals, 
which develop when the vertical temperature gradient of the gas disc is positive and its adiabatic index is larger than one, so long as the buoyancy frequency of the fluid at the planet location matches with the planet orbital frequency \citep[][]{zhu+2012, lubow+2014}. As demonstrated by \citet[][]{bae+2021}, buoyancy spirals are more prone to develop in the outer disc at relatively high elevations over the midplane where the thermal relaxation times are low due to infrequent gas--dust collisions. These two substructures differ in that Lindblad spirals carry mainly azimuthal and radial velocity perturbations, while vertical motions predominate in buoyancy spirals. \citet[][]{bae+2021} also demonstrate that the pitch angle of buoyancy spirals is generally smaller than that of Lindblad spirals, especially near the radial location of the perturbing planet. Another distinguishing feature between the two signatures is that the magnitude of the velocity perturbations in Lindblad spirals decreases with increasing height over the disc midplane, while it increases with increasing height in buoyancy spirals.  \\

\begin{figure*}
   \centering
     \includegraphics[width=0.85\textwidth]{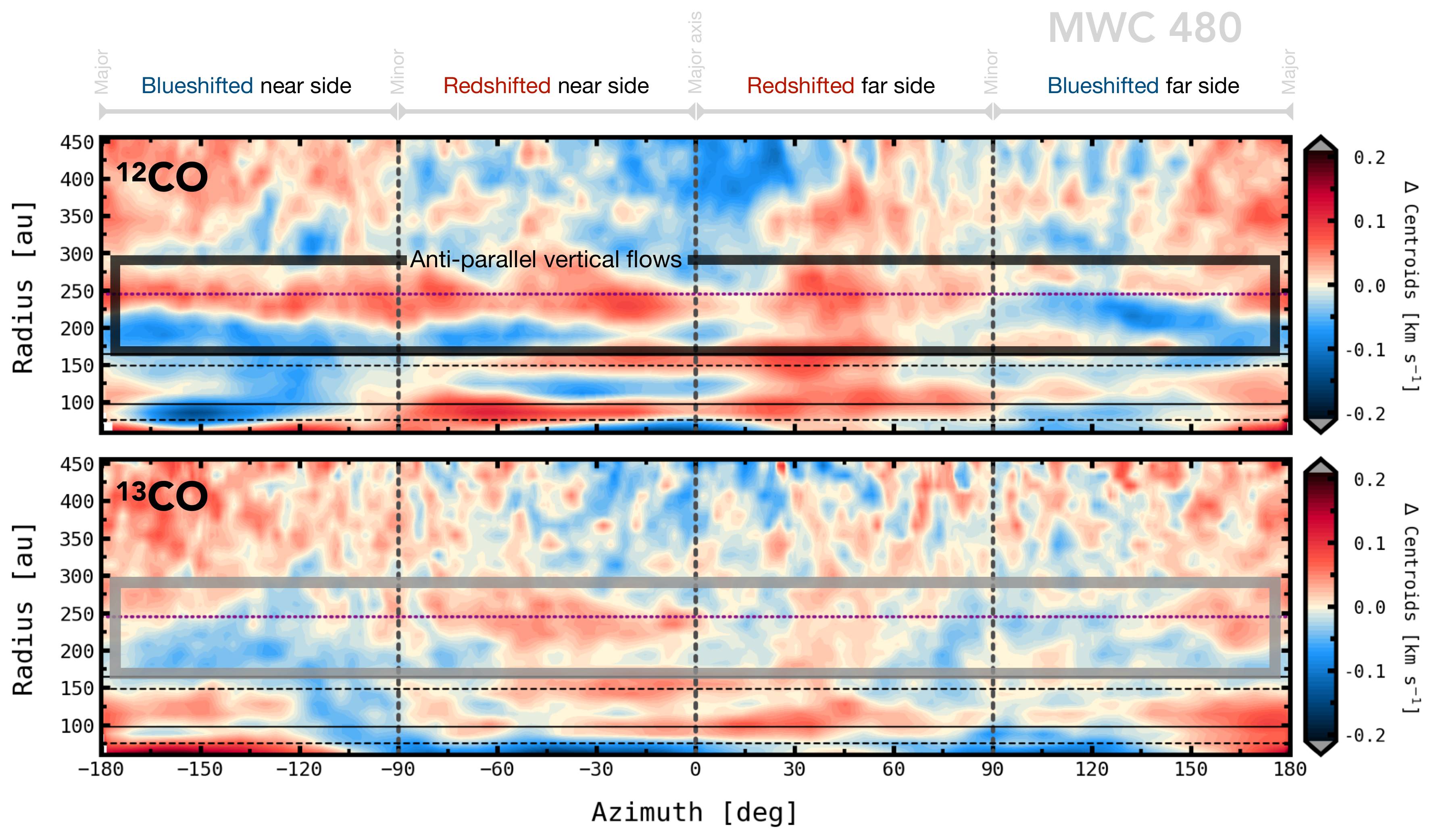}
     \includegraphics[width=0.85\textwidth]{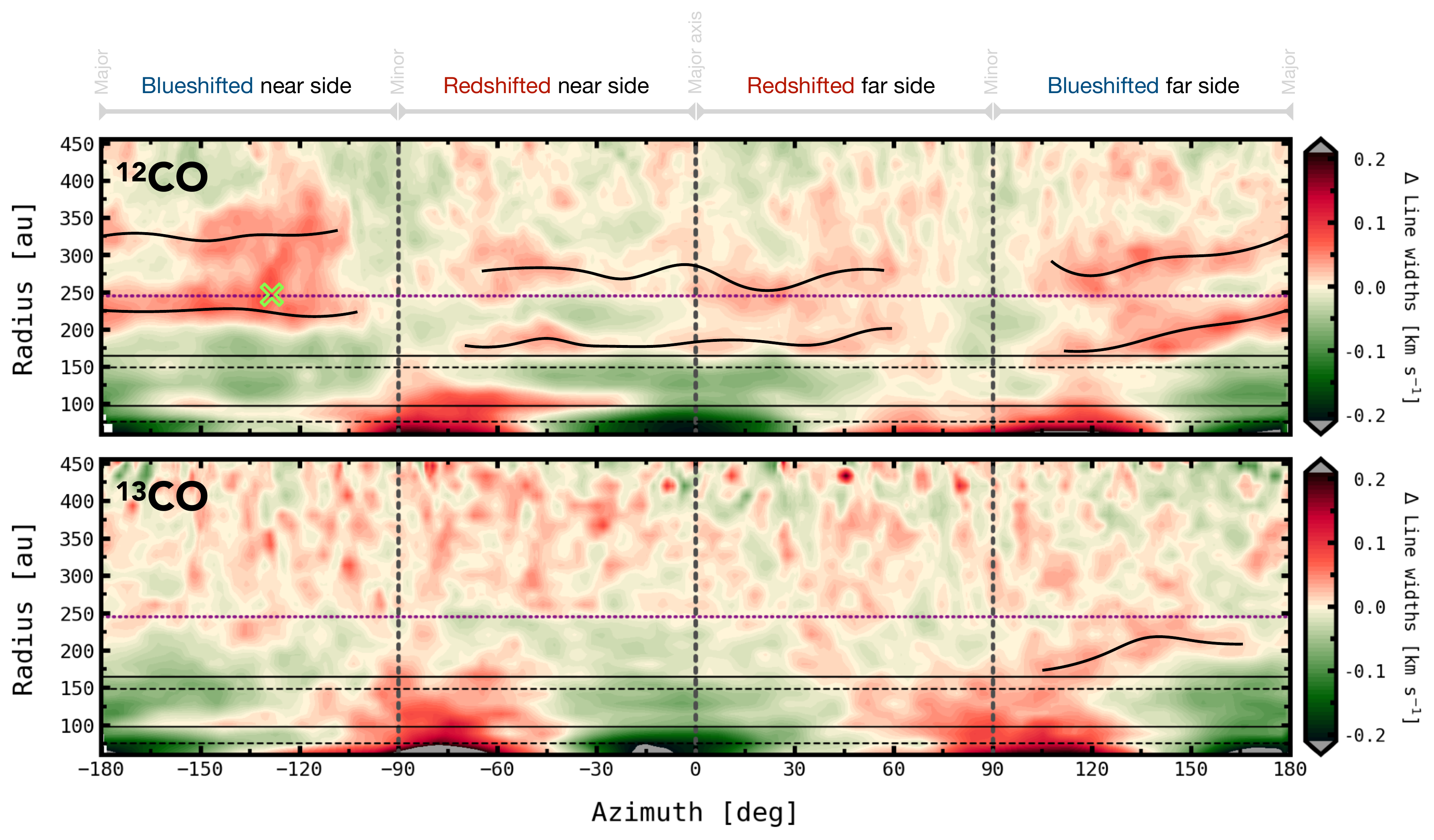}

      \caption{Azimuthal deprojection of velocity and line width residuals obtained for the disc of \mwc{} as observed in \twCO{} and \thCO{}. The horizontal dashed and solid lines indicate the radial location of dust gaps and rings, respectively. The purple dotted line is the radial location of kink-like features observed in the datacube channels. The vertical lines denote the azimuthal location of the disc main axes as projected on the sky. The black lines overlaid in the line width residuals highlight the location of extended line width enhancements, likely associated to non-thermal motions. The green cross marks the location of the planet candidate reported in Sect. \ref{sec:extended_perturbations} based on \twCO{} line width enhancements. Cartesian versions of these maps can be found in Figs. \ref{fig:residuals_12co_mwc480} and \ref{fig:residuals_kinematics_13co}.
              }
         \label{fig:polar_deproj_mwc480}
\end{figure*}

\noindent\textit{\mwc{}}. In Figure \ref{fig:polar_deproj_mwc480}, we show polar deprojections of velocity and line width residuals obtained in \twCO{} and \thCO{} for the disc around \mwc{}. As illustrated in the top panel, the \twCO{} disc exhibits strong and azimuthally extended $\sim\! 50$\,m\,s$^{-1}$ velocity perturbations that seem to be concentric and confined to narrow $\sim\!30-50$\,au radial sections. 
As we subsequently demonstrate in Sect. \ref{sec:gas_substructure} through the analysis of azimuthal averages of velocity residuals,
the velocity perturbations spanning between the dust ring B165 and $R\sim\!300$\,au, enclosed by rectangles in Fig. \ref{fig:polar_deproj_mwc480}, represent contiguous anti-parallel vertical flows of material in the atmosphere of the \mwc{} disc. One of these meridional flows points upwards with respect to the disc midplane, at $R=193$\,au, and the other downwards, at $R=245$\,au. The azimuthal coherence of these substructures suggests that vertical motions must indeed be dominant in this region of the disc. However, the fact that they are not fully axisymmetric indicates that radial and/or azimuthal perturbations to the gas velocities are also present.

Another interesting feature worthy of highlighting in this region of the disc is that the strong vertical perturbations observed in \twCO{} are either vanished or loosely evident in \thCO{}, as illustrated in the second top panel of Fig. \ref{fig:polar_deproj_mwc480}. The large difference between the heights traced by \twCO{} and \thCO{} may be the reason of this modulation. While \twCO{} is located at a $z/R\sim0.2$, both \thCO{} and \eiCO{} barely reach $z/R\sim0.1$ scale-heights (Fig. \ref{fig:attributes_all_co}, bottom row). At these altitudes, it is foreseeable that other velocity components may become dominant, making the magnitude and patterns of the kinematic signatures look different to those in \twCO{}. This result supports the hypothesis of \citet[][]{teague+2021}, who proposed that the meridional velocities observed in \twCO{} alongside the tightly wound morphology of peak intensity substructures could be the result of buoyancy spirals driven by a planet at $R=245$\,au in this disc. As mentioned earlier, \citet[][]{bae+2021} demonstrated that the magnitude of the vertical motions across the vertical extent of these type of spirals is expected to decrease as the disc midplane is approached, which is in line with our finding. 

Furthermore, as illustrated in the bottom panels of Fig. \ref{fig:polar_deproj_mwc480}, we detect line width enhancements of the order of $\sim\!20-50$\,m\,s$^{-1}$, close to the interfaces between the aforementioned vertical flows, between $\sim\!175-325$\,au. These enhancements appear more strongly on the blueshifted side of the disc, and again mainly at elevated layers traced by \twCO{}. Interestingly, these substructures are not as extended nor as symmetric in azimuth as their velocity counterparts at similar radii. In fact, they seem to open linearly to wider orbits as a function of azimuth on the blueshifted far side, as if they were part of a spiral structure. The radial coincidence of these line width increments possibly driven by radially localised turbulent motions, and the vertical flows detected in velocity residuals is yet another signature that supports the presence of a massive planet in this region of the disc. Line width enhancements have recently been reported as potential indirect signatures of planets in the discs of \hd{} \citep[][]{izquierdo+2022}, and TW\,Hydrae \citep[][]{teague+2022}.
Indeed, using 3D hydrodynamics and radiative transfer simulations, \citet[][]{dong+2019} showed that a massive planet can trigger turbulent vertical motions on and around its orbit, which should be observable by quantifying the broadening of line profiles, as long as the planet mass and the angular and spectral resolution of the observations are high enough. Considering this and the fact that the vertical velocity signatures are in good agreement with those expected for a planet-driven buoyancy spiral, we propose a planet candidate in the disc of \mwc{} at an orbital radius of $R=245$\,au, namely at the radial location proposed by \citet[][]{teague+2021}, but at an azimuth of $\phi=-128^\circ$, where the peak in \twCO{} line width residuals is found. The location of this planet candidate is marked as a green cross in Fig. \ref{fig:polar_deproj_mwc480}, bottom panel. See Fig. \ref{fig:residuals_12co_mwc480}, top panel, for a Cartesian view of the line width perturbations in this disc.

Additionally, to understand how the retrieved velocity perturbations vary with the assumed vertical structure, we take this source as an example to investigate the impact of using an irregular emission surface on the retrieved velocity residuals. For this experiment, we fix the \twCO{} model surface to the non-parametric emitting layer extracted by \citet{paneque+2022}, and re-run \discminer{} to find a new set of parameters for the remaining attributes summarised in Table \ref{table:attributes_parameters}. For better comparison, we also fix the disc inclination in both the smooth and irregular-surface model to the value found by \citet{liu+2019} from mm continuum modelling. As illustrated in Figure \ref{fig:surface_residuals_mwc480}, we find that using an irregular surface leads to higher intensity and velocity residuals overall. In particular, we note that the irregular-surface model displays quadruple patterns in the velocity residuals at large radii ($R>400$\,au), but also around more specific radial separations (e.g. at $R=300$\,au and $R=400$\,au). The former suggests that the surface tapering found by the geometrical, non-parametric method is possibly over-estimated due to the influence of the increasingly strong sub-Keplerian rotation at large radial separations. The latter, on the other hand, may indicate that some of the vertical bumps and troughs retrieved by the geometrical method are instead modulated by the radially confined velocity perturbations present in this disc, and not the consequence of actual variations in the disc surface density. Hence, we only consider smooth parametric surfaces found by \discminer{} for the rest of the analysis. \\

\noindent\textit{\hd{}}. In the disc of \hd{} there are particularly prominent extended perturbations, most of them driven by pressure modulations due to the presence of gas gaps which trigger deviations from Keplerian rotation in the azimuthal component of the velocity field \citep[see e.g.][]{teague+2018a, zhang+2021}. These azimuthal fluctuations are seen as ring-like structures with a sign flip around the disc minor axis owing to the projection of the velocity field along the line of sight of the observer (see Fig. \ref{fig:residuals_kinematics_12co}, top middle panel). As illustrated in Fig. 5 of \citet[][]{teague+2019}, super-Keplerian azimuthal velocities would appear redshifted on the redshifted side of the disc, and blueshifted on the blueshifted side, and the other way around for sub-Keplerian velocities. Strong sub-Keplerian azimuthal velocities are also seen at large radii ($R>350$\,au) induced by a sharp pressure gradient possibly driven by the tapered shape of the surface density profile expected near the edge of the disc \citep[][]{dullemond+2020, zhang+2021}. Additionally, there is an elongated spiral-like perturbation around an orbital radius of $R=260$\,au spanning over a large azimuthal extent between $\phi\!\sim\!90^\circ$ to $-60^\circ$, which has been recently linked to a massive planet at the same radius \citep[][]{teague+2021, izquierdo+2022, calcino+2022}, originally proposed by \citet[][]{pinte+2018a} after observation of an intensity kink in \twCO{} channels. Interestingly, this elongated signature is more strongly seen at high elevations traced by \twCO{} emission. In \thCO{}, the perturbation is just barely seen and hindered by the background noise, whereas in \eiCO{} the azimuthal component of the perturbation seems to dominate, suggesting that this spiral is also likely triggered by buoyancy resonances as in the disc of \mwc{}, and that the perturbations near the disc midplane could be associated with a shallow density gap carved by the planet candidate. \\

\begin{figure*}
    \centering
     \includegraphics[width=0.85\textwidth]{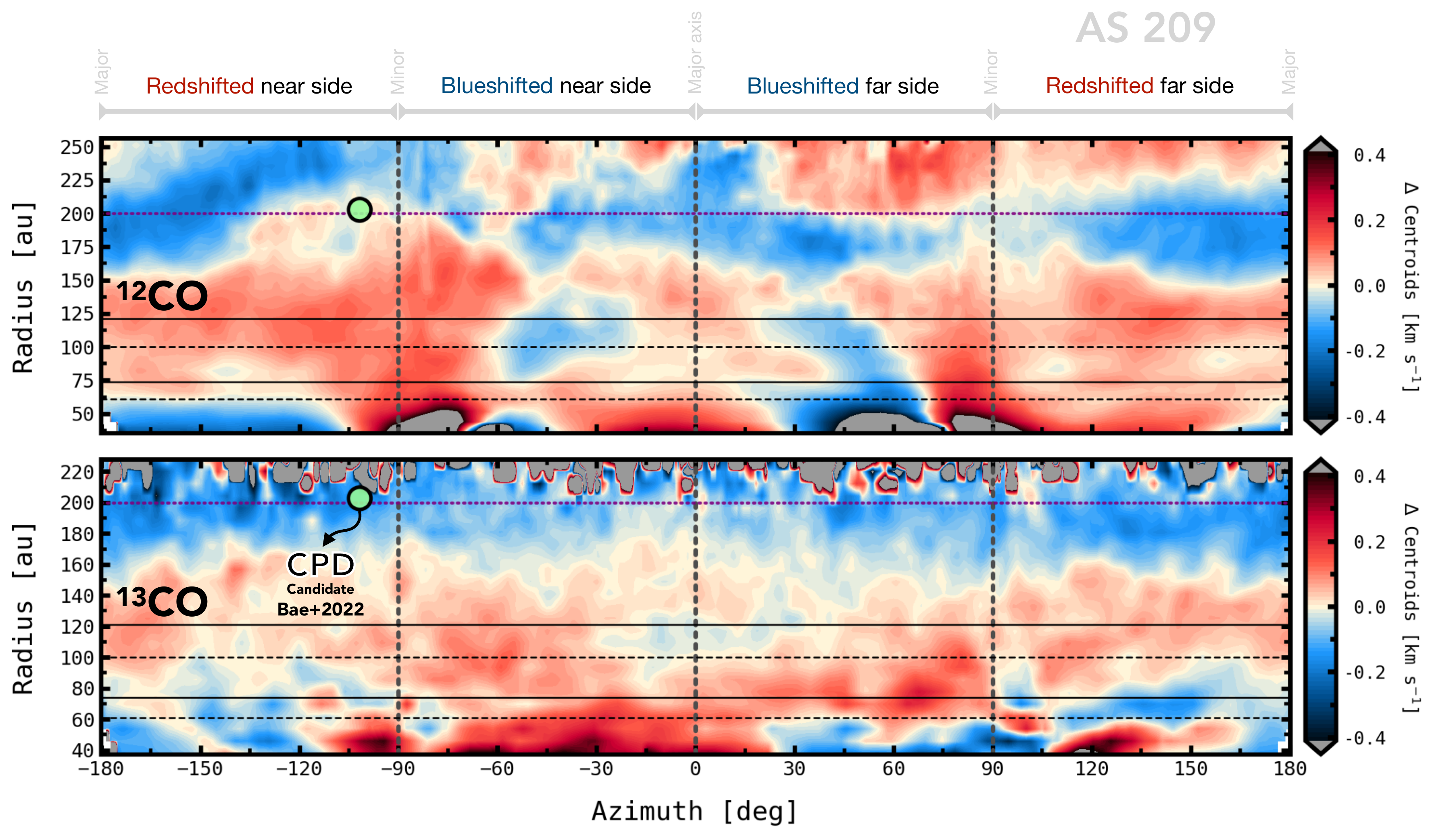}
     \includegraphics[width=0.85\textwidth]{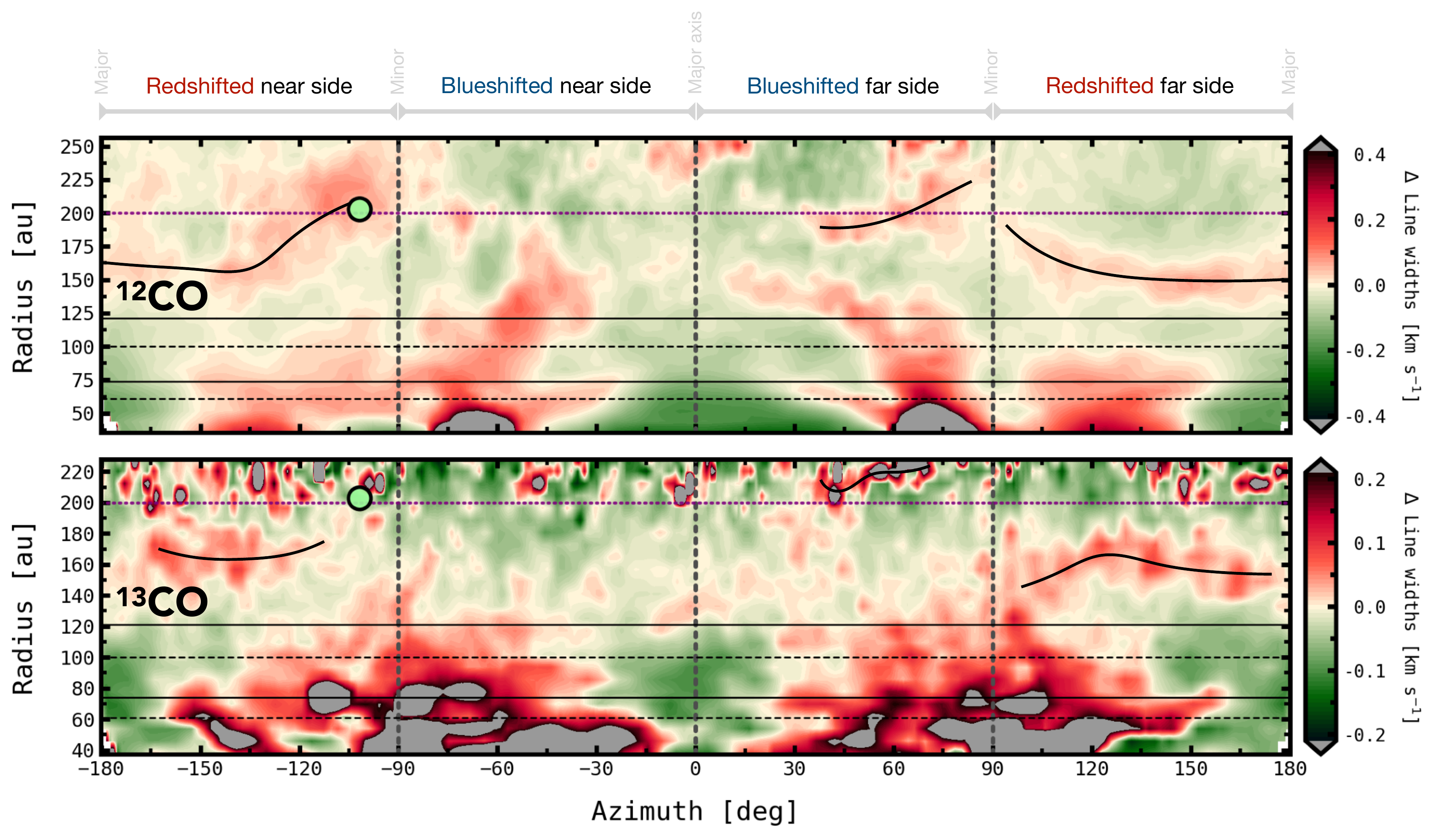}
      \caption{As Fig. \ref{fig:polar_deproj_mwc480} but for the disc of \as{}. The green circle marks the location of the CPD proposed by \citet[][]{bae+2022}. Cartesian versions of these maps can be found in Figs. \ref{fig:residuals_kinematics_12co}, \ref{fig:residuals_kinematics_13co} and \ref{fig:residuals_linewidth_12co}.
              }
         \label{fig:polar_deproj_as209}
\end{figure*} 

\noindent\textit{\as{}}. The disc of \as{} displays downward and upward vertical flows around contiguous annuli centred at $R\!\sim\!130$\,au and $R\!\sim\!190$\,au, respectively. As illustrated in Figure \ref{fig:polar_deproj_as209}, and as opposed to \mwc{}, the vertical flows in this disc are clearly seen in \twCO{} and \thCO{} simultaneously, meaning that these motions span coherently across large vertical extents and are therefore less compatible with planet-driven buoyancy spirals.
Instead, 
non-planetary mechanisms such as the vertical shear instability, which can trigger vertical motions across a wide range of scale heights \citep[see e.g.][]{barraza-alfaro+2021}, or magnetically driven winds via ambipolar diffusion, studied by 
\citet{galloway+2023} for this specific source, are favoured in this case.
 
Even though we do not detect any localised velocity perturbation in the disc of \as{} near the CPD proposed by \citet{bae+2022} at a radius of $R=203$\,au, we note that the azimuthally coherent morphology of the large-scale upward velocity flow traced by \twCO{}, at $R\!\sim\!190$\,au, is disrupted around the azimuth of the planet candidate where it extends to wider orbits forming an arc-like pattern that surrounds the suggested location of the planet. Furthermore, we spot a prominent $\sim\!50$\,m\,s$^{-1}$ line width enhancement in \twCO{} at the same location of the CPD candidate that supports the presence of this object. As illustrated in Fig. \ref{fig:polar_deproj_as209}, bottom panels, this feature is $\sim\!50$\,au wide around the vicinity of the CPD, and becomes narrower at larger azimuthal separations, extending coherently for almost $180^\circ$ near the CPD orbit.
The same azimuthally extended substructure is found in \thCO{} line widths between $R=140-180$\,au radii. However, the line width bump observed in \twCO{} around the immediate vicinity of the CPD no longer remains in this tracer, suggesting that the enhanced velocity dispersion near the planet candidate occurs primarily in the atmosphere of the disc. The peak of the \twCO{} line width residuals surrounding the CPD candidate is located at an orbital radius of $R=210$\,au, and an azimuth of $\phi=-113^\circ$ in the disc reference frame. 

\begin{figure*}
   \centering
   \includegraphics[width=1.0\textwidth]{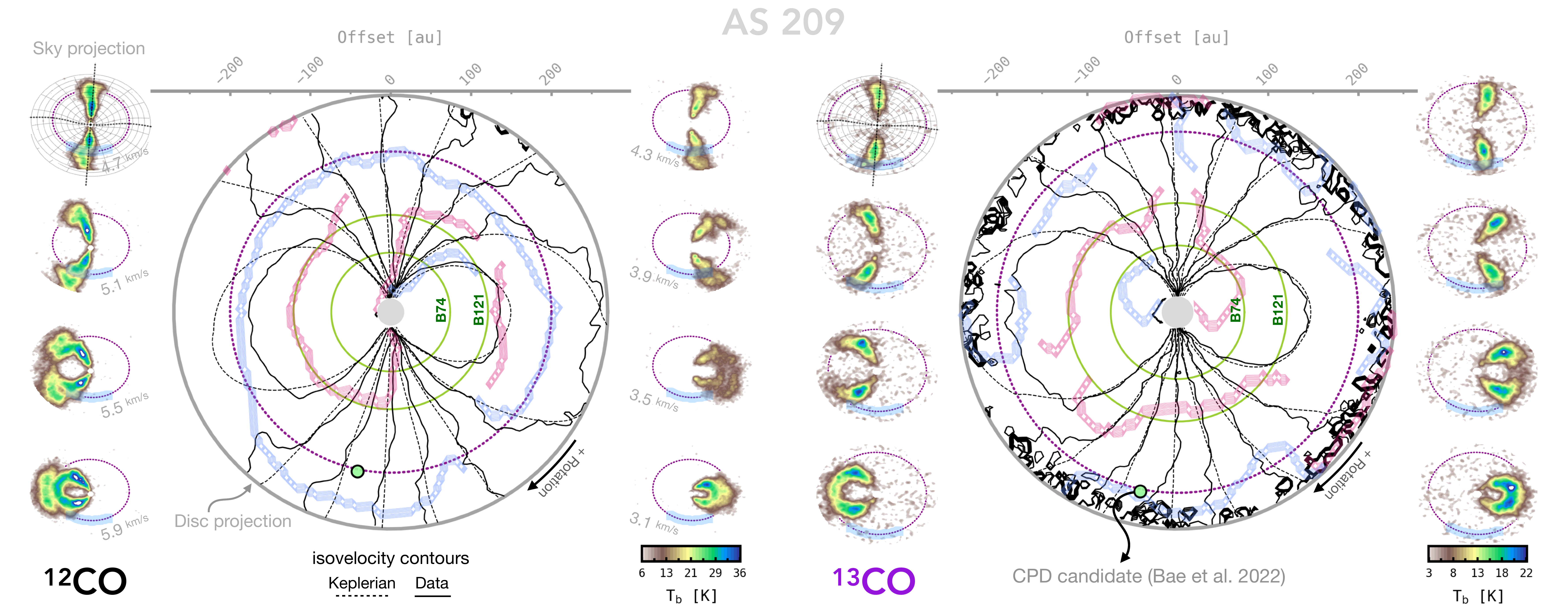}
     \caption{Illustrating isovelocity contours and the presence of coherent velocity substructures in the disc of \as{} as probed by \twCO{} and \thCOfull{} line emission. Black solid and dashed lines show isovelocity contours from the data and the best-fit Keplerian model, respectively. For reference, the isovelocity levels correspond to the same velocity channels of the intensity maps displayed around the main panels. Coherent sub-- and super Keplerian substructures identified by \textsc{FilFinder} \citep[][]{koch+2015} in velocity residuals (Figs. \ref{fig:residuals_kinematics_12co} and \ref{fig:residuals_kinematics_13co}) are overlaid as blue and red thick lines. The green circle marks the location of the CPD candidate proposed by \citet[][]{bae+2022} in the disc reference frame. We note that the intensity kinks most prominently seen in the south of the sky in \twCO{} channels, between 3.5 and 5.5\,km\,s$^{-1}$, are not driven by localised velocity perturbations, but instead are the consequence of a large-scale blue-shifted flow that spans over the entire azimuth of the disc at around $\sim\!200$\,au, and appears at both scale-heights traced by \twCO{} and \thCO{}. In Sect. \ref{sec:gas_substructure_velocity_flows}, we show that such a coherent signature is the result of upward vertical flows.
     }
    \label{fig:skeleton_as209}
\end{figure*}

We stress that the seemingly localised kink-like feature observed in the \twCO{} channel maps of \as{} in the south of the sky does not originate from a localised kinematic perturbation. Instead, it is driven by the upward meridional flow which extends across the entire azimuth of the disc as demonstrated in this Section. Figure \ref{fig:skeleton_as209} shows isovelocity contours extracted from the data and model channel maps, and highlights the location of vertical flows to illustrate that the kinks observed in intensity channels are far from probing localised features in the kinematics. Even though kinks have been associated with the presence of planet-driven perturbations \citep[][]{pinte+2018a}, such as Lindblad spirals dominated by radial and azimuthal velocities \citep[][]{rabago+2021, bollati+2021}, we emphasise that any type of velocity perturbation, strong enough along the line of sight, can cause kinks in intensity channels regardless of its physical origin and/or predominant velocity component. 

Finally, in Figure \ref{fig:absorption_as209} we illustrate that the azimuthal modulation of the intensity field driven by foreground cloud absorption in this source does not translate into variations in the velocity perturbations that can be quantified. The reason is that the relatively low inclination and flat vertical structure of this disc make the emitted line profiles to be nearly symmetric in frequency, and thus uniform intensity variations do not lead to differences in the retrieved velocities. \\

\noindent\textit{\im{}}. The disc of \im{} exhibits strong velocity perturbations near the projected minor axis, potentially dominated by azimuthal velocities (Fig. \ref{fig:residuals_kinematics_12co}, bottom left panel). However, due to the strong diffuse emission in-between \twCO{} layers not accounted by the \discminer{} model, these signatures may not represent physical disturbances and should be interpreted with care. On the other hand, although \im{} is thought to host a massive gas disc \citep[$\sim\!0.1$\,M$_\odot$,][]{lodato+2022}, it is still unclear whether it is gravitationally unstable. This disc displays complex, nearly symmetric, spiral structures in the mm continuum, typical of GI, but with varying pitch angle as function of radius which would be in better agreement with the presence of an embedded companion \citep[see discussions by][]{huang+2018, verrios+2022}. We do not detect clear spiral signatures in the kinematics of any of the CO isotopologues, which would favour a gravitationally (at least marginally) stable scenario in this source. 
Instead, we identify arc-like fluctuations both in velocity and peak intensity residuals on the blueshifted side of the outer disc between $R\!\sim\!300-700$\,au. \\

\noindent\textit{\gm{}}. The disc of \gm{} displays large-scale velocity perturbations both in the atmosphere and at layers closer to the midplane (Figs. \ref{fig:residuals_kinematics_12co} and \ref{fig:residuals_kinematics_13co}). It is known that this system is potentially in late interaction with material infalling from the remnant envelope 
\citep[][]{huang+2021}, which leads to the presence of intensity and velocity spirals and trails, or filamentary structures, that appear most notably at high elevations where the interaction is strongest. The prominent wedge of redshifted velocities between $\phi\!\sim\!45-90^\circ$ in the disc reference frame (Fig. \ref{fig:residuals_kinematics_12co}), and of line width enhancements in the same region (Fig. \ref{fig:residuals_linewidth_12co}), is in good agreement with that inflow of material. This wedge is also seen as an excess of integrated \twCO{} intensity in Fig. 17 of \citet[][]{huang+2021} after subtraction of a Keplerian mask. We note that multiple spiral-like features are also present in peak intensity residuals of this system as illustrated in Fig. \ref{fig:residuals_peakint_12co}. However, in \thCO{} we only tentatively identify spirals in the kinematics, highlighted by solid lines in Fig. \ref{fig:residuals_kinematics_13co}. 
In a non-planetary scenario, these can be the consequence of gravitational instabilities developing in a disc \citep[][]{kratter+2016}. However, \citet[][]{schwarz+2021} demonstrate that in spite of the possibly high mass reservoir available in this disc \citep[$\sim\!0.2$\,M$_\odot$, see also][]{lodato+2022}, the gas temperature is high enough across much of the disc extent to prevent gravitational instabilities to develop. Since we only detect spiral features convincingly in intensity and velocity residuals at high elevations traced by \twCO{}, we argue that they are more likely the result of infalling material perturbing the gas disc surface density and temperature rather than a consequence of GI. \\

\section{Gas substructures as sites to search for planets} \label{sec:gas_substructure}

In addition to the localised features and the non-axisymmetric perturbations manifested as spirals and filamentary structures in the disc kinematics (see Sect. \ref{sec:kinematics}), embedded planets can also induce axisymmetric variations in the gas surface density and temperature of the host disc in the form of annular gaps and cavities. These substructures translate into pressure gradients capable of producing observable, axisymmetric velocity modulations, both in the rotational component of the velocity field \citep[][]{kanagawa+2015}, and in the vertical direction through the meridional circulation of material from the disc atmosphere onto the midplane and vice-versa \citep[][]{morbidelli+2014}. Moreover, these features are also expected to trigger localised fluctuations in the radial profiles of molecular peak intensities \citep[see e.g.][]{facchini+2018_gaps} and line widths \citep{izquierdo+2021}, which can be useful to understand whether the underlying pressure modulations are dominated by density and/or temperature variations. How do these observables (anti$-$)correlate with each other, if at all, and where do they sit radially in this sample of discs? In this Section, we focus on the analysis of azimuthally averaged radial profiles of the observable quantities explored in this work to provide a general view of the physical structure and dynamics of the discs in the sample, as well as the connection between annular gas substructures and the planet candidates proposed in Sect. \ref{sec:kinematics} and summarised in Table \ref{table:planet_locations}.

\subsection{Azimuthal and meridional velocity flows}
\label{sec:gas_substructure_velocity_flows}

\begin{figure*}
   \centering  
      \includegraphics[width=0.98\textwidth]{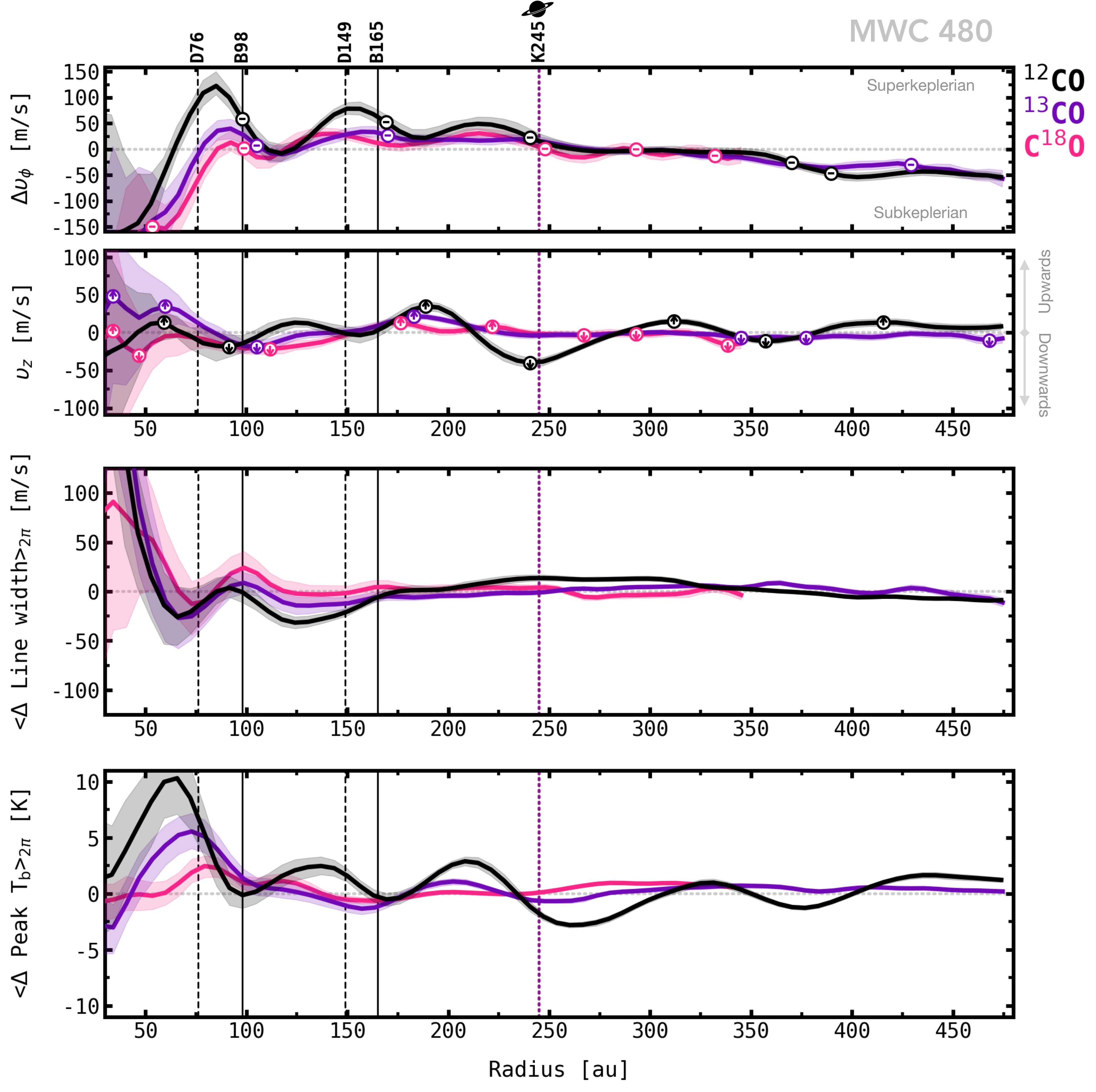}
      \caption{Azimuthally averaged profiles of velocity (top), line width (middle) and peak intensity (bottom) residuals obtained for \twCO{}, \thCO{}, and \eiCO{}, for the disc around \mwc{}. The radial location of millimetre dust gaps and rings is illustrated as dashed and solid lines, respectively. The radial distance of the most prominent kink apparent in \twCO{} channel maps is shown by the dotted purple line. Strong pressure bumps traced by minimal velocity gradients are marked with minuses in the top panel. Peak meridional flows hinting at gas moving away and towards the disc midplane are highlighted with arrows pointing up and down in the second top panel. The planet marker indicates the orbital radius of the planet candidate proposed at $R=245$\,au (see Table \ref{table:planet_locations}).
              }
         \label{fig:averaged_residuals_mwc480}
    
   \end{figure*}


\begin{figure*}
   \centering  
      \includegraphics[width=0.98\textwidth]{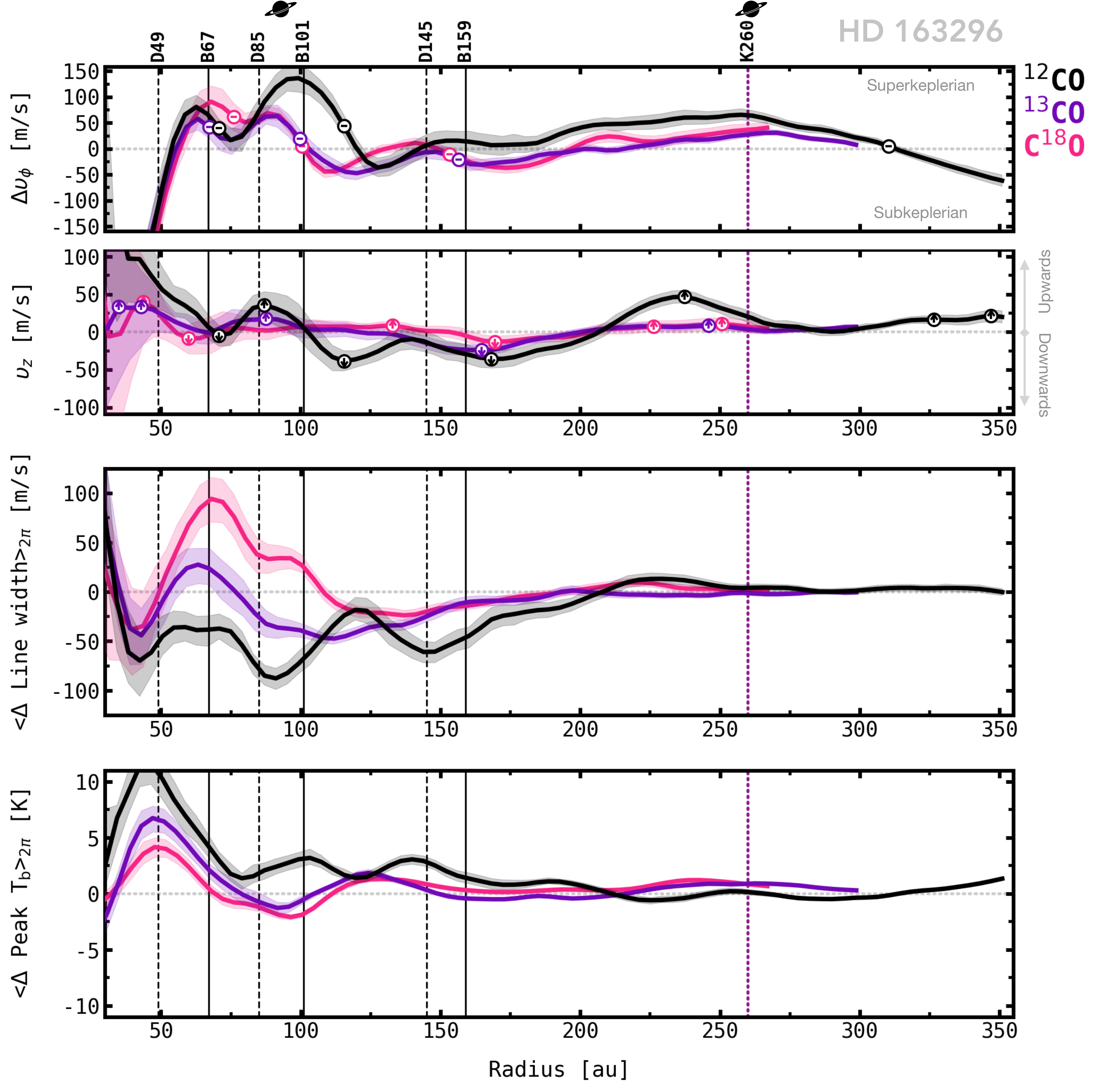}
      \caption{As Fig. \ref{fig:averaged_residuals_mwc480} but for the disc around \hd{}. The planet markers indicate the orbital radii of the planet candidates associated with the localised velocity perturbations P94 and P261 (see Table \ref{table:planet_locations}).  
              }
         \label{fig:averaged_residuals_hd163296}
         
   \end{figure*}


In Sect. \ref{sec:azimuthal_meridional_velocities}, we demonstrate that it is possible to readily obtain an estimate of the azimuthal and vertical components of the velocity perturbation field by computing azimuthal averages of different kinds of velocity residual maps. These azimuthal and meridional velocity flows are presented in the top panels of Figure \ref{fig:averaged_residuals_mwc480}, for the disc of \mwc{}, and in Figures \ref{fig:averaged_residuals_hd163296}, \ref{fig:averaged_residuals_as209}, \ref{fig:averaged_residuals_imlup} and \ref{fig:averaged_residuals_gmaur}, for the discs of \hd{}, \as{}, \im{}, and \gm{}, respectively.
Along both types of velocity profiles, we mark the location of pressure maxima, traced by negative velocity gradients in azimuthal velocities, and the direction of the meridional flow on local maxima and minima in vertical velocities. To assess the effect of continuum subtraction on these velocity profiles, we performed the same analysis on intensity cubes of all three CO isotopologues for \mwc{} without continuum subtraction and, as illustrated in Fig. \ref{fig:wcont_vprofiles_mwc480}, we found no differences with respect to the continuum-subtracted cubes that could affect any of the results summarised in this Section. \\

\noindent\textit{\mwc{}}. From Fig. \ref{fig:averaged_residuals_mwc480}, we note that in the disc of \mwc{} the positive and negative radial gradients in the azimuthal component of the velocity, retrieved for all CO isotopologues, are in excellent correlation with the radial location of gaps and rings observed in the mm continuum, respectively. This result suggests that the pressure modulations present in the gas component of this system have been there long enough for the mm-sized dust grains to be trapped in annular structures at the currently observed locations. 
It is also worth highlighting the prominent meridional circulation of material flowing up and down with respect to the disc midplane in this system. In the radial section comprised between the dust ring B165 and a radius of $R\!\sim\!300$\,au, we identify strong vertical motions with amplitudes of up to $50$\,m\,s$^{-1}$ in the disc atmosphere, traced by \twCO{} at $R=193$\,au and $R=245$\,au. This is in line with the vertical velocities found by \citet{teague+2021} in \twCO{} using a different extraction method (see e.g. Fig. 5 in their paper).
The fact that the line-of-sight velocities along this structure are highly coherent in the azimuthal coordinate, as shown in Fig. \ref{fig:polar_deproj_mwc480}, and that its underlying pitch angle is very low as a function of radius, allows us to estimate with good precision the magnitude of the underlying vertical motions using the decomposition method introduced in Sect. \ref{sec:azimuthal_meridional_velocities}. 
Furthermore, it is also notable that although all CO isotopologues trace similar variations in the azimuthal velocity component, only \twCO{} (tracing $z/r\!\sim\!0.2$ scale heights, Fig. \ref{fig:attributes_all_co}) displays strong fluctuations in the vertical component while \thCO{} and \eiCO{} (tracing $z/r<0.1$) remain nearly unperturbed. As discussed in Sect. \ref{sec:extended_perturbations}, this vertical modulation of vertical velocities strongly supports the presence of buoyancy spirals driven by an embedded planet, originally proposed by \citet[][]{teague+2021} through the observation of tightly wound substructures in \twCO{} temperatures. In Sect. \ref{sec:extended_perturbations}, we also use line width information to propose a planet candidate at an orbital radius of $R=245$\,au and an azimuth of $\phi=-128^\circ$ in this disc.  
\\

\noindent\textit{\hd{}}. In the top panel of Figure \ref{fig:averaged_residuals_hd163296} we illustrate that most of the gas pressure bumps present in the disc of this source, traced by negative radial gradients in the azimuthal component of the velocity field, are also co-spatial with the radial location of dust rings observed in the mm continuum. This suggests that the latter are indeed the consequence of pressure traps as confirmed by \citet[][]{rosotti+2020}.
In fact, the even better match between the location of these rings and pressure bumps traced by \thCO{} and \eiCO{}, which are $\sim\!50\%$ closer than \twCO{} to the disc midplane (see Fig. \ref{fig:attributes_all_co}, bottom row), further supports this finding. Similarly, the positive velocity gradients identified in the same velocity component at the radial locations of $R\!\sim\!50, 85, 135$\,au, in all CO isotopologues, are in excellent agreement with the gas gaps proposed by \citet[][]{isella+2016} based on radiative transfer models of this disc. 
Also, from the second top panel in Fig. \ref{fig:averaged_residuals_hd163296}, we note that the meridional circulation of material is very pronounced in this system too. 
Upward and downward gas flows are particularly prominent around the location of millimetre dust and gas substructures between $R\!\sim\!50-160$\,au, and co-spatial with pressure minima and maxima, respectively. Further out, there is an upward velocity flow centred at $R\!\sim\!240$\,au near the planet candidate associated with the K260 kink. This feature is strongly detected in the disc atmosphere traced by \twCO{}, but decays rapidly towards lower scale heights probed by \thCO{} and \eiCO{}. As discussed in Sect. \ref{sec:localised_perturbations}, such a prominent vertical dependence could be interpreted as due to planet-driven buoyancy spirals which, as opposed to Lindblad spirals, are expected to develop more significantly in the disc atmosphere and become weaker as the disc midplane is approached. We note that this radial profile of \twCO{} vertical velocities is equivalent to that of \citet{teague+2019nat, teague+2021}, but with opposite sign. We have cross-checked with the corresponding authors via private communication and agreed that the profile presented in Fig. \ref{fig:averaged_residuals_hd163296}, second top panel, is correct.

\textit{\as{}}. The average residual velocities computed for the disc of \as{} in the three CO isotopologues are the most dissimilar of the sample (see Fig. \ref{fig:averaged_residuals_as209}). Due to the fact that the model background velocity of all molecules is different $-$because stellar masses and orientation parameters are allowed to vary freely in all models$-$ a velocity offset with respect to the Keplerian model background may be expected. However, taking systematic offsets aside, the three isotopologue velocity profiles are topologically different. On top of that, only the pressure bumps traced by \twCO{} coincide, within a beam, with the radial location of the millimetre dust rings, B74 and B121. 
The fact that the \thCO{} and \eiCO{} velocities behave differently, suggests that the radial pressure gradients in this disc fluctuate strongly as the midplane is approached. This was already suggested by \citet{teague+2018b}, who proposed a vertically varying pressure profile to explain the radial offset between scattered light and continuum dust rings observed in \as{}, assuming that both are driven by pressure traps. Nevertheless, it is still puzzling that the pressure bumps in \thCO{} and \eiCO{} are even further from the location of millimetre dust rings in spite of being closer than \twCO{} to the disc midplane\footnote{\twCO{} traces layers between $z/R\!\sim\!0.1-0.2$, whereas \thCO{} and \eiCO{} trace $z/R<0.1$ scale heights (see Fig. \ref{fig:attributes_all_co}).}. A plausible scenario could be that some of the velocity components near the midplane of the disc are far from axisymmetric, in which case the azimuthally averaged velocities would not provide a good description of the actual azimuthal component of the velocity as it would be contaminated by non-axisymmetric radial and/or vertical motions. We identify, however, a strong positive gradient in both the azimuthal velocities of \twCO{} and \thCO{} between $R=85-110$\,au, around the millimetre dust gap D100, which \citet[][]{fedele+2018} modelled by invoking a low-mass planet ($<0.1\,\Mj$) at an orbital radius of $R=103$\,au. On the other hand, this disc displays two fairly prominent vertical flows of material, one approaching the midplane at a radius of $R\!\sim\!130$\,au, and another moving away from the midplane at $R\!\sim\!190$\,au, with a comparable magnitude of the order of $50-100$\,m\,s$^{-1}$ traced by both \twCO{} and \thCO{}. The latter signature coincides within a beam size with the radial location where a CPD candidate was proposed earlier by \citet[][]{bae+2022} at $\sim\!200$\,au, and finds a plausible interpretation in the work of \citet{galloway+2023}, who conclude that this coherent upward flow could be the consequence of magnetically driven winds dominated by ambipolar diffusion, triggered by the possibly low densities around the orbit of the proposed planet+CPD system. Further details on the kinematic signatures identified around this planet candidate are summarised in Sect. \ref{sec:extended_perturbations}. \\

\noindent\textit{\im{}}. The millimetre dust gap D116 and ring B133 of this disc co-locate very well with pressure minima and maxima traced by azimuthal velocity flows in all CO isotopologues (Fig. \ref{fig:averaged_residuals_imlup}), namely, across a wide range of vertical layers ($z/r\!\sim\!0.1-0.3$, Fig. \ref{fig:attributes_all_co}). The dust ring B220, however, does not seem to align with any pressure bump in the gas, and therefore a different mechanism other than pressure trapping is likely acting on this radial section of the disc midplane. Additionally, we detect a downward meridional flow of material between $R\!\sim\!120-135$\,au, both in \twCO{} and \thCO{}, overlapping with the millimetre dust gap D116 and ring B133. Although no localised perturbation is detected in this region of the disc (see Fig. \ref{fig:folded_velocities_nondetections}), we note that this meridional gas flow is near the planet candidate proposed by \citet[][]{pinte+2020} and \citet[][]{verrios+2022} at $R=117$\,au. \\

\noindent\textit{\gm{}}. As illustrated in Fig. \ref{fig:averaged_residuals_gmaur}, the strong interaction between this disc and material possibly infalling from a remnant envelope \citep[see e.g.][]{huang+2021}, makes both azimuthal and vertical flows at radii $R>150$\,au to change substantially with varying scale height. At radii interior to this region, the azimuthal component of the velocity follows similar modulations for all CO isotopologues as opposed to the vertical component, which still exhibits large variations in the $z-$direction. We note that the millimetre dust ring B86 is highly correlated with a gas pressure bump present at all scale heights, traced by negative gradients in the azimuthal component of the velocities in all three CO isotopologues. Also, the most prominent pressure minima are well aligned with the radial location of millimetre dust gaps D68 and D142. Conversely, no clear pressure bump is found at the location of the B163 ring, suggesting that this dust substructure is either not the result of pressure trapping processes or that the external material is already biasing the derivation of rotation velocities as of this radius.

\begin{figure*}
\centering
   \includegraphics[width=0.95\textwidth]{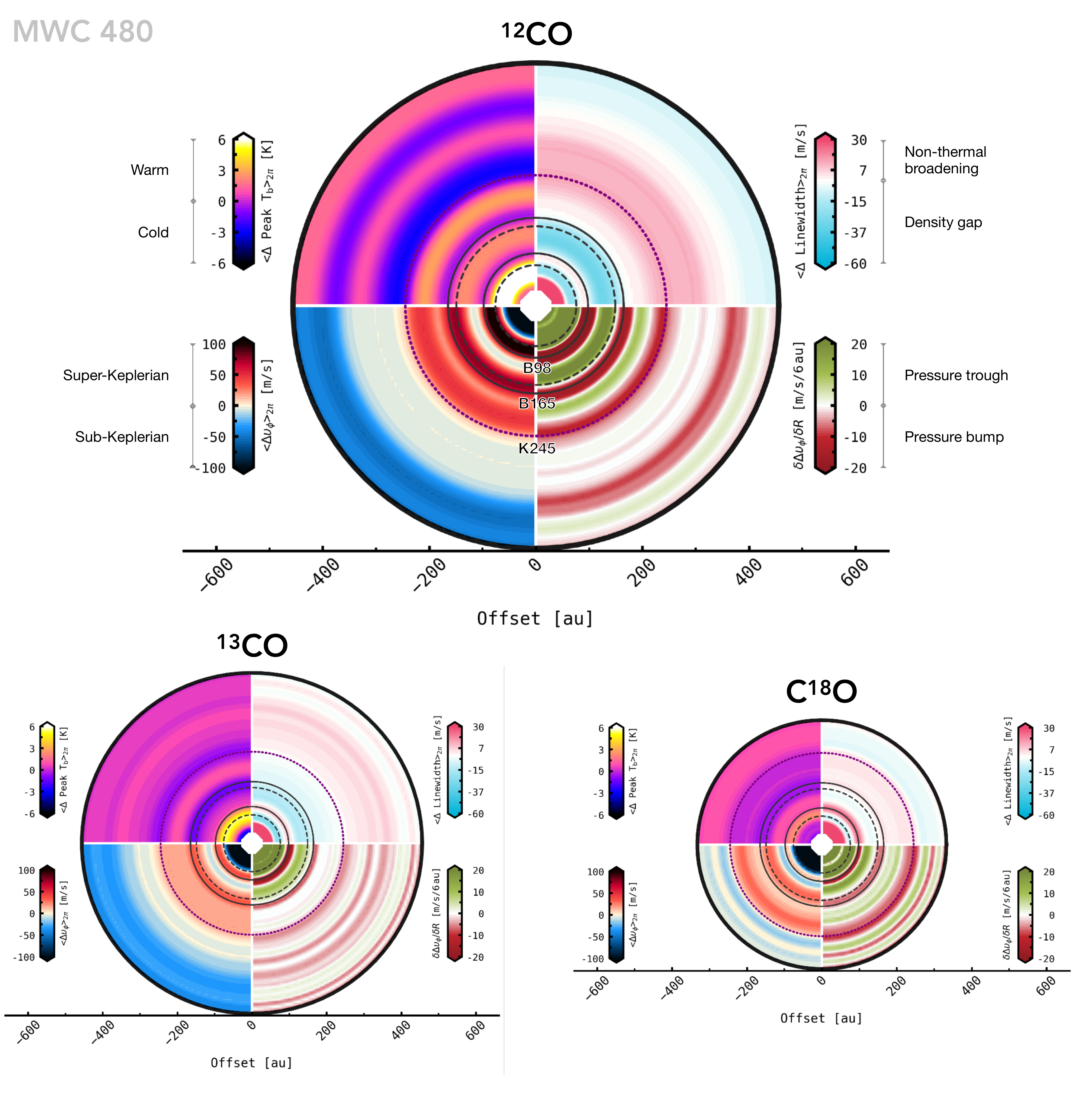} 
     \caption{Summary of azimuthally averaged residuals extrapolated onto a 2D grid to aid visualisation of (anti--)correlations between the different line profile observables extracted from the disc of \mwc{}. The radial location of dust gaps and rings is illustrated as dashed and solid lines, respectively. The radial distance of the most prominent kink apparent in \twCO{} channel maps is marked by the dotted purple line.
              }
         \label{fig:pie_average_residuals_mwc480}
\end{figure*} 

\subsection{Surface density gaps traced by azimuthal velocity modulations and line widths}


\setlength{\tabcolsep}{5.0pt} 

\begin{table*}
\centering
{\renewcommand{\arraystretch}{2.0}

\caption{Summary of radial substructures derived from azimuthally averaged residuals for the disc of \mwc{}.} \label{table:radial_subst_mwc480}

\begin{tabular}{ rcccccl } 

\toprule
\toprule
Radius & Dust$^a$ & Vertical$^b$ & $\Delta\upsilon_\phi$ & T$_{\rm b}$ & L$_w$ & Notes \\ 
\midrule

76 & \bluearrow{} & \Bluearrow{} & \Redplus{} & \Redarrow{} & \Bluearrow{} & \makecell[l]{Warm gas gap, detected in kinematics and line widths of all tracers.} \\

149 & \bluearrow{} & \bluearrow{}/\bluearrow{} & /\redplus{}/ & \redarrow{}\bluearrow{}\bluearrow{} & / & \makecell[l]{Strong vertical temperature gradient.} \\

\midrule

98 & \redarrow{} & / & \Blueminus{} & \Bluearrow{} & \Redarrow{} & \makecell[l]{Cold pressure bump detected in kinematics, leading to dust ring. \\ Enhanced line broadening in all tracers indicative of gas over-density.} \\

165 & \redarrow{} & / & \Blueminus{} & \Bluearrow{} & \Redarrow{} & \makecell[l]{Cold pressure bump detected in kinematics, leading to dust ring. \\ Enhanced line broadening in all tracers indicative of gas over-density.} \\

\midrule

125 & / & / & \Redplus{} & \Redarrow{} & \Bluearrow{} & \makecell[l]{Warm gas gap, detected in kinematics and line widths of all tracers.} \\ 

200 & / & / & \redplus{}/\redplus{} & \Redarrow{} & / & \makecell[l]{Warm, shallow pressure minimum, mainly seen in \twCO{} kinematics.} \\

245 & / & \Bluearrow{} & \Blueminus{} & \bluearrow{}\bluearrow{}/ & \redarrow{}// & \makecell[l]{Cold gas pressure bump possibly driven by density enhancement.} \\ 

360 & / & \bluearrow{}\bluearrow{}/ & \blueminus{}\blueminus{}/ & \bluearrow{}// & / & \makecell[l]{Shallow pressure bump, mainly seen in \twCO{} kinematics.} \\ 

\bottomrule

\end{tabular}
 \caption*{\textbf{Note.} Red and blue arrows highlight peaks and troughs, respectively. Red pluses and blue minuses refer to positive and negative velocity gradients, indicative of radial gradients in the gas pressure. A circled plus/minus, or double-lined arrow, means that all tracers agree with the same signature, otherwise, each symbol corresponds to \twCO{}, \thCO{} and \eiCO{}, from left to right. $^a$The reference radii are the centres of dust gaps and rings reported in mm wavelengths by \citet{law+2021_maps3} (unless indicated otherwise). $^b$Vertical modulations from \citet{paneque+2022} are also shown for comparison. Other interesting features found at different radial locations are grouped in the third row of the table.} 
 
  }
\end{table*}
\setlength{\tabcolsep}{5.0pt} 

\begin{table*}
\centering
{\renewcommand{\arraystretch}{1.0}

 \caption{As Table \ref{table:radial_subst_mwc480} but for the disc of \hd{}.}
  \label{table:radial_subst_hd163296}
  
\begin{tabular}{ rcccccl } 

\toprule
\toprule
Radius & Dust$^a$ & Vertical$^b$ & $\Delta\upsilon_\phi$ & T$_{\rm b}$ & L$_w$ & Notes \\ 
\midrule

45 & \bluearrow{} & \Bluearrow{} & \Redplus{} & \Redarrow{} & \Bluearrow{} & Warm gas gap. \makecell[l]{Detected in kinematics and line widths of all tracers.} \vspace{0.15cm} \\

86 & \bluearrow{} & \Bluearrow{} & \Redplus{} & \Bluearrow{} & \Bluearrow{} & Cold gas gap. \makecell[l]{Detected in kinematics and line widths of all tracers.} \vspace{0.15cm} \\

141 & \bluearrow{} & /\bluearrow{}\bluearrow{} & \redplus{}\redplus{}/ & \redarrow{}// & \bluearrow{}// & \makecell[l]{Shallow, warm gas gap. Detected in \twCO{} and \thCO{} kinematics, \\ and in \twCO{} line widths.} \vspace{0.15cm} \\

270 & \bluearrow{}$^{\rm optical}$ & \bluearrow{}\bluearrow{}/ & /\redplus{}\redplus{} & / & / & \makecell[l]{Vertical bumps and positive velocity gradients near the location of \\ kinematic kink at $R=260$\,au \citep{pinte+2018a, izquierdo+2022}} \vspace{0.15cm} \\

\midrule

67 & \redarrow{} & / & \Blueminus{} & / & \redarrow{}// & \makecell[l]{Pressure bumps detected in kinematics, leading to dust ring. \\ Line broadening in \twCO{} indicative of higher density.} \vspace{0.15cm} \\

101 & \redarrow{} & / & /\blueminus{}\blueminus{} & \redarrow{}\bluearrow{}\bluearrow{} & /\redarrow{}\redarrow{} & \makecell[l]{Pressure bumps detected in kinematics, leading to dust ring. \\ High line broadening in \eiCO{} hinting at enhanced turbulent motions.} \vspace{0.15cm} \\

159 & \redarrow{} & / & \Blueminus{} & / & / & \makecell[l]{Pressure bumps detected in kinematics, leading to dust ring.} \\

\midrule

120 & / & / & \blueminus{}/\redplus{} & \bluearrow{}\redarrow{}\redarrow{} & \redarrow{}\bluearrow{}\bluearrow{} & \makecell[l]{Funnel-like structure causing radial offset between pressure bumps \\ observed in \twCO{} compared to those in \thCO{} and \eiCO{} (see Sect. \ref{sec:gas_substructure})} \\ 

\bottomrule

\end{tabular}

  }
\end{table*}

As demonstrated in Fig. 6 of \citetalias{izquierdo+2021}, the presence of gaps in the surface density of a disc has an observable impact on the radial distribution of line widths from molecular emission. In particular, due to the reduced optical depth in a density gap, azimuthally averaged line widths from optically thick tracers are expected to be significantly smaller than those in the unperturbed background of the disc. Thus, if the gas temperature is high within the gap, this implies that if a positive radial gradient in azimuthal velocities and a minimum in line widths are detected simultaneously, we can assert that the pressure minimum associated with the gap is strongly dominated by low surface density values and not by its temperature structure. As illustrated by \citet[][]{rab+2020} (see e.g. their Fig. 7), if the high temperature of the gap was the dominant component, the azimuthal velocity flows would instead display negative radial gradients at the gap centre.

In Figure \ref{fig:pie_average_residuals_mwc480}, we make explicit the correlation between line width minima (right, top panel) and pressure troughs traced by peak velocity gradients (right, bottom), as a function of the orbital radius for the disc around \mwc{} in the three CO isotopologues. We note that the pressure troughs (in green) co-locate with line width minima (in light blue), and that the pressure bumps coincide with the location of dust rings and less shallow line widths. This analysis suggests the presence of at least two gas gaps centred at a radius of $R=76$\,au and $R=125$\,au, dominated by low surface densities, and indicates that the dust rings detected in previous mm continuum observations of this system, B98 and B165, are indeed the consequence of pressure traps. We also note that the same signatures hold in both \thCO{} and \eiCO{}, although less clearly than in \twCO{}. This is expected because for increasingly optically thinner tracers the dependence of line widths on the gas surface density is weaker \citep[][]{hacar+2016}.
Despite the fact that there are peaks in the intensity of all three CO isotopologues, likely associated with elevated temperatures at the radial location of these gaps, the simultaneous detection of line width and pressure troughs indicates that the low surface density in both regions dominates the shape of the pressure gradient responsible for the observed kinematic modulations.

Finally, we note that positive gradients in the azimuthal component of the velocity field agree excellently with the location of line width minima in all discs, especially in \twCO{} and \thCO{} emission. In Tables \ref{table:radial_subst_mwc480}, \ref{table:radial_subst_hd163296}, \ref{table:radial_subst_as209}--\ref{table:radial_subst_gmaur}, we present a summary of the radial locations of maxima and minima in velocity gradients and line widths for all sources. Conversely, in the same tables we report that intensity peaks and troughs show no clear correlation or anti-correlation with respect to the latter two observables. As discussed by \citet[][]{turner+2012} and \citet{rab+2020}, peaks and dips in the brightness temperature do not necessarily align with the location of gas density rings and gaps (see Fig. \ref{fig:pie_average_residuals_mwc480}, top left vs top right panel). Therefore, we emphasise that a simultaneous analysis of the kinematics and widths of the molecular line profile revealed by emission of (at least marginally) optically thick tracers allows understanding the gas disc substructures and thus the potential sites of planet formation with greater robustness.

\subsection{Strong vertical dependence of radial pressure in the disc of \hd{}, around the planet candidate P94}

Taking a closer look to the residual velocity profiles of the disc of \hd{} in Fig. \ref{fig:averaged_residuals_hd163296}, it is easy to note that between $R=86$\,au and $R=141$\,au, the rotation velocities traced by \twCO{} vary substantially in magnitude and extent with respect to those in \thCO{} and \eiCO{}. As these isotopologues emit at different altitudes above the midplane, this is evidence of the vertical dependence of the radial pressure gradient of the gas disc. However, the origin of these pressure variations remains unclear as yet. Do density and temperature gradients contribute similarly to the observed pressure gradients, or does one of them predominate?

The other two average profiles, computed from peak intensity and line width residuals, are helpful to gauge the relative contribution of density and temperature fluctuations to the radial pressure gradient of the disc. Consider the radial section of the disc between $R\sim75$\,au and $R\sim100$\,au. In this region, the residual peak brightness temperature of \twCO{} (Fig. \ref{fig:averaged_residuals_hd163296}, bottom panel) displays a positive gradient, followed by a negative gradient until it reaches a temperature trough at $R\sim120$\,au. The \twCO{} line widths (middle panel), on the other hand, exhibit a pronounced positive gradient between the line width minimum near D86 and the line width maximum at $125$\,au. Assuming that fluctuations in the brightness temperatures and line widths of \twCO{} are tightly linked to local variations in the gas temperature and density, respectively \citep[see e.g.][]{hacar+2016,izquierdo+2022}, the coincidence of a temperature trough and a line width peak around $120$\,au suggests that the pressure maximum, revealed by the disc kinematics, is mainly driven by a gas ring at this radius.

Conversely, \thCO{} and \eiCO{} show the opposite behaviour in both the residual brightness temperatures and line widths within the same radial section. Nevertheless, the same argument made for the pressure bump traced by \twCO{} at $R=120$\,au holds for these two molecules, but this time closer-in at $R=100$\,au. The pressure maximum revealed by the negative velocity gradient of both isotopologues at this radius coexists with a line width peak and a temperature trough, indicating that this pressure bump is most likely driven by gas density enhancements\footnote{We note that the line width peak of both \thCO{} and \eiCO{} at $R=100$\,au is rather subtle compared to that of \twCO{} at $R=120$\,au, and yet the decreasing velocity modulation is not negligible in either case. The reason is that the line width residuals of these optically thinner tracers are not as sensitive to density fluctuations as they are for \twCO{} \citep[][]{hacar+2016}. Instead, local temperature and turbulence may also play a key role, making it harder to distinguish between the different contributions to the broadening of the line profile, unless detailed radiative transfer modelling is introduced.}. The radial shift between this pressure maximum, and the one traced by \twCO{}, is in agreement with the findings of \citet[][]{rab+2020}, who used thermochemical models to show that the relative contribution of temperature and density fluctuations to the pressure gradient may vary substantially at different radii and heights in this disc. 

Morphologically, we are possibly witnessing a density substructure that is opening along the vertical direction, like a funnel, around the millimetre dust gap, D85, and a gas gap at the same radial location, possibly carved by the planet candidate, P94, proposed by \citet[][]{izquierdo+2022} and ratified in this work through localised velocity and line width perturbations detected in multiple CO isotopologues. This geometry would explain why the radial separation between the pressure minimum around D86, due to a gas gap, and the pressure maximum at $R=120$\,au traced by \twCO{}, is almost twice as large as the separation between pressure minima and maxima in the same radial section at the elevations traced by \thCO{} and \eiCO{}.

\subsection{Non-axisymmetric variations in the radial profiles of the disc around \mwc{}} \label{sec:non_axisymmetric_mwc480}

\begin{figure*}
   \centering
   \includegraphics[width=1.0\textwidth]{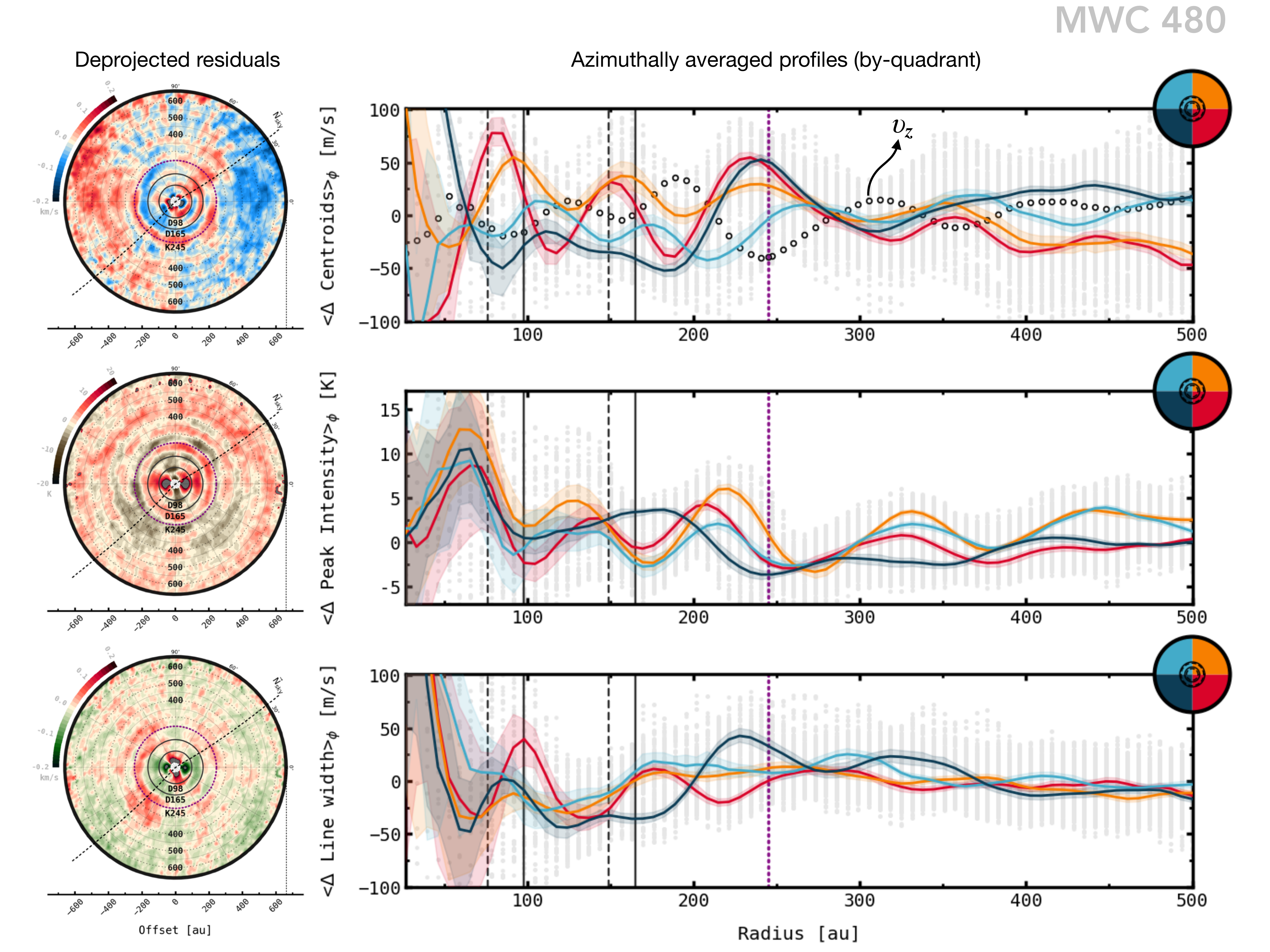} 
      \caption{Illustrating the presence of non-axisymmetric features in the disc of \mwc{}, as observed in \twCO{}. \textit{Left column}: Velocity, peak intensity, and line width residuals deprojected on the disc reference frame.
      \textit{Right column}: Azimuthally averaged residuals extracted per quadrant to highlight azimuthal variations in the different observables. The colour of each radial profile indicates the quadrant of the disc where it was computed according to the colouring code in the top right corner. Quadrants to the right of the disc vertical axis correspond to the redshifted side of the disc.
              }
         \label{fig:quadrant_average_residuals_mwc480}
   \end{figure*}

Not all variations in the line profile observables explored in this study are axisymmetric. There are numerous signatures that appear spatially localised (as presented in Sect. \ref{sec:localised_perturbations}), and others that span over large azimuthal and radial portions, at times forming arcs along concentric annuli, but also filament-like structures that stretch irregularly across the discs (as in Sect. \ref{sec:extended_perturbations}). 

To quantify and better visualise these asymmetries, it is helpful to compute azimuthal averages over narrower azimuthal ranges instead of considering $2\pi$ averages. This is illustrated in Figure \ref{fig:quadrant_average_residuals_mwc480} for the disc of \mwc{}, where we show azimuthal averages of the three residual observables obtained for \twCO{}, in the four quadrants delimited by the disc main axes as projected in the sky. It is clear that the average residual profiles, especially those from centroid velocities, vary significantly from one quadrant to another. However, we note that between the outermost millimetre dust ring, B165, and a radius of $R\sim350$\,au, the average velocities in each quadrant of the disc seem to fluctuate less with respect to each other than at closer-in annuli. 
It is in this region where the azimuthal coherence of the perturbation reveals the three-dimensional nature of the velocity field, seemingly dominated by vertical motions as illustrated in Fig. \ref{fig:averaged_residuals_mwc480}. Likewise, the radial sign flips are the consequence of contiguous anti-parallel vertical flows, which could be related to the presence of planet-driven buoyancy spirals \citep{bae+2021,teague+2021}, as discussed earlier in Sect. \ref{sec:extended_perturbations}. 

Furthermore, as demonstrated in line width residuals in the top left panel of Fig. \ref{fig:residuals_linewidth_12co}, the disc around \mwc{} displays prominent arc-like enhancements in the \twCO{} line widths, more strongly seen on the blueshifted side around orbital radii between $R\sim225$ and $R\sim325$\,au, which we associate with the presence of an embedded planet at $R=245$\,au radius and $\phi=-128^\circ$ azimuth (see Sect. \ref{sec:extended_perturbations}). With the by-quadrant analysis presented in Fig. \ref{fig:quadrant_average_residuals_mwc480}, we infer that the line width asymmetries at $R\sim225$\,au, in the third quadrant of the disc, are $\sim\!50$\,m\,s$^{-1}$ higher than those in the mirroring fourth quadrant. Moreover, we note that these line width enhancements are anti-correlated with the temperature peaks found at the same radial and azimuthal location, suggesting that thermal broadening could not be the main driver of these arc-like features in line widths. Instead, turbulent motions should thus be favoured as the cause of such signatures. 

\section{Discussion} \label{sec:discussion}

\subsection{Morphology of the P94 perturbation in \hd{}} \label{sec:discussion_hd163296}

The detection of localised velocity signatures in different tracers around the planet candidate P94 allows us to map the vertical structure of the associated velocities. We note that both the perturbation detected in \twCO{}, or P94, and the one detected in \thCO{}, or P81, exhibit a Doppler shift pattern expected from the gravitational and hydrodynamic interaction of a planet with the disc fluid in its immediate vicinity \citep[see e.g.][]{rabago+2021}. However, the morphology of these two signatures is clearly different between each other. The redshifted part of the P94 perturbation is in an outer orbit with respect to the centre of the proposed planet at $R=94$\,au, while the blueshifted part is in an inner orbit (see Fig. \ref{fig:summary_hd163296}). The structure of the P81 perturbation, on the contrary, manifests as a redshifted part which coincides azimuthally with that of P94, but it is located interior to the orbital radius of the blueshifted component at its peak. Although not significantly detected by our clustering algorithm, we identify a similar pattern in \eiCO{}. This suggests that the perturbation sitting at relatively low scale-heights, traced either by \thCO{} or \eiCO{}, is dominated by the radial component of the velocity, which would thus be in agreement with the velocity structure expected for planetary wakes along Lindblad spirals near the disc midplane \citep[see Fig. 4 of][]{rabago+2021}. A similar kinematic feature suggestive of a planet-driven Lindblad spiral was recently reported by \citet[][]{teague+2022} in the disc of TW\,Hydrae. 

\subsection{Line broadening as a tracer of massive planets} \label{sec:discussion_linewidths} 

The simultaneous detection of localised velocity and line width perturbations pointing to the presence of both coherent motions and enhanced velocity dispersions around the planet candidate P94 in the disc of \hd{} opens the door to the use of yet another observable --non-thermal line broadening-- to search for embedded planets in discs, as originally predicted by \citet{dong+2019}. Consequently, we employ these two observables to propose the most plausible azimuthal location of a protoplanet candidate possibly driving buoyancy spiral perturbations in the disc of \mwc{}.
In the disc of \as{}, however, while we identify strong line width enhancements around the vicinity of the CPD candidate reported by \citet[][]{bae+2022} at $R=203$\,au, we do not detect velocity signatures that can be associated with localised perturbations linked to the proposed object. Nevertheless, we highlight that this disc is known to host a prominent gas gap revealed by integrated intensity maps at an orbital distance of $\sim\!200$\,au \citep[][]{guzman+2018,law+2021_maps3}, leaving the planet candidate with little surrounding material and hence making the detection of coherent small-scale velocity variations a harder challenge.
The presence of strong line broadening along with the absence of localised coherent velocities around the CPD candidate suggests that the gas motions in this region are predominantly turbulent.

\subsection{Localised velocity perturbations are not so common}

The fact that highly localised velocity perturbations expected from the interaction between discs and giant planets are only found in one of the five discs in the sample may indicate that massive planets do not commonly form at large separations from the central star, and if they do, they should migrate rapidly towards lower orbits. Alternatively, it is likely that the influence of other physical mechanisms such as gravitational and hydrodynamic instabilities along with the presence of pressure substructures, as well as the effect of multiple embedded planets and stellar companions, is sufficient to trigger deviations from Keplerian rotation strong enough to hide the localised features driven by single massive planets. A systematic characterisation of composite hydrodynamic simulations including several physical mechanisms at once will be key in the field of disc kinematics to unambiguously distinguish between planetary and non-planetary signatures retrieved from high resolution observations of protoplanetary discs.

\subsection{Correlation between pressure bumps and mm dust rings}

Using multiple molecular lines from the same sample of discs, \citet[][]{jiang+2022} found no clear (anti--)correlations between line intensities and the location of dust rings and gaps observed in mm continuum data, leaving an open debate on the mechanisms that led to the formation of the millimetre dust substructures present in these discs.
However, our analysis of azimuthal velocity flows indicates that at least nine of the eleven millimetre dust rings covered in this study do coexist with maxima in the radial gas pressure profile traced by different CO isotopologues. This finding indicates that there is a clear relationship between the gas pressure variations, more effectively captured by velocity modulations, and the substructures observed in the mm continuum, suggesting that gas pressure traps largely dominate the formation of dust rings in these sources.

\section{Conclusions} \label{sec:conclusions}

In this work we apply the \discminer{} modelling and analysis techniques on high angular and velocity resolution archival data of \twCO{}, \thCO{}, and \eiCO{} $J=2-1$ line emission from five discs around the young stars \mwc{}, \hd{}, \as{}, \im{}, and \gm{}, observed as part of the ALMA large program MAPS \citep[][]{oberg+2021}. To study deviations from Keplerian rotation and, more generally, to identify fluctuations in the underlying intensity and velocity fields, we first produce best-fit channel maps for all discs and CO isotopologues, assuming they are smooth in intensity and Keplerian in rotation velocity. Next, we compare morphological properties of the data and the resulting model line profiles, such as amplitude, width, and velocity shift, to search for localised and extended fluctuations hinting at velocity and density substructures in the discs. Finally, we apply a clustering algorithm to search for localised velocity and line width perturbations possibly driven by massive unseen planets, and study the presence of extended kinematic substructures that may also be the consequence of embedded planets. Our main findings are as follows,

\subsection{Giant planets around \hd{}, \mwc{}, \as{}}
\begin{itemize}
    \item Localised velocity features are only detected in the disc of \hd{}, where they are robustly seen in \twCO{} and \thCO{} simultaneously, and tentatively in \eiCO{}. These signatures are identified and confirmed around the location of the planet candidate P94, proposed by \citet[][]{izquierdo+2022} using \twCO{} observations with lower spectral resolution. Instead, large-scale kinematic and line width substructures are more common in the other discs. In the discs of \mwc{} and \as{}, we spot numerous ring--, arc-- and spiral-like signatures in all line profile observables, spanning over large azimuthal extents, which may also be driven by unseen planets. \smallskip 
    \item Some of the large-scale features observed in the kinematics of the discs around \mwc{} and \as{} are compatible with anti-parallel vertical flows of material. In \mwc{}, these meridional flows take place primarily in the atmosphere of the disc, while in \as{} they propagate freely from layers in the atmosphere down to regions closer to the disc midplane, as traced by \twCO{} and \thCO{} emission. This varying modulation of meridional velocities as function of scale height suggests that such signatures are likely linked to different driving mechanisms in each disc. Planet buoyancy spirals are favoured in \mwc{}, whereas MHD winds or vertical shear instabilities provide a better interpretation of the kinematic signatures observed in \as{}. \smallskip
    \item No localised velocity or intensity signatures are found around the location of the CPD candidate proposed by \citet[][]{bae+2022} in the disc of \as{} in any of the CO isotopologues. We emphasise that kinks extracted through visual inspection of intensity channels should not be attributed to localised kinematic features without prior inspection of the underlying velocity field. The presence of vertical flows in this disc, spanning over large azimuthal extents, are the cause of the strong kinks observed in \twCO{} channels around the orbital radius of the CPD candidate. \smallskip
    \item We detect both localised and extended line width enhancements in at least two CO isotopologues near the location of the planet candidate P94 in \hd{}, and around the vicinity of the CPD candidate proposed by \citet[][]{bae+2022} in \as{}. These features would be in agreement with turbulent motions triggered by massive embedded planets as theorised by \citet{dong+2019}. \smallskip
    \item Extended line width enhancements are also seen in the atmosphere of the disc around \mwc{}, in the same radial section where anti-parallel vertical flows are detected, namely between $\sim\!200-300$\,au. Likewise, this could be a hint of turbulent motions driven by a young massive planet embedded in this region of the disc. Using this information, we propose a planet candidate at an orbital radius of $R=245$\,au and an azimuth of $\phi=-128^\circ$ in this system. \smallskip
    \item Although not statistically representative, a ratio of four massive planets found at large orbital radii in five protoplanetary discs is noteworthy. If this ratio holds up in future surveys, this could indicate that gravitational instabilities and/or rapid planetary migration are common mechanisms of evolution and planet formation in circumstellar discs.
\end{itemize}

\subsection{Non-detections in the discs of \im{} and \gm{}}

\begin{itemize}
    \item The discs of \im{} and \gm{} do not exhibit localised nor large-scale velocity perturbations that can be associated with the presence of massive planets.
    \item Kinematic analysis of these discs does not reveal clear indications of spiral arms typical of gravitational instabilities either, despite the possibly high mass reservoir available in both systems \citep[][]{lodato+2022}. Strong spiral features are only detected in the temperature structure of the disc of \gm{} most notably at high elevations traced by \twCO{}, which are likely the result of interaction between the disc atmosphere and infalling material from the remnant envelope rather than a consequence of gravitational instabilities. \smallskip
    \item The fact that no clear hints of massive planets are found in the probably most massive discs in the sample is interesting but not statistically conclusive. A systematic study of observable signatures driven by embedded planets in massive environments, extracted from hydrodynamic simulations, would help understand if the disc mass has a significant impact on the detectability of planet-driven perturbations.
\end{itemize}

\subsection{Gas gaps, and relationship between pressure bumps and dust substructures}
\begin{itemize}
    \item Azimuthal averages of standard and absolute velocity residuals are sufficient to disentangle the vertical and azimuthal components of the velocity perturbations, respectively. Azimuthal velocity flows are prominent in all discs, with magnitudes of the order of $50-150$\,m\,s$^{-1}$, suggestive of strong pressure variations as function of orbital radius. Meridional flows are robustly detected in all discs, with average amplitudes lower than $50$\,m\,s$^{-1}$. No strong correlation between gas gaps, traced by positive azimuthal velocity gradients, and meridional flows is found.  \smallskip 
    \item Most of the pressure bumps and minima detected in all discs appear to be dominated by surface density fluctuations. Also, we identify prominent vertical variations in the radial pressure profile of the disc around \hd{}, in the same radial section where the planet candidate P94 is proposed. \smallskip
    \item Nine out of eleven millimetre dust continuum rings in the sample are co-located with pressure bumps traced by negative rotation velocity gradients, indicating that pressure traps are a common mechanism for the formation of dust substructures in these discs. \smallskip
    \item A simultaneous analysis of velocity and line width variations in (at least marginally) optically thick tracers provides a robust assessment of the underlying surface density substructures responsible for the configuration of axisymmetric azimuthal and vertical flows in gas discs.
\end{itemize}

Our study demonstrates that a comprehensive analysis of line profile observables, including velocity, line width and intensity, from multiple molecular tracers, provides unparalleled access to the three-dimensional physical structure of circumstellar discs, and is indispensable towards a robust detection and characterisation of planet formation sites.

\begin{acknowledgements}
The authors thank the anonymous referee for their insightful comments and discussions that helped us improve the presentation of results and the clarity of the methodologies introduced in this work.
This paper makes use of the following ALMA data:
ADS/JAO. ALMA\#2018.1.01055.L.
ALMA is a partnership of ESO (representing its member
states), NSF (USA) and NINS (Japan), together with NRC
(Canada), MOST and ASIAA (Taiwan), and KASI (Re- public
of Korea), in cooperation with the Republic of Chile. The
Joint ALMA Observatory is operated by ESO, AUI/NRAO
and NAOJ. The National Radio Astronomy Observatory is a
facility of the National Science Foundation operated under
cooperative agreement by Associated Universities, Inc.
This work was partly supported by the Italian Ministero dell Istruzione, Universit\`a e Ricerca through the grant Progetti Premiali 2012 – iALMA (CUP C$52$I$13000140001$), 
by the Deutsche Forschungs-gemeinschaft (DFG, German Research Foundation) - Ref no. FOR $2634$/$1$ TE $1024$/$1$-$1$, 
and by the DFG cluster of excellence Origins (www.origins-cluster.de). 
This project has received funding from the European Union's Horizon 2020 research and innovation programme under the Marie Sklodowska-Curie grant agreement No 823823 (DUSTBUSTERS) and from the European Research Council (ERC) via the ERC Synergy Grant {\em ECOGAL} (grant 855130). SF is funded by the European Union under the European Union's Horizon Europe Research \& Innovation Programme 101076613 (UNVEIL). GR is funded by the European Union (ERC DiscEvol, project number 101039651). GR acknowledges support from the Netherlands Organisation for Scientific Research (NWO, program number 016.Veni.192.233) and from an STFC Ernest Rutherford Fellowship (grant number ST/T003855/1). EvD acknowledges support from the ERC grant agreement No. 101019751 MOLDISK. Views and opinions expressed are however those of the author(s) only and do not necessarily reflect those of the European Union or the European Research Council Executive Agency. Neither the European Union nor the granting authority can be held responsible for them.
\end{acknowledgements}

%
%
\bibliographystyle{aa}
\bibliography{references}

\begin{appendix}

\section{Supporting tables} \label{sec:appendix_tables}

\setlength{\tabcolsep}{5.0pt} 

\begin{table*}
\centering
{\renewcommand{\arraystretch}{0.0}
\captionsetup{justification=centering}
 \caption{Best-fit model parameters obtained for channel maps of the disc of \mwc{}.}
  \label{table:pars_mwc480}
\begin{tabular}{ lrlccc } 

\toprule
 
 &  &  & \multicolumn{3}{c}{\mwc{} ($162.0$\,pc)} \\
\cmidrule(lr){4-6}
Attribute & Parameter & Unit & \twCO{} & \thCO{} & \eiCO{} \\
\midrule
Orientation & \makecell[r]{$i$ \\ PA \\ $x_c$ \\ $y_c$} & \makecell[l]{$[^\circ]$ \\\relax $[^\circ]$ \\\relax [au] \\\relax [au]} & \makecell[c]{$-38.6$ $[-37.0]$ \\ $327.9$ $[328.2]$ \\ $-0.9$ $[-3.2]$ \\ $-1.8$ $[0.7]$}& \makecell[c]{$-37.3$ $[-37.0]$ \\ $327.6$ $[327.6]$ \\ $-3.7$ $[-0.6]$ \\ $-1.9$ $[-1.0]$}& \makecell[c]{$-36.1$ $[-37.0]$ \\ $327.6$ $[327.6]$ \\ $-3.6$ $[2.3]$ \\ $0.1$ $[-3.2]$}\\
\midrule
Velocity & \makecell[r]{$M_\star$ \\ $\upsilon_{\rm sys}$} & \makecell[l]{[M$_\odot$] \\\relax [km s$^{-1}$]} & \makecell[c]{$1.97$ $[2.07]$ \\ $5.10$ $[5.10]$}& \makecell[c]{$2.08$ $[2.12]$ \\ $5.09$ $[5.10]$}& \makecell[c]{$2.22$ $[2.16]$ \\ $5.09$ $[5.10]$}\\
\midrule
Upper surface & \makecell[r]{$z_0$ \\ $p$\\ $R_t$ \\$q$} & \makecell[l]{[au] \\\relax [--] \\\relax [au] \\\relax [--]} & \makecell[c]{$52.7$ \\ $1.65$ \\ $108.6$ \\ $0.55$} & \makecell[c]{$12.4$ \\ $1.51$ \\ $307.7$ \\ $1.32$} & \makecell[c]{$5.4$ \\ $2.29$ \\ $82.9$ \\ $1.22$}  \vspace{0.07cm} \\
Lower surface & \makecell[r]{$z_0$ \\ $p$\\ $R_t$ \\$q$} & \makecell[l]{[au] \\\relax [--] \\\relax [au] \\\relax [--]} & \makecell[c]{$14.2$ \\ $1.11$ \\ $803.9$ \\ $2.86$}& \makecell[c]{$8.9$ \\ $1.08$ \\ $510.9$ \\ $3.60$}& \makecell[c]{$--$ \\ $--$ \\ $--$ \\ $--$}\\
\midrule
Peak intensity & \makecell[r]{$I_0$ \\ $p$ \\$q$} & \makecell[l]{[Jy px$^{-1}$] \\\relax [--] \\\relax [--]} & \makecell[c]{$2.61$ \\ $-2.66$ \\ $2.16$} & \makecell[c]{$4.08$ \\ $-2.29$ \\ $2.03$} & \makecell[c]{$0.23$ \\ $-0.63$ \\ $0.56$} \vspace{0.07cm} \\
Line width & \makecell[r]{$L_w$ \\ $p$ \\ $q$} & \makecell[l]{[km s$^{-1}$] \\\relax [--] \\\relax [--]} & \makecell[c]{$0.32$ \\ $-0.29$ \\ $-0.36$} & \makecell[c]{$0.16$ \\ $-0.33$ \\ $-0.45$} & \makecell[c]{$0.29$ \\ $-0.84$ \\ $-0.11$} \vspace{0.07cm} \\
Line slope & \makecell[r]{$L_s$ \\ $p$} & \makecell[l]{[--] \\\relax [--]} & \makecell[c]{$1.68$ \\ $0.31$} & \makecell[c]{$1.50$ \\ $0.16$} & \makecell[c]{$1.58$ \\ $0.09$} \vspace{0.05cm} \\
\bottomrule

\end{tabular}
\vspace{0.05cm}
\caption*{\textbf{Note.} Parameters in brackets result from fixing inclination to the value found for the millimetre continuum in \citet{liu+2019}.}
  }

\end{table*}
\setlength{\tabcolsep}{5.0pt} 

\begin{table*}
\centering
{\renewcommand{\arraystretch}{0.0}
\captionsetup{justification=centering}
 \caption{Best-fit model parameters obtained for channel maps of the disc of \hd{}.}
  \label{table:pars_hd163296}
\begin{tabular}{ lrlccc } 

\toprule
 
 &  &  & \multicolumn{3}{c}{\hd{} ($101.5$\,pc)} \\
\cmidrule(lr){4-6}
Attribute & Parameter & Unit & \twCO{} & \thCO{} & \eiCO{} \\
\midrule
Orientation & \makecell[r]{$i$ \\ PA \\ $x_c$ \\ $y_c$} & \makecell[l]{$[^\circ]$ \\\relax $[^\circ]$ \\\relax [au] \\\relax [au]} & \makecell[c]{$46.7$ $[46.7]$ \\ $312.8$ $[312.8]$ \\ $-2.2$ $[-1.4]$ \\ $0.3$ $[0.6]$}& \makecell[c]{$45.1$ $[46.7]$ \\ $312.8$ $[312.8]$ \\ $-1.6$ $[-1.9]$ \\ $0.4$ $[-0.3]$}& \makecell[c]{$45.9$ $[46.7]$ \\ $312.8$ $[312.8]$ \\ $-1.0$ $[0.4]$ \\ $-0.4$ $[-0.3]$}\\
\midrule
Velocity & \makecell[r]{$M_\star$ \\ $\upsilon_{\rm sys}$} & \makecell[l]{[M$_\odot$] \\\relax [km s$^{-1}$]} & \makecell[c]{$1.92$ $[1.91]$ \\ $5.77$ $[5.77]$}& \makecell[c]{$1.98$ $[1.91]$ \\ $5.77$ $[5.76]$}& \makecell[c]{$1.96$ $[1.92]$ \\ $5.76$ $[5.76]$}\\
\midrule
Upper surface & \makecell[r]{$z_0$ \\ $p$\\ $R_t$ \\$q$} & \makecell[l]{[au] \\\relax [--] \\\relax [au] \\\relax [--]} & \makecell[c]{$29.6$ \\ $1.29$ \\ $435.4$ \\ $1.66$} & \makecell[c]{$16.4$ \\ $1.20$ \\ $479.9$ \\ $1.86$} & \makecell[c]{$68.1$ \\ $2.56$ \\ $23.2$ \\ $0.51$}  \vspace{0.07cm} \\
Lower surface & \makecell[r]{$z_0$ \\ $p$\\ $R_t$ \\$q$} & \makecell[l]{[au] \\\relax [--] \\\relax [au] \\\relax [--]} & \makecell[c]{$16.7$ \\ $1.16$ \\ $559.0$ \\ $3.50$}& \makecell[c]{$12.3$ \\ $1.28$ \\ $440.5$ \\ $3.71$}& \makecell[c]{$7.2$ \\ $1.87$ \\ $346.2$ \\ $2.94$}\\
\midrule
Peak intensity & \makecell[r]{$I_0$ \\ $p$ \\$q$} & \makecell[l]{[Jy px$^{-1}$] \\\relax [--] \\\relax [--]} & \makecell[c]{$0.57$ \\ $-1.96$ \\ $1.67$} & \makecell[c]{$1.65$ \\ $-2.61$ \\ $2.38$} & \makecell[c]{$6.11$ \\ $-4.00$ \\ $2.50$} \vspace{0.07cm} \\
Line width & \makecell[r]{$L_w$ \\ $p$ \\ $q$} & \makecell[l]{[km s$^{-1}$] \\\relax [--] \\\relax [--]} & \makecell[c]{$0.32$ \\ $-0.20$ \\ $-0.47$} & \makecell[c]{$0.22$ \\ $-0.32$ \\ $-0.43$} & \makecell[c]{$0.07$ \\ $0.06$ \\ $-0.68$} \vspace{0.07cm} \\
Line slope & \makecell[r]{$L_s$ \\ $p$} & \makecell[l]{[--] \\\relax [--]} & \makecell[c]{$1.88$ \\ $0.16$} & \makecell[c]{$1.67$ \\ $0.15$} & \makecell[c]{$1.58$ \\ $0.07$} \vspace{0.05cm} \\
\bottomrule

\end{tabular}
\vspace{0.05cm}
\caption*{\textbf{Note.} Parameters in brackets result from fixing inclination to the value found for the millimetre continuum in \citet{huang+2018_incl}.}
  }

\end{table*}
\setlength{\tabcolsep}{5.0pt} 

\begin{table*}
\centering
{\renewcommand{\arraystretch}{0.0}
\captionsetup{justification=centering}
 \caption{Best-fit model parameters obtained for channel maps of the disc of \as{}.}
  \label{table:pars_as209}
\begin{tabular}{ lrlccc } 

\toprule
 
 &  &  & \multicolumn{3}{c}{\as{} ($121.0$\,pc)} \\
\cmidrule(lr){4-6}
Attribute & Parameter & Unit & \twCO{} & \thCO{} & \eiCO{} \\
\midrule
Orientation & \makecell[r]{$i$ \\ PA \\ $x_c$ \\ $y_c$} & \makecell[l]{$[^\circ]$ \\\relax $[^\circ]$ \\\relax [au] \\\relax [au]} & \makecell[c]{$-37.3$ $[-35.0]$ \\ $85.6$ $[85.2]$ \\ $0.0$ $[0.6]$ \\ $-1.8$ $[-1.6]$}& \makecell[c]{$-33.5$ $[-35.0]$ \\ $85.8$ $[85.9]$ \\ $-1.5$ $[-0.7]$ \\ $-2.9$ $[-0.3]$}& \makecell[c]{$-32.9$ $[-35.0]$ \\ $85.5$ $[85.4]$ \\ $-1.3$ $[-1.8]$ \\ $-3.7$ $[-6.7]$}\\
\midrule
Velocity & \makecell[r]{$M_\star$ \\ $\upsilon_{\rm sys}$} & \makecell[l]{[M$_\odot$] \\\relax [km s$^{-1}$]} & \makecell[c]{$1.14$ $[1.27]$ \\ $4.64$ $[4.63]$}& \makecell[c]{$1.35$ $[1.26]$ \\ $4.66$ $[4.66]$}& \makecell[c]{$1.41$ $[1.28]$ \\ $4.66$ $[4.66]$}\\
\midrule
Upper surface & \makecell[r]{$z_0$ \\ $p$\\ $R_t$ \\$q$} & \makecell[l]{[au] \\\relax [--] \\\relax [au] \\\relax [--]} & \makecell[c]{$36.7$ \\ $3.20$ \\ $105.9$ \\ $1.30$} & \makecell[c]{$10.2$ \\ $1.57$ \\ $190.8$ \\ $7.04$} & \makecell[c]{$7.0$ \\ $3.56$ \\ $153.9$ \\ $7.05$}  \vspace{0.07cm} \\
Lower surface & \makecell[r]{$z_0$ \\ $p$\\ $R_t$ \\$q$} & \makecell[l]{[au] \\\relax [--] \\\relax [au] \\\relax [--]} & \makecell[c]{$7.4$ \\ $2.03$ \\ $322.4$ \\ $7.37$}& \makecell[c]{$25.3$ \\ $3.63$ \\ $96.0$ \\ $1.78$}& \makecell[c]{$--$ \\ $--$ \\ $--$ \\ $--$}\\
\midrule
Peak intensity & \makecell[r]{$I_0$ \\ $p$ \\$q$} & \makecell[l]{[Jy px$^{-1}$] \\\relax [--] \\\relax [--]} & \makecell[c]{$0.072$ \\ $-1.84$ \\ $0.84$} & \makecell[c]{$0.132$ \\ $-1.41$ \\ $1.31$} & \makecell[c]{$0.004$ \\ $-1.32$ \\ $0.39$} \vspace{0.07cm} \\
Line width & \makecell[r]{$L_w$ \\ $p$ \\ $q$} & \makecell[l]{[km s$^{-1}$] \\\relax [--] \\\relax [--]} & \makecell[c]{$0.33$ \\ $-0.42$ \\ $-0.11$} & \makecell[c]{$0.17$ \\ $-0.24$ \\ $-0.25$} & \makecell[c]{$0.27$ \\ $1.83$ \\ $0.38$} \vspace{0.07cm} \\
Line slope & \makecell[r]{$L_s$ \\ $p$} & \makecell[l]{[--] \\\relax [--]} & \makecell[c]{$1.71$ \\ $0.14$} & \makecell[c]{$1.46$ \\ $0.48$} & \makecell[c]{$0.82$ \\ $1.52$} \vspace{0.05cm} \\
\bottomrule

\end{tabular}
\vspace{0.1cm}
\caption*{\textbf{Note.} Parameters in brackets result from fixing inclination to the value found for the millimetre continuum in \citet{huang+2018_incl}.}

  }
\end{table*}
\setlength{\tabcolsep}{5.0pt} 

\begin{table*}
\centering
{\renewcommand{\arraystretch}{0.0}
\captionsetup{justification=centering}
 \caption{Best-fit model parameters obtained for channel maps of the disc of \im{}.}
  \label{table:pars_imlup}
\begin{tabular}{ lrlccc } 

\toprule
 
 &  &  & \multicolumn{3}{c}{\im{} ($158.0$\,pc)} \\
\cmidrule(lr){4-6}
Attribute & Parameter & Unit & \twCO{} & \thCO{} & \eiCO{} \\
\midrule
Orientation & \makecell[r]{$i$ \\ PA \\ $x_c$ \\ $y_c$} & \makecell[l]{$[^\circ]$ \\\relax $[^\circ]$ \\\relax [au] \\\relax [au]} & \makecell[c]{$-52.3$ $[-47.5]$ \\ $145.0$ $[144.5]$ \\ $-2.9$ $[2.1]$ \\ $0.8$ $[-1.5]$}& \makecell[c]{$-49.5$ $[-47.5]$ \\ $144.2$ $[144.3]$ \\ $-0.0$ $[0.5]$ \\ $-2.7$ $[-3.7]$}& \makecell[c]{$-49.9$ $[-47.5]$ \\ $144.2$ $[144.1]$ \\ $0.9$ $[-3.8]$ \\ $-2.2$ $[-0.7]$}\\
\midrule
Velocity & \makecell[r]{$M_\star$ \\ $\upsilon_{\rm sys}$} & \makecell[l]{[M$_\odot$] \\\relax [km s$^{-1}$]} & \makecell[c]{$1.01$ $[1.14]$ \\ $4.52$ $[4.52]$}& \makecell[c]{$1.11$ $[1.16]$ \\ $4.52$ $[4.52]$}& \makecell[c]{$1.13$ $[1.18]$ \\ $4.50$ $[4.50]$}\\
\midrule
Upper surface & \makecell[r]{$z_0$ \\ $p$\\ $R_t$ \\$q$} & \makecell[l]{[au] \\\relax [--] \\\relax [au] \\\relax [--]} & \makecell[c]{$27.0$ \\ $1.38$ \\ $858.0$ \\ $1.13$} & \makecell[c]{$20.6$ \\ $1.28$ \\ $560.5$ \\ $1.68$} & \makecell[c]{$34.2$ \\ $2.81$ \\ $60.9$ \\ $0.62$}  \vspace{0.07cm} \\
Lower surface & \makecell[r]{$z_0$ \\ $p$\\ $R_t$ \\$q$} & \makecell[l]{[au] \\\relax [--] \\\relax [au] \\\relax [--]} & \makecell[c]{$20.9$ \\ $1.18$ \\ $936.5$ \\ $4.31$}& \makecell[c]{$18.5$ \\ $1.21$ \\ $682.4$ \\ $1.87$}& \makecell[c]{$21.3$ \\ $2.58$ \\ $104.5$ \\ $0.65$}\\
\midrule
Peak intensity & \makecell[r]{$I_0$ \\ $p$ \\$q$} & \makecell[l]{[Jy px$^{-1}$] \\\relax [--] \\\relax [--]} & \makecell[c]{$2.48$ \\ $-3.71$ \\ $2.94$} & \makecell[c]{$1.94$ \\ $-2.93$ \\ $2.79$} & \makecell[c]{$9.82$ \\ $-5.46$ \\ $2.90$} \vspace{0.07cm} \\
Line width & \makecell[r]{$L_w$ \\ $p$ \\ $q$} & \makecell[l]{[km s$^{-1}$] \\\relax [--] \\\relax [--]} & \makecell[c]{$0.51$ \\ $-0.31$ \\ $-0.21$} & \makecell[c]{$0.79$ \\ $-0.84$ \\ $0.33$} & \makecell[c]{$0.59$ \\ $-0.80$ \\ $0.11$} \vspace{0.07cm} \\
Line slope & \makecell[r]{$L_s$ \\ $p$} & \makecell[l]{[--] \\\relax [--]} & \makecell[c]{$1.82$ \\ $0.18$} & \makecell[c]{$1.41$ \\ $0.23$} & \makecell[c]{$1.30$ \\ $0.31$} \vspace{0.05cm} \\
\bottomrule

\end{tabular}
\vspace{0.1cm}
\caption*{\textbf{Note.} Parameters in brackets result from fixing inclination to the value found for the millimetre continuum in \citet{huang+2018_incl}.}
  }

\end{table*}
\setlength{\tabcolsep}{5.0pt} 

\begin{table*}
\centering
{\renewcommand{\arraystretch}{0.0}
\captionsetup{justification=centering}
 \caption{Best-fit model parameters obtained for channel maps of the disc of \gm{}.}
  \label{table:pars_gmaur}
\begin{tabular}{ lrlccc } 

\toprule
 
 &  &  & \multicolumn{3}{c}{\gm{} ($159.0$\,pc)} \\
\cmidrule(lr){4-6}
Attribute & Parameter & Unit & \twCO{} & \thCO{} & \eiCO{} \\
\midrule
Orientation & \makecell[r]{$i$ \\ PA \\ $x_c$ \\ $y_c$} & \makecell[l]{$[^\circ]$ \\\relax $[^\circ]$ \\\relax [au] \\\relax [au]} & \makecell[c]{$50.8$ $[53.2]$ \\ $53.8$ $[54.0]$ \\ $-0.1$ $[1.6]$ \\ $-7.7$ $[-4.4]$}& \makecell[c]{$54.0$ $[53.2]$ \\ $57.1$ $[56.6]$ \\ $-2.4$ $[-1.9]$ \\ $-2.5$ $[-4.3]$}& \makecell[c]{$49.1$ $[53.2]$ \\ $57.4$ $[57.1]$ \\ $2.2$ $[-3.1]$ \\ $4.7$ $[-3.2]$}\\
\midrule
Velocity & \makecell[r]{$M_\star$ \\ $\upsilon_{\rm sys}$} & \makecell[l]{[M$_\odot$] \\\relax [km s$^{-1}$]} & \makecell[c]{$1.01$ $[0.98]$ \\ $5.63$ $[5.63]$}& \makecell[c]{$1.11$ $[1.10]$ \\ $5.62$ $[5.62]$}& \makecell[c]{$1.30$ $[1.15]$ \\ $5.60$ $[5.60]$}\\
\midrule
Upper surface & \makecell[r]{$z_0$ \\ $p$\\ $R_t$ \\$q$} & \makecell[l]{[au] \\\relax [--] \\\relax [au] \\\relax [--]} & \makecell[c]{$53.3$ \\ $1.72$ \\ $177.5$ \\ $0.53$} & \makecell[c]{$15.8$ \\ $1.52$ \\ $475.2$ \\ $2.50$} & \makecell[c]{$70.2$ \\ $0.70$ \\ $33.0$ \\ $0.51$}  \vspace{0.07cm} \\
Lower surface & \makecell[r]{$z_0$ \\ $p$\\ $R_t$ \\$q$} & \makecell[l]{[au] \\\relax [--] \\\relax [au] \\\relax [--]} & \makecell[c]{$23.3$ \\ $1.50$ \\ $770.5$ \\ $0.96$}& \makecell[c]{$13.1$ \\ $1.50$ \\ $488.9$ \\ $3.84$}& \makecell[c]{$--$ \\ $--$ \\ $--$ \\ $--$}\\
\midrule
Peak intensity & \makecell[r]{$I_0$ \\ $p$ \\$q$} & \makecell[l]{[Jy px$^{-1}$] \\\relax [--] \\\relax [--]} & \makecell[c]{$1.33$ \\ $-3.33$ \\ $2.49$} & \makecell[c]{$0.26$ \\ $-2.95$ \\ $1.82$} & \makecell[c]{$0.87$ \\ $-0.60$ \\ $2.60$} \vspace{0.07cm} \\
Line width & \makecell[r]{$L_w$ \\ $p$ \\ $q$} & \makecell[l]{[km s$^{-1}$] \\\relax [--] \\\relax [--]} & \makecell[c]{$0.28$ \\ $0.29$ \\ $-0.76$} & \makecell[c]{$0.20$ \\ $-0.09$ \\ $-0.45$} & \makecell[c]{$0.13$ \\ $-0.77$ \\ $-0.61$} \vspace{0.07cm} \\
Line slope & \makecell[r]{$L_s$ \\ $p$} & \makecell[l]{[--] \\\relax [--]} & \makecell[c]{$1.90$ \\ $-0.02$} & \makecell[c]{$1.51$ \\ $0.17$} & \makecell[c]{$1.42$ \\ $0.11$} \vspace{0.05cm} \\
\bottomrule

\end{tabular}
\vspace{0.1cm}
\caption*{\textbf{Note.} Parameters in brackets result from fixing inclination to the value found for the millimetre continuum in \citet{huang+2020}.}
  }

\end{table*}


\setlength{\tabcolsep}{5.0pt} 

\begin{table*}
\centering
{\renewcommand{\arraystretch}{2.0}

\caption{As Table \ref{table:radial_subst_mwc480} but for the disc of \as{}.}
  \label{table:radial_subst_as209}
  
\begin{tabular}{ rcccccl } 

\toprule
\toprule
Radius & Dust$^a$ & Vertical$^b$ & $\Delta\upsilon_\phi$ & T$_{\rm b}$ & L$_w$ & Notes \\ 
\midrule

61 & \bluearrow{} & / & \redplus{}\blueminus{}/ & / & \bluearrow{}\bluearrow{}/ & \makecell[l]{Candidate gas gap with strong vertical pressure variations.} \\

100 & \bluearrow{} & / & \redplus{}\redplus{}/ & \redarrow{}// & \Bluearrow{} & \makecell[l]{Gas gap detected in kinematics and line widths. \\Positive velocity gradients more closely centred at $R=90$\,au.} \\

\midrule

74 & \redarrow{} & / & \blueminus{}\redplus{}/ & \bluearrow{}\bluearrow{}/ & \redarrow{}\redarrow{}/ & \makecell[l]{Pressure bump and minimum traced by \twCO{} and \thCO{}. \\ Indication of strong vertical pressure variations.} \\

121 & \redarrow{} & / & \blueminus{}/\redplus{} & / & / & \makecell[l]{Strong vertical pressure variations traced by the kinematics.} \\

\midrule

140 & / & / & \blueminus{}/\redplus{} & \bluearrow{}\redarrow{}\redarrow{} & / & \makecell[l]{Strong vertical pressure variations traced by the kinematics.} \\ 

160 & / & / & /\blueminus{}\blueminus{} & / & /\redarrow{}\bluearrow{} & \makecell[l]{Pressure bump traced by kinematics near the midplane.} \\ 

200 & / & / & \blueminus{}\blueminus{}/ & \bluearrow{}\redarrow{}/ & \redarrow{}\bluearrow{}/ & \makecell[l]{Pressure bump detected in the kinematics, mainly in \twCO{}.} \\

\bottomrule

\end{tabular}

  }
\end{table*}

\setlength{\tabcolsep}{5.0pt} 

\begin{table*}
\centering
{\renewcommand{\arraystretch}{2.0}

\caption{As Table \ref{table:radial_subst_mwc480} but for the disc of \im{}.}
  \label{table:radial_subst_imlup}
  
\begin{tabular}{ rcccccl } 

\toprule
\toprule
Radius & Dust$^a$ & Vertical$^b$ & $\Delta\upsilon_\phi$ & T$_{\rm b}$ & L$_w$ & Notes \\ 
\midrule

116 & \bluearrow{} & / & \Redplus{} & \Redarrow{} & \bluearrow{}\bluearrow{}/ & \makecell[l]{Warm gas gap detected in kinematics and line widths.} \\

209 & \bluearrow{} & / & \Redplus{} & \redarrow{}// & \bluearrow{}// & \makecell[l]{Tentative gas gap from kinematics and \twCO{} line widths.} \\

\midrule

133 & \redarrow{} & / & \Blueminus{} & / & / & \makecell[l]{Pressure bump detected in kinematics of all COs.} \\

220 & \redarrow{} & / & \Redplus{} & \redarrow{}// & / & \makecell[l]{Dust ring co-located with pressure minimum, suggesting \\ formation mechanism different from pressure trapping.} \\

\midrule

279 & / & \bluearrow{} & \blueminus{}/\blueminus{} & / & / & \makecell[l]{Pressure bump coincident with vertical modulation.} \\ 

393 & / & \bluearrow{} & \blueminus{}\blueminus{}/ & / & \bluearrow{}// & \makecell[l]{Pressure bump coincident with vertical modulation.} \\ 

\bottomrule

\end{tabular}
  
  }
\end{table*}

\setlength{\tabcolsep}{5.0pt} 

\begin{table*}
\centering
{\renewcommand{\arraystretch}{2.0}

\caption{As Table \ref{table:radial_subst_mwc480} but for the disc of \gm{}.}
  \label{table:radial_subst_gmaur}
  
\begin{tabular}{ rcccccl } 

\toprule
\toprule
Radius & Dust$^a$ & Vertical$^b$ & $\Delta\upsilon_\phi$ & T$_{\rm b}$ & L$_w$ & Notes \\ 
\midrule

68 & \bluearrow{} & / & \Redplus{} & /\redarrow{}\redarrow{} & \bluearrow{}/\bluearrow{} & \makecell[l]{Gas gap detected in kinematics and line widths.} \\

142 & \bluearrow{} & / & \redplus{}\redplus{}/ & \redarrow{}// & / & \makecell[l]{Tentative gas gap from \twCO{} and \thCO{} kinematics.} \\

\midrule

86 & \redarrow{} & / & \Blueminus{} & \Bluearrow{} & \redarrow{}/\redarrow{} & \makecell[l]{Strong pressure bump, gas density ring co-located \\ with dust ring.} \\

163 & \redarrow{} & / & / & / & //\redarrow{} & \makecell[l]{Dust ring not clearly correlated with pressure bump.} \\

\bottomrule

\end{tabular}
  
  }
\end{table*}

\section{Supporting figures} \label{sec:appendix_figures}

\begin{figure*}
   \centering
    \includegraphics[width=0.8\textwidth]{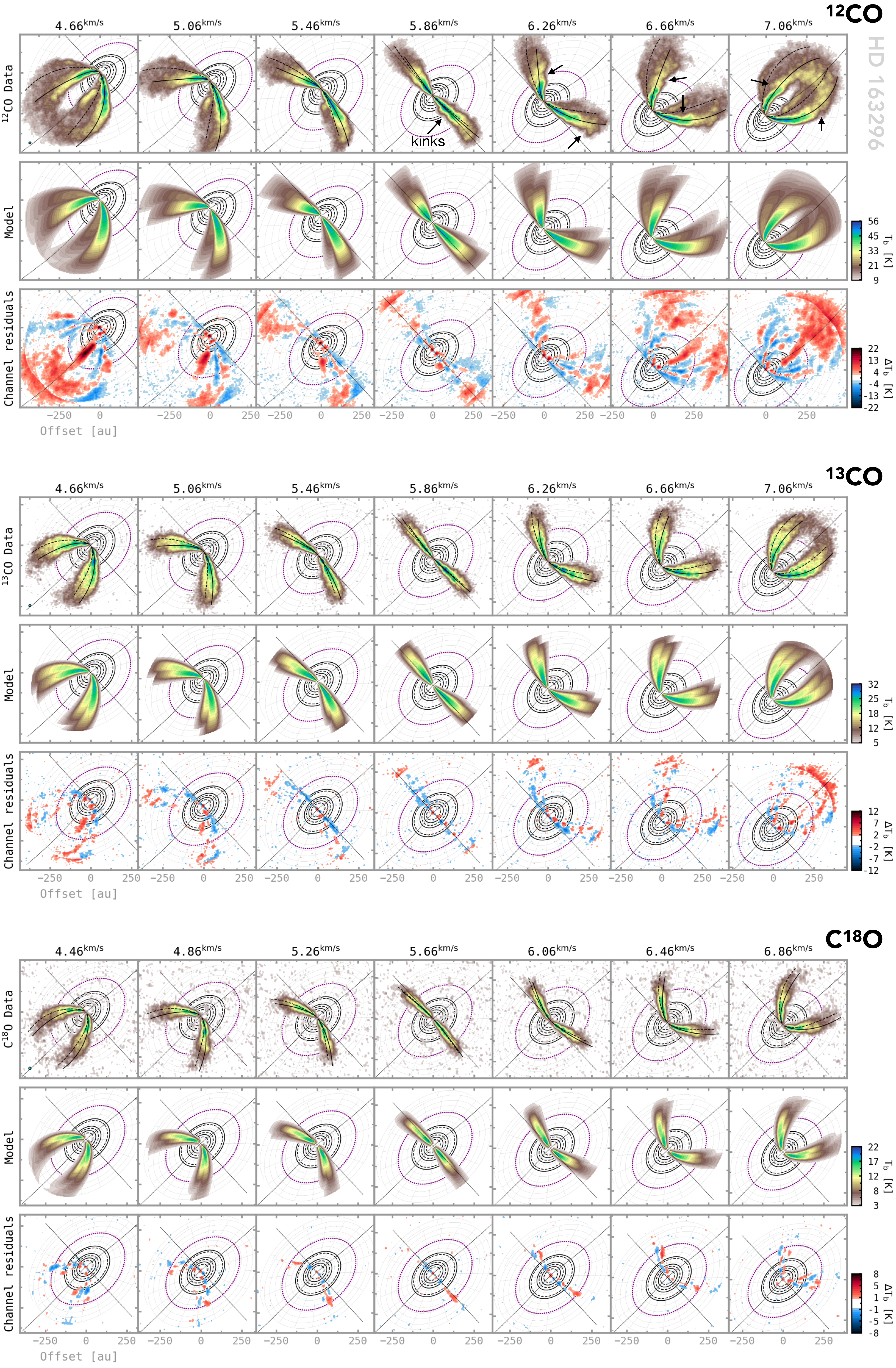}
      \caption{As Fig. \ref{fig:channel_maps_mwc480} but for the disc of \hd{}. 
              }
         \label{fig:channel_maps_hd163296}
   \end{figure*}

\begin{figure*}
   \centering
    \includegraphics[width=0.8\textwidth]{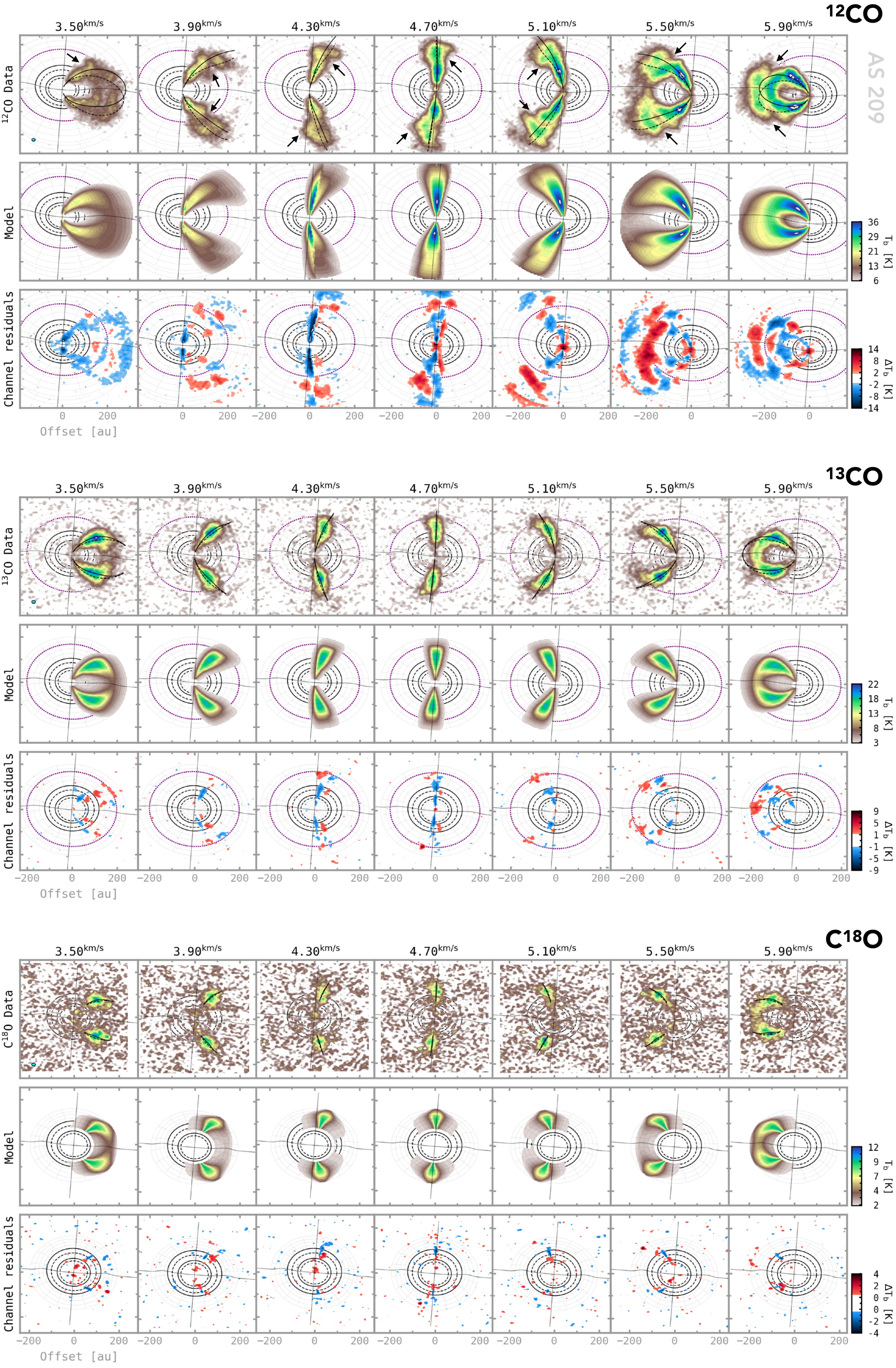}
      \caption{As Fig. \ref{fig:channel_maps_mwc480} but for the disc of \as{}. 
              }
         \label{fig:channel_maps_as209}
   \end{figure*}

\begin{figure*}
   \centering
    \includegraphics[width=0.8\textwidth]{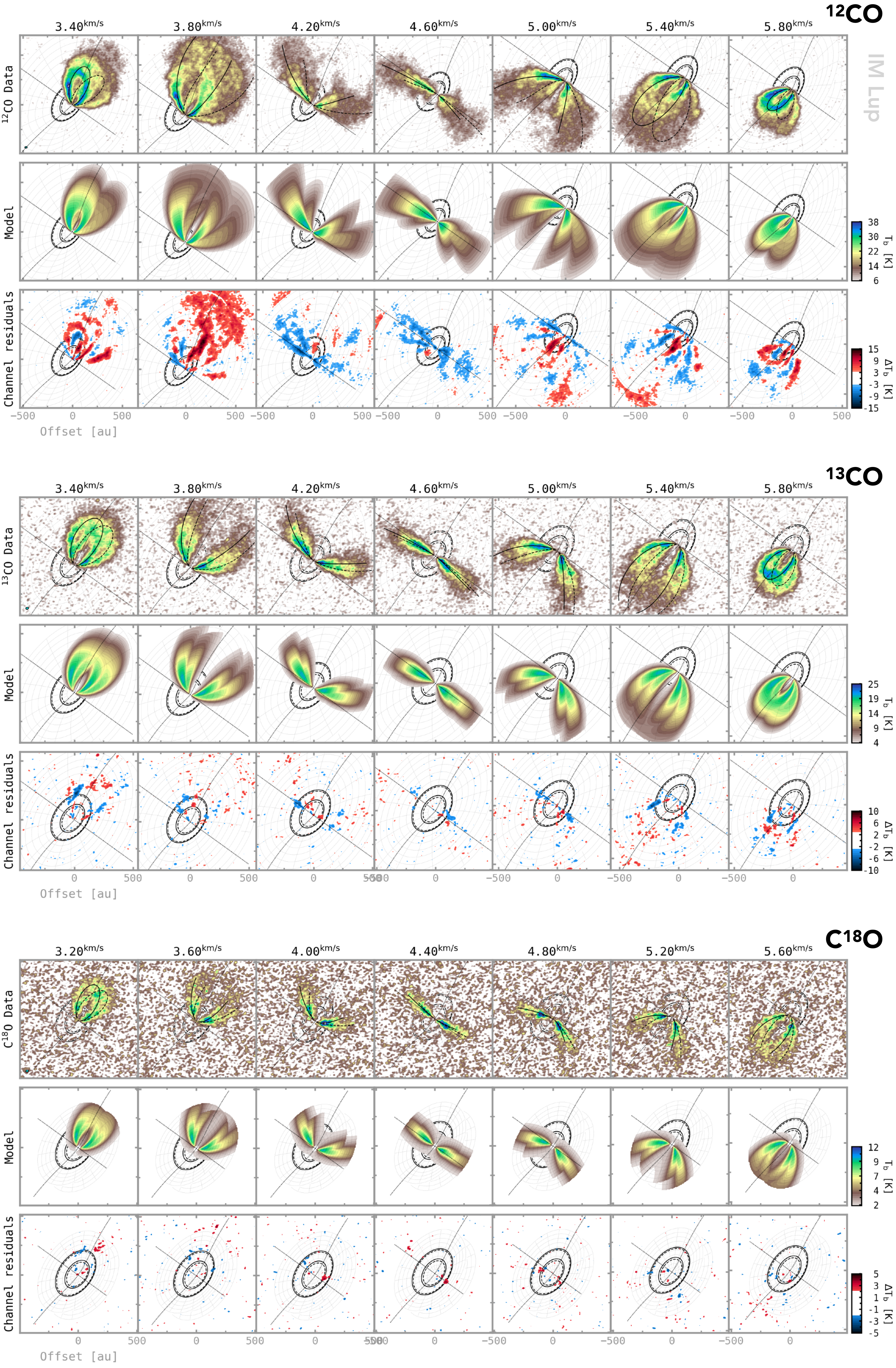}
      \caption{As Fig. \ref{fig:channel_maps_mwc480} but for the disc of \im{}. 
              }
         \label{fig:channel_maps_imlup}
   \end{figure*}

\begin{figure*}
   \centering
    \includegraphics[width=0.8\textwidth]{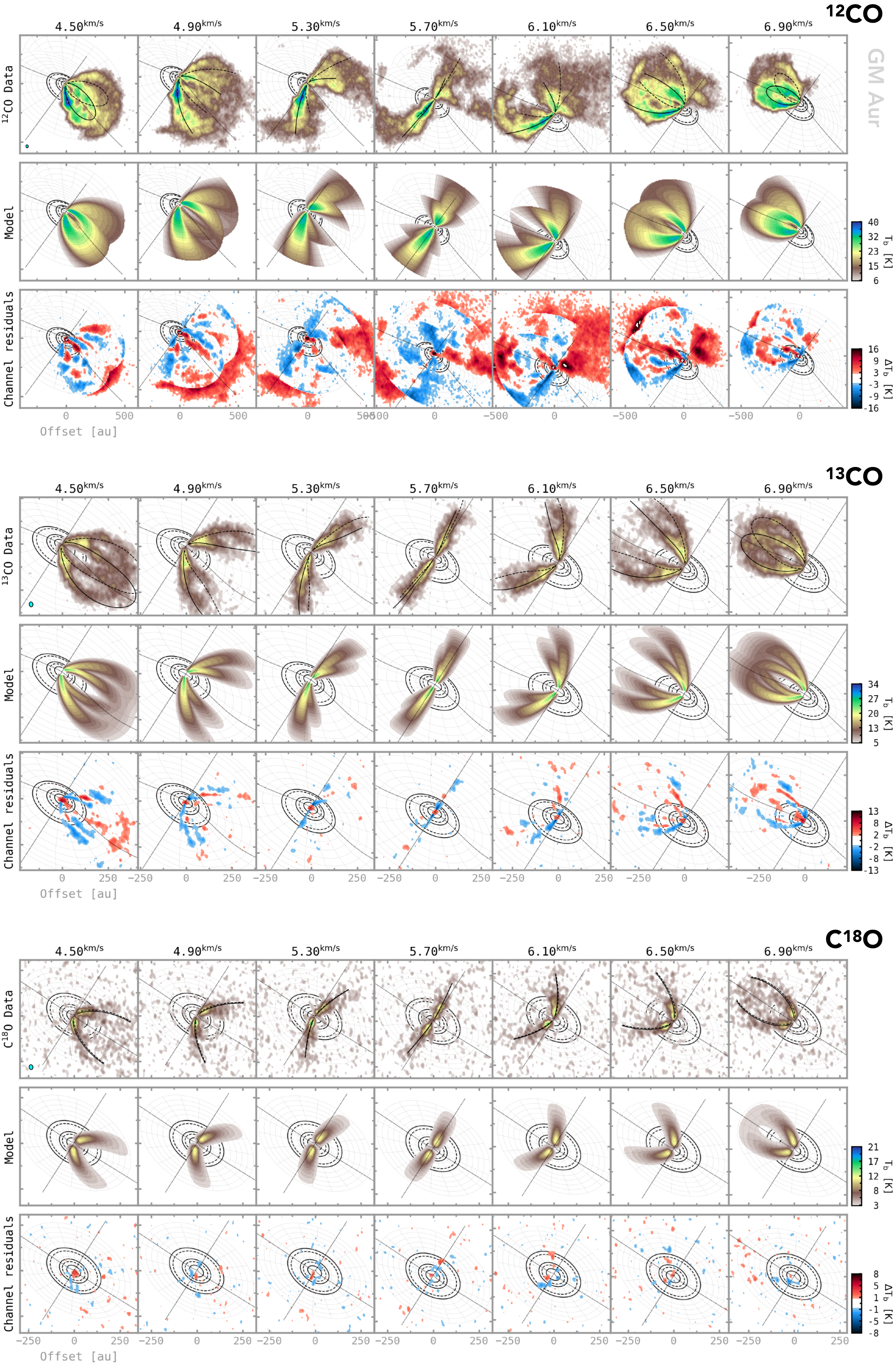}
      \caption{As Fig. \ref{fig:channel_maps_mwc480} but for the disc of \gm{}. 
              }
         \label{fig:channel_maps_gmaur}
   \end{figure*}

Here we provide complementary figures that support some of the results of the work, or figures that were not presented in the main text for reasons of space. In Figures \ref{fig:channel_maps_hd163296}--\ref{fig:channel_maps_gmaur}, we compare data and model channel maps for all CO isotopologues and discs but \mwc{}, whose channel maps were already shown in Fig. \ref{fig:channel_maps_mwc480}. In Figure \ref{fig:attributes_all_co}, we present rotation curves and azimuthally averaged profiles of peak intensities and line widths, measured on the data alone for all isotopologues, considering the disc vertical structure which is also shown. In Figure \ref{fig:discminer_workflow}, we illustrate the \discminer{} modelling flow and analysis capabilities. We show residual maps of \thCO{} and \eiCO{} centroid velocities in Figures \ref{fig:residuals_kinematics_13co} and \ref{fig:residuals_kinematics_c18o}. Figures \ref{fig:residuals_linewidth_12co}, \ref{fig:residuals_linewidth_13co}, and \ref{fig:residuals_linewidth_c18o}, illustrate \twCO{}, \thCO{}, and \eiCO{} line width residuals, respectively, and Figures \ref{fig:residuals_peakint_12co}, \ref{fig:residuals_peakint_13co}, and \ref{fig:residuals_peakint_c18o} show peak intensity residuals for the same tracers. In Figure \ref{fig:surface_residuals_mwc480}, we show the effect of using parametric and non-parametric emission surfaces on the retrieved intensity and velocity residuals from the \twCO{} disc of \mwc{}. In Figure \ref{fig:absorption_as209}, we illustrate the impact of assuming an azimuthally symmetric and asymmetric intensity model (accounting for foreground absorption) on the resulting intensity and velocity residuals from the \twCO{} disc of \as{}. In Figures \ref{fig:averaged_residuals_as209}, \ref{fig:averaged_residuals_imlup}, and \ref{fig:averaged_residuals_gmaur}, we present radial profiles of azimuthal and vertical velocity components, as well as azimuthally averaged line width and peak intensity residuals for the discs of \as{}, \im{} and \gm{}, respectively, for all isotopologues. A version of the same plot can be found in the main text in Fig. \ref{fig:averaged_residuals_mwc480} for \mwc{}, and in Fig. \ref{fig:averaged_residuals_hd163296} for \hd{}. Finally, in Figure \ref{fig:wcont_vprofiles_mwc480}, we show the effect of continuum subtraction on the velocity profiles extracted for the disc of \mwc{}.

\begin{figure*}
   \centering
   \includegraphics[width=1.0\textwidth]{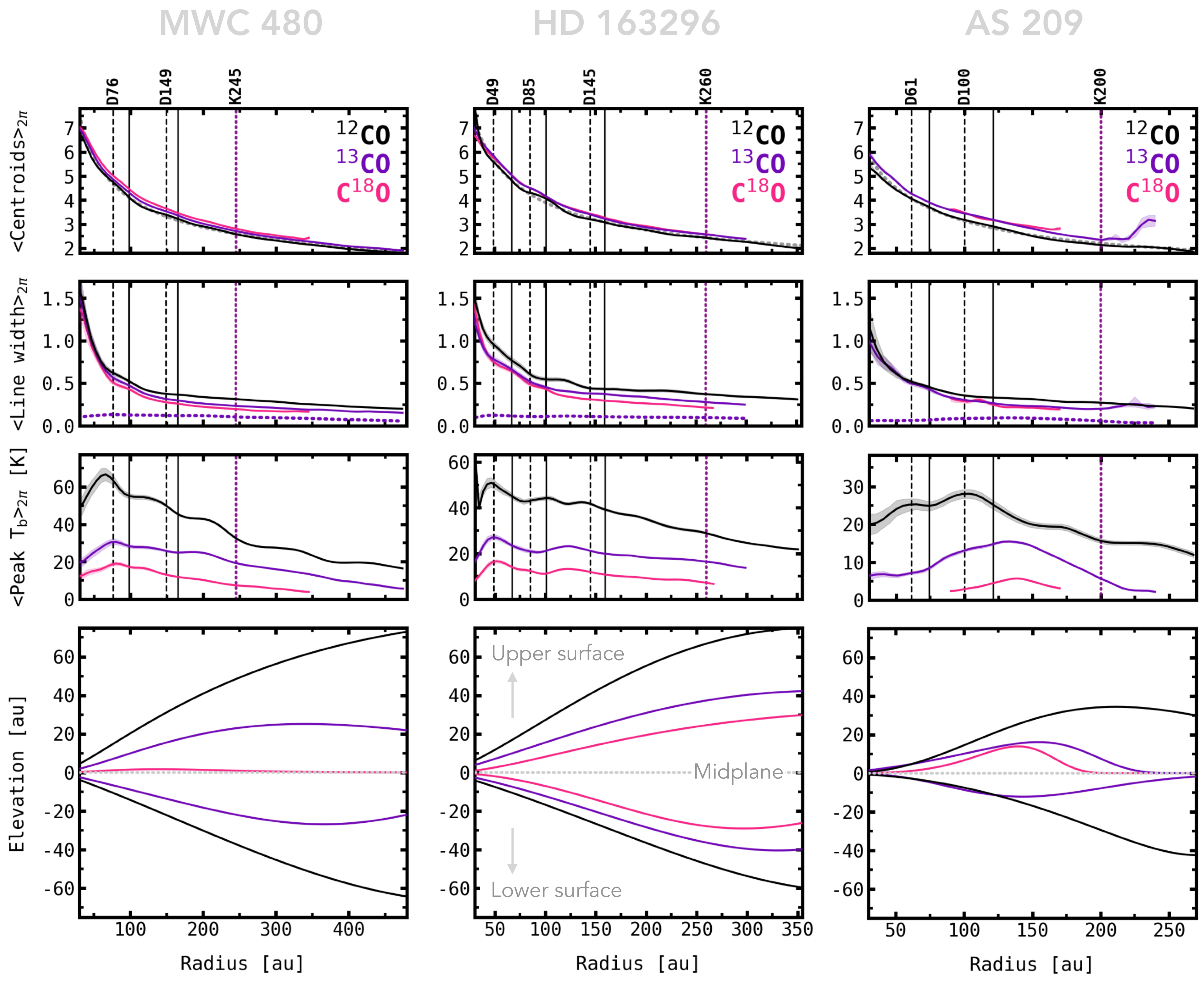}
      \caption{Additional observables obtained from the \discminer{} analysis of the discs around \mwc{}, \hd{}, \as{}, (\im{} and \gm{} in next page) as traced by \twCO{}, \thCO{}, and \eiCO{}. \textit{Top row}: Rotation curves computed from azimuthal averages of deprojected line-of-sight velocities. For reference, the grey dashed line in each panel is the Keplerian rotation curve of the best-fit model found for \twCO{}. 
      \textit{Second row}: Azimuthally averaged profiles of line widths, defined as the standard deviation of Gaussians fitted to the data line profiles. For reference, the purple dashed line in each panel represents the thermal broadening computed at the \thCO{} peak brightness temperature.
      \textit{Third row}: Azimuthally averaged profiles of peak intensities converted to brightness temperatures using the Rayleigh-Jeans law. 
      \textit{Bottom row}: Elevation of upper and lower emission surfaces. Apart from intensity profiles, all panels in each row share the same $x$ and $y-$\!axis extent. Vertical dashed, solid, and dotted lines mark the radial location of dust gaps, rings, and kinks observed in the channel maps of \twCO{}, respectively.
              }
         \label{fig:attributes_all_co}
   \end{figure*}

\begin{figure*}
\ContinuedFloat
   \centering
   \includegraphics[width=0.7\textwidth]{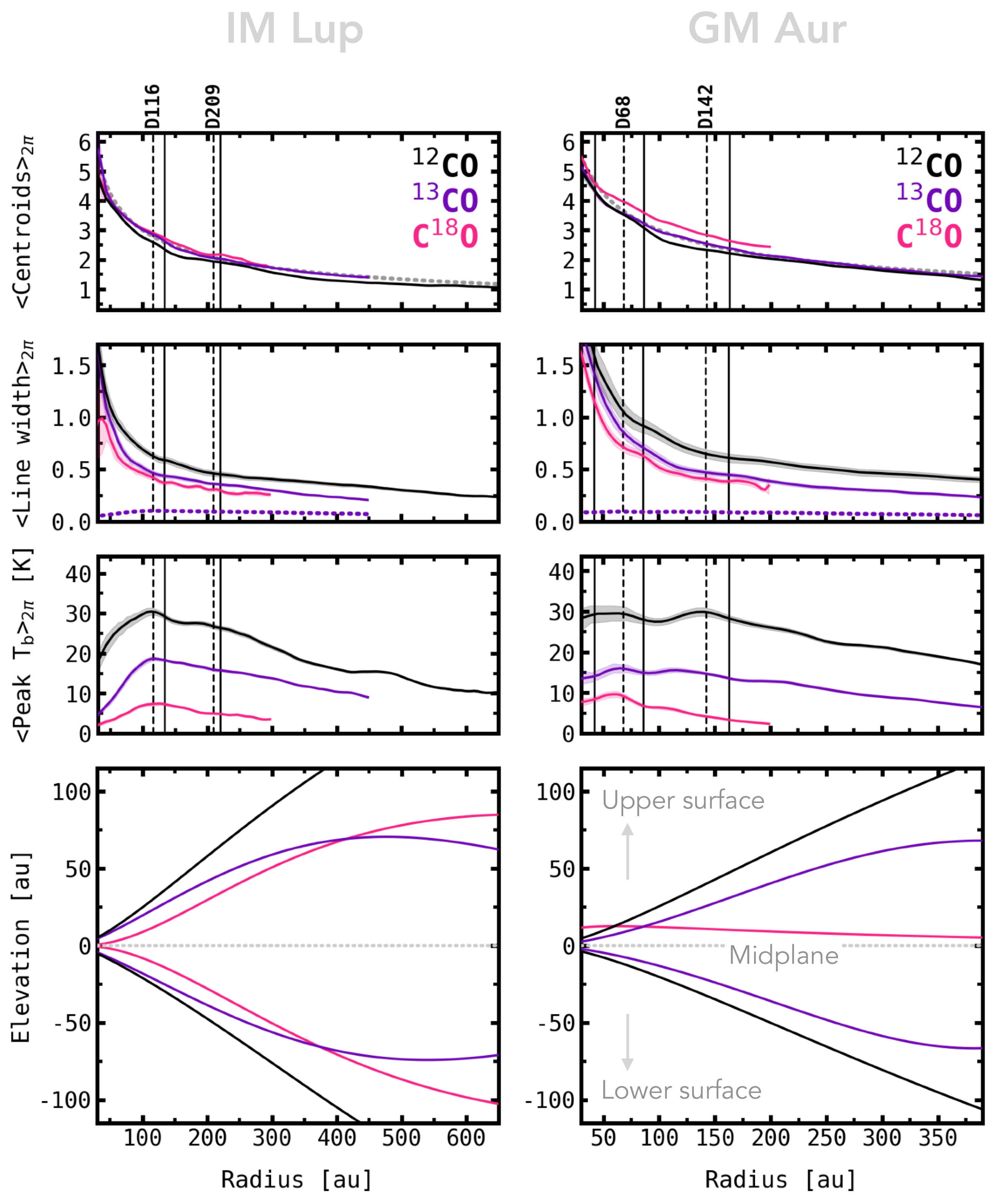}
      \caption{-- \textit{continued}
              }
   \end{figure*}

\begin{figure*}
   \centering
   \includegraphics[width=1.0\textwidth]{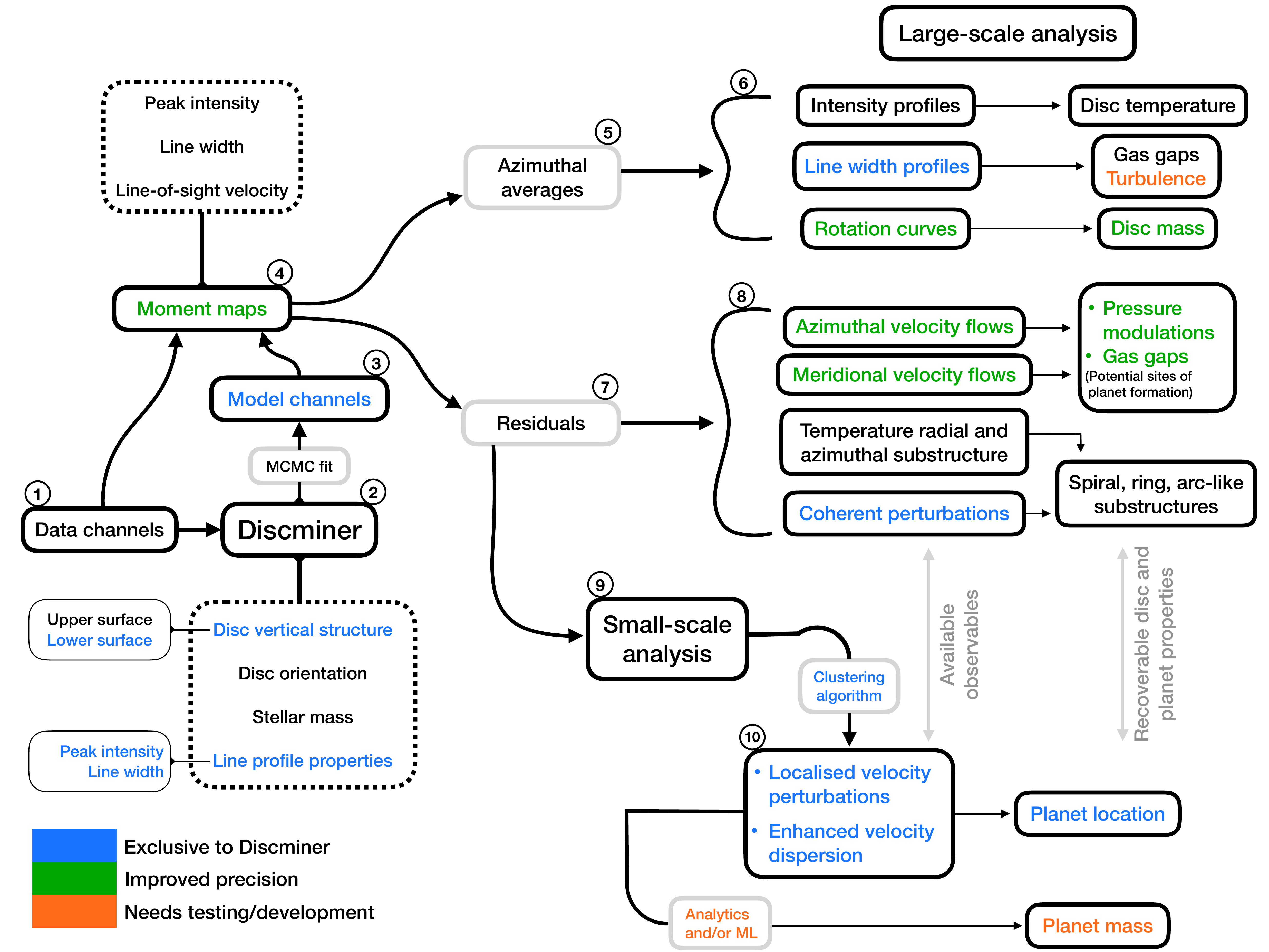}
      \caption{Schematic representation of the \discminer{} modelling flow and capabilities. \discminer{} produces best-fit channel maps of the input data cube by modelling the disc vertical structure and orientation, stellar mass, and attributes that shape the line profile morphology, assuming that the disc emission is smooth and Keplerian. The model and data channels are then compared through three types of moment maps. Large-scale analyses of these maps allow to look into the disc dynamical and thermodynamic structure, but are also useful to reveal potential signatures driven by planetary and non-planetary mechanisms, such as meridional flows or spiral-like perturbations. Small-scale analysis of residuals from these maps leads to the retrieval of the potential location of embedded planets through the detection of localised velocity perturbations and sites of enhanced velocity dispersion.
              }
         \label{fig:discminer_workflow}
   \end{figure*}

   \begin{figure*}
   \centering
   \includegraphics[width=0.77\textwidth]{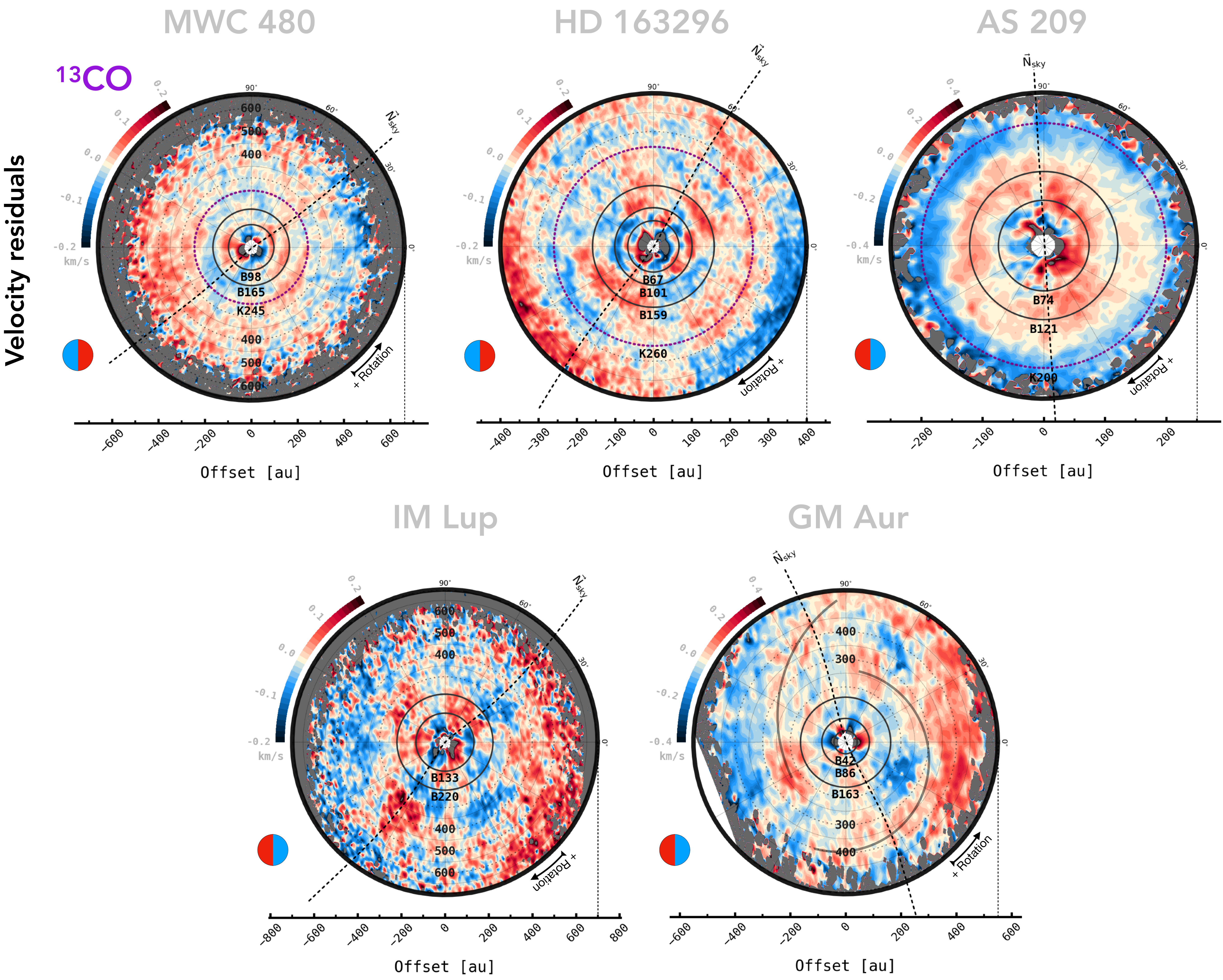}
    \caption{As Fig. \ref{fig:residuals_kinematics_12co} but for \thCO{}. This isotopologue traces vertical layers between $0.1<z/r<0.15$ in this sample of discs.
              }
         \label{fig:residuals_kinematics_13co}
   \end{figure*}

   \begin{figure*}
   \centering
   \includegraphics[width=0.77\textwidth]{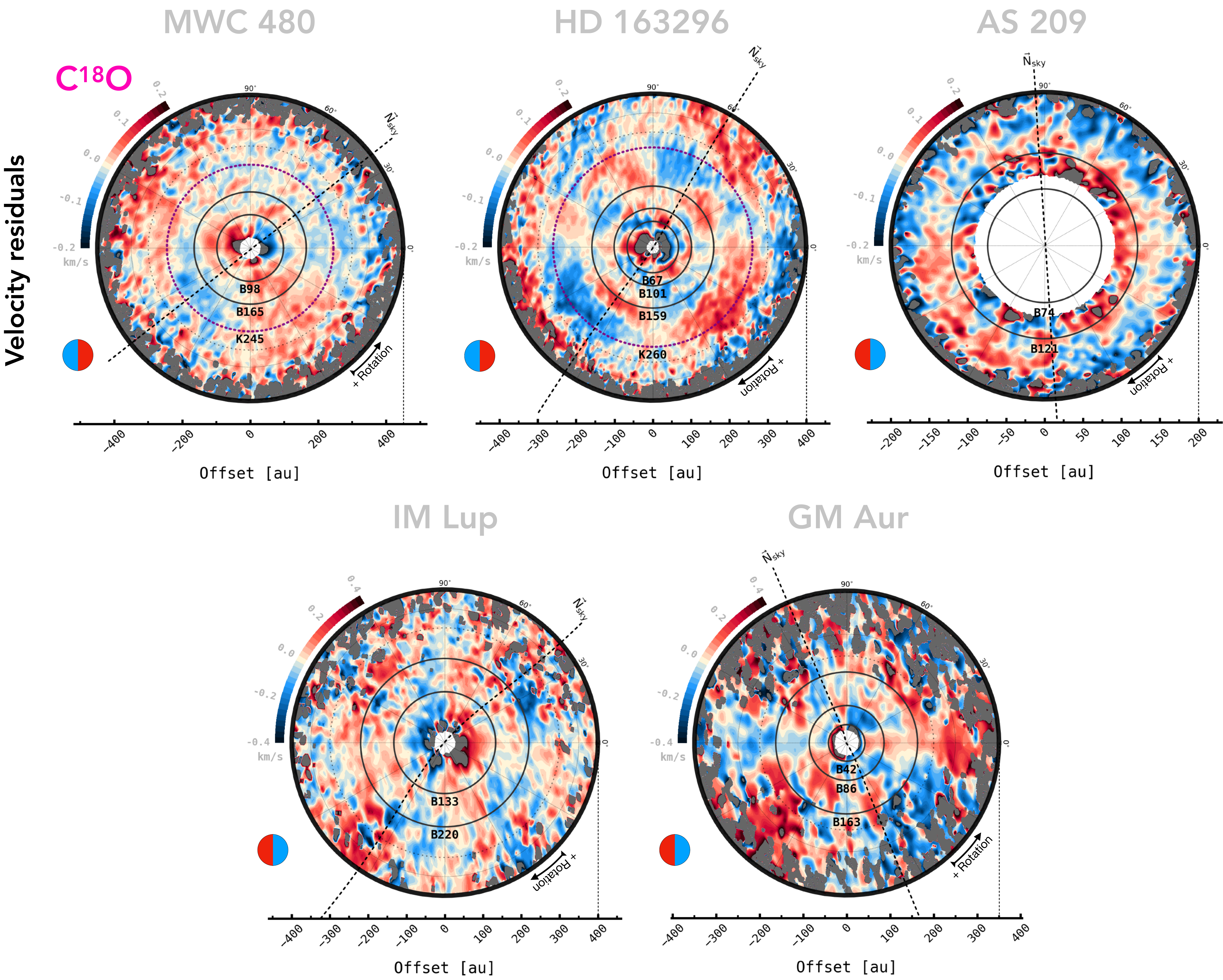}
    \caption{As Fig. \ref{fig:residuals_kinematics_12co} but for \eiCO{}. This isotopologue traces vertical layers between $0.0<z/r<0.1$ in this sample of discs.
              }
         \label{fig:residuals_kinematics_c18o}
   \end{figure*}

   \begin{figure*}
   \centering
   \includegraphics[width=0.77\textwidth]{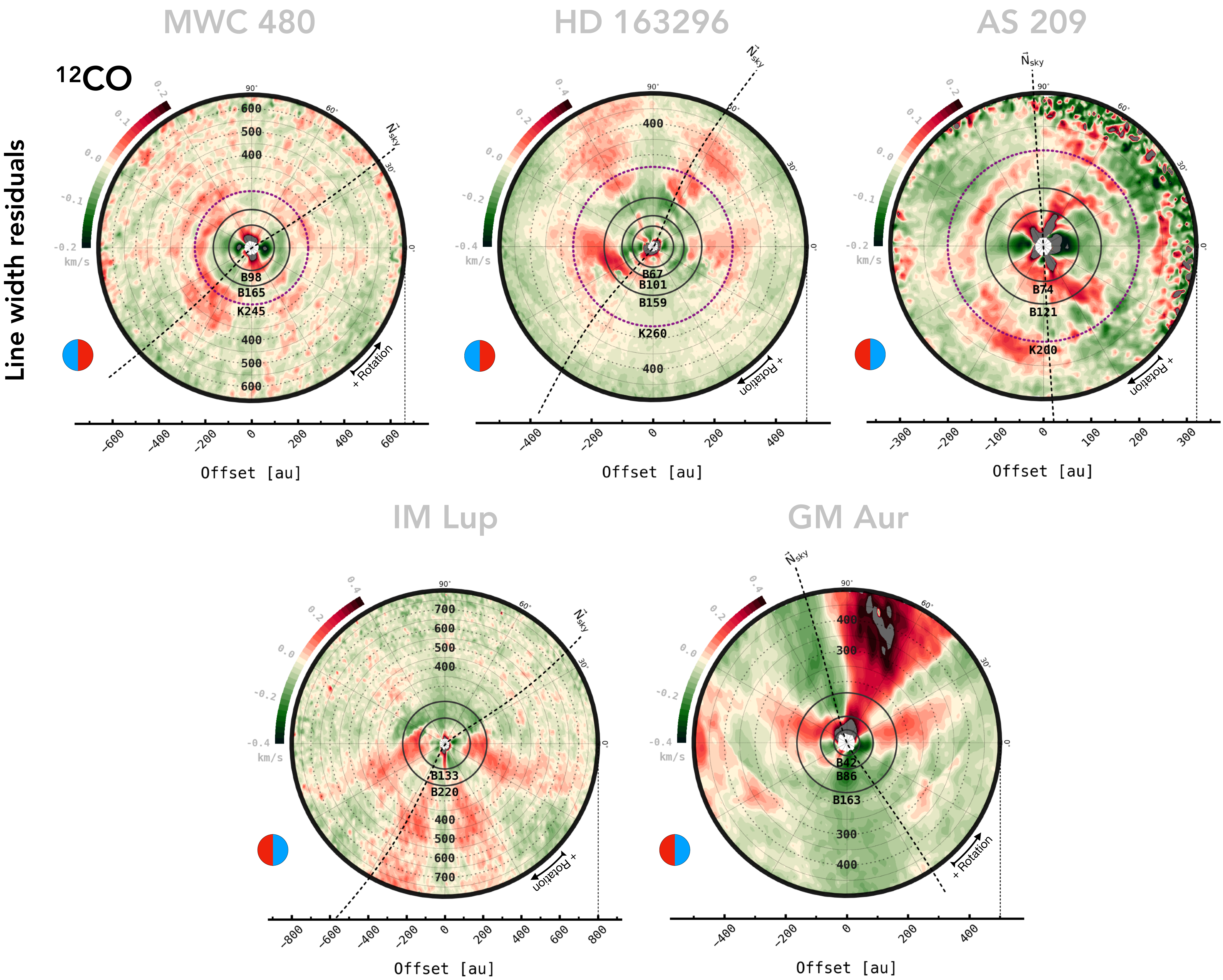}
    \caption{As Fig. \ref{fig:residuals_kinematics_12co} but for line width residuals.
              }
         \label{fig:residuals_linewidth_12co}
   \end{figure*}

   \begin{figure*}
   \centering
   \includegraphics[width=0.77\textwidth]{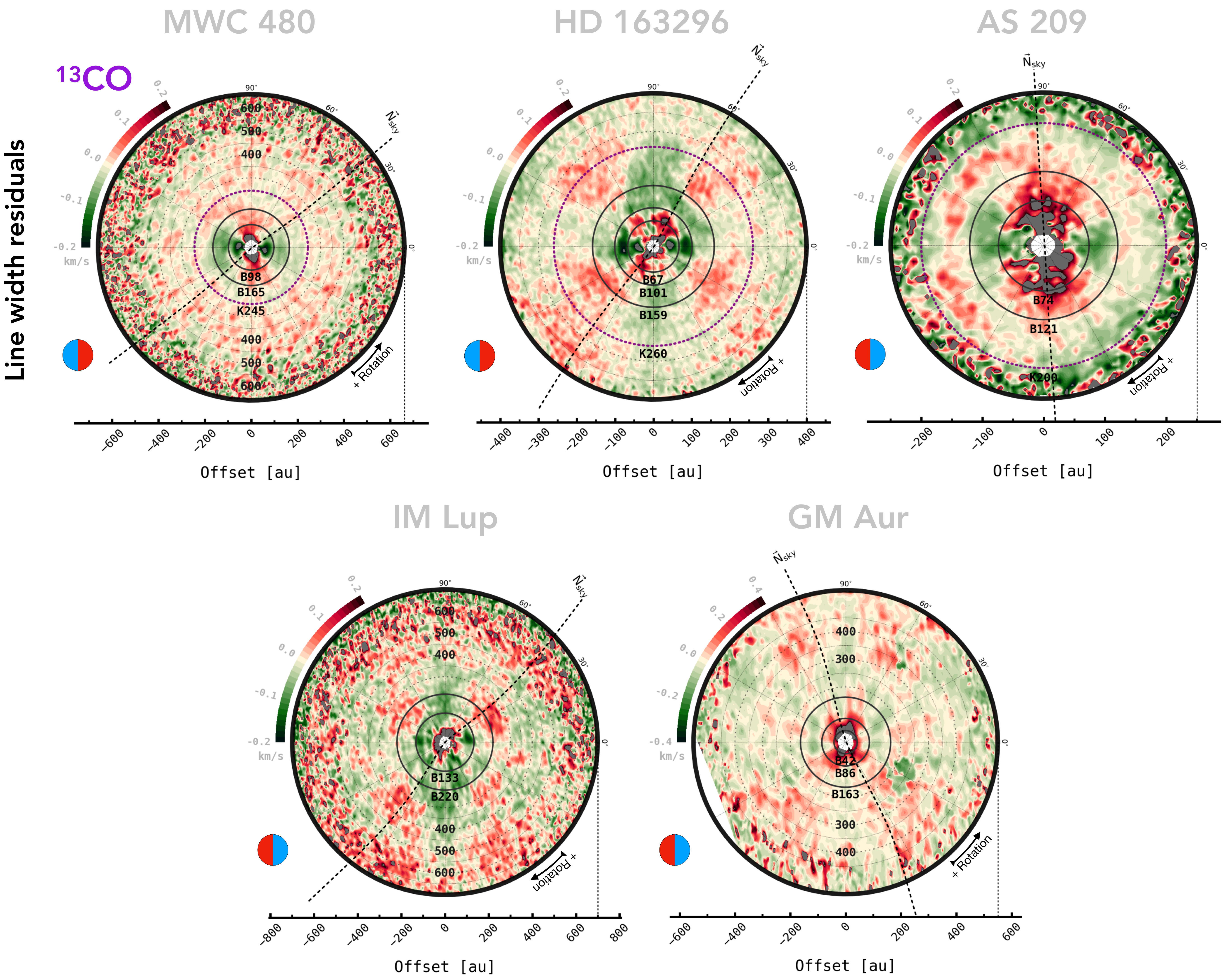}
    \caption{As Fig. \ref{fig:residuals_kinematics_12co} but for line width residuals from \thCO{}.
              }
         \label{fig:residuals_linewidth_13co}
   \end{figure*}

   \begin{figure*}
   \centering
   \includegraphics[width=0.77\textwidth]{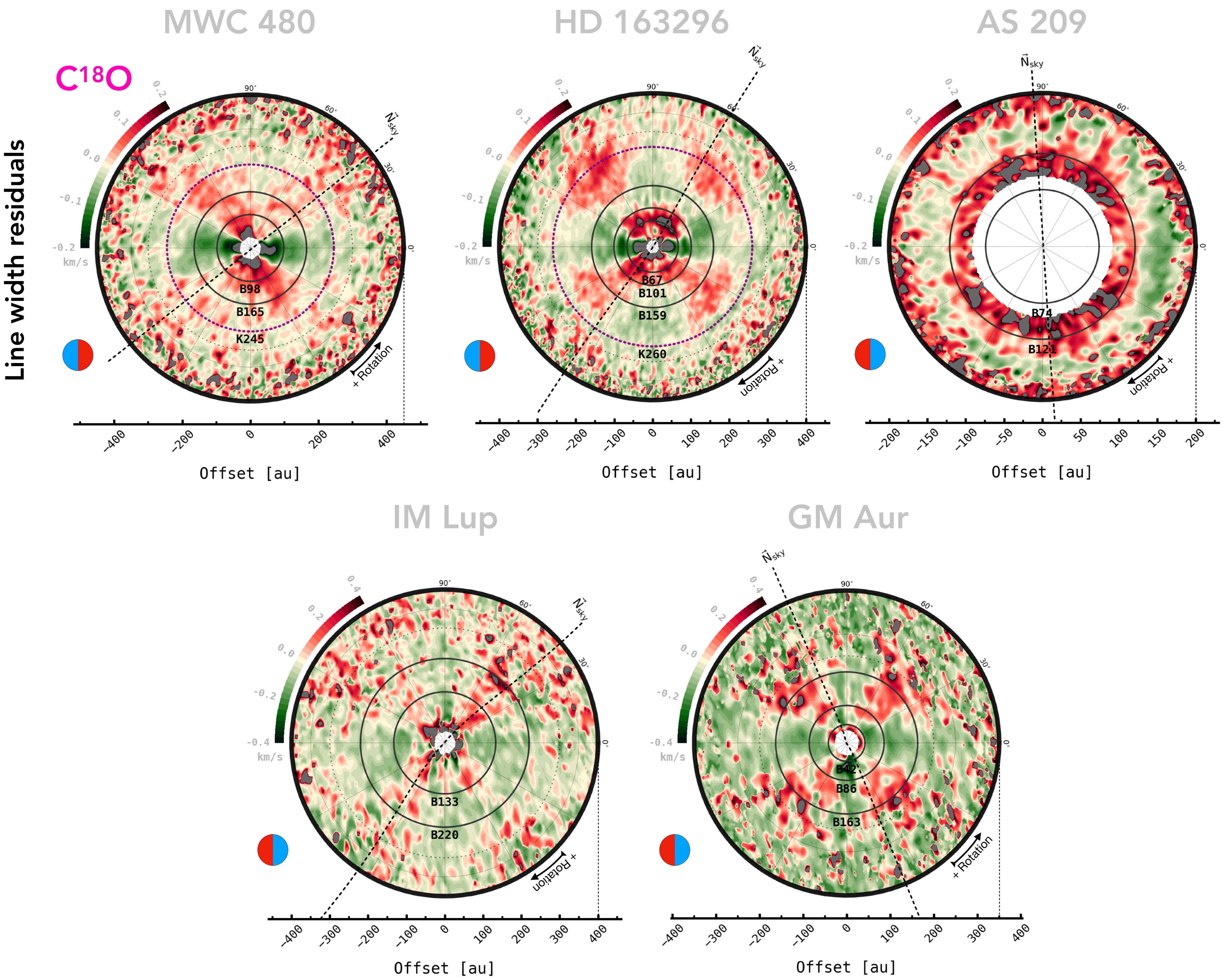}
    \caption{As Fig. \ref{fig:residuals_kinematics_12co} but for line width residuals from \eiCO{}.
              }
         \label{fig:residuals_linewidth_c18o}
   \end{figure*}

   \begin{figure*}
   \centering
   \includegraphics[width=0.77\textwidth]{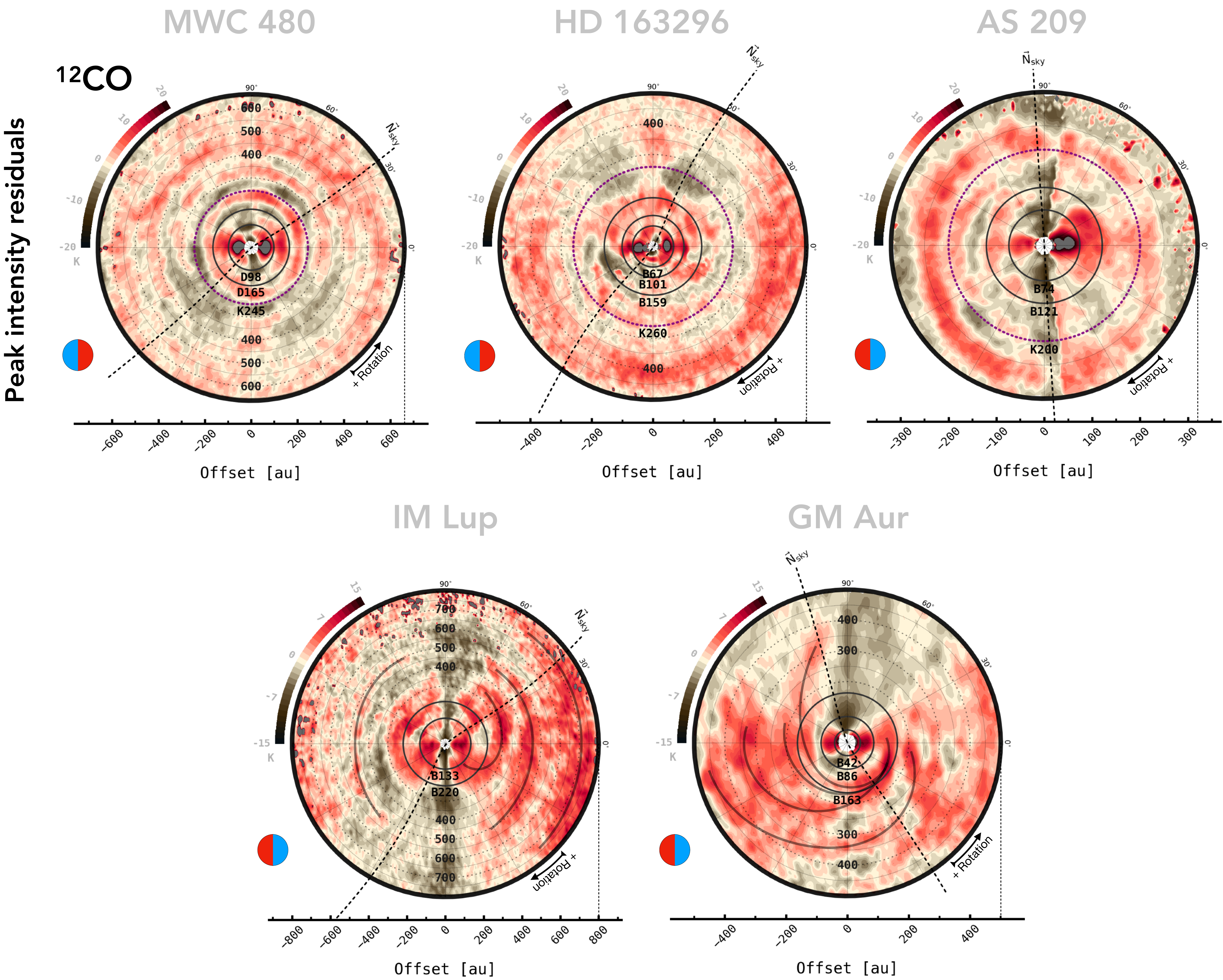}
    \caption{As Fig. \ref{fig:residuals_kinematics_12co} but for peak intensity residuals.
              }
         \label{fig:residuals_peakint_12co}
   \end{figure*}

   \begin{figure*}
   \centering
   \includegraphics[width=0.77\textwidth]{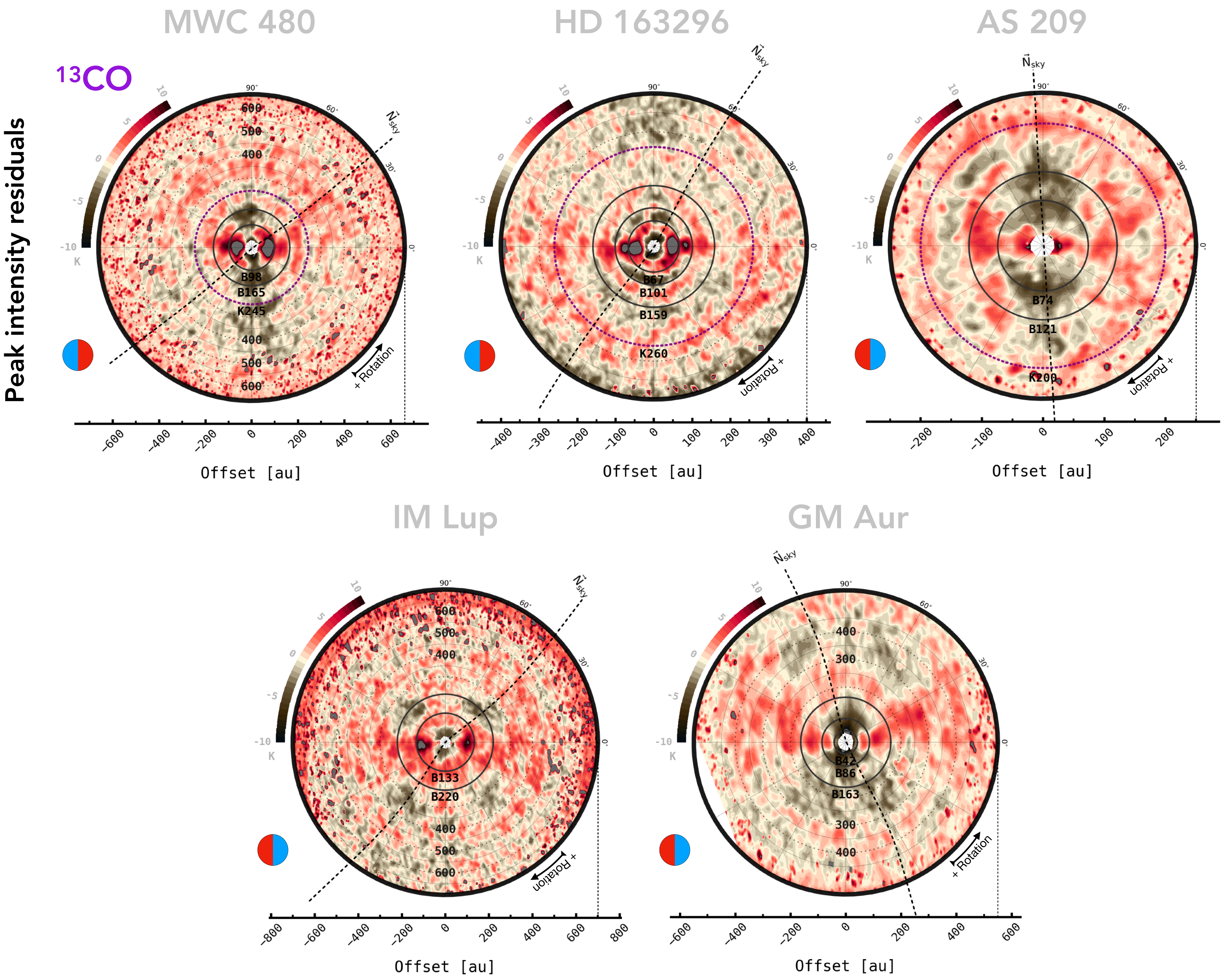}
    \caption{As Fig. \ref{fig:residuals_kinematics_12co} but for peak intensity residuals from \thCO{}.
              }
         \label{fig:residuals_peakint_13co}
   \end{figure*}

   \begin{figure*}
   \centering
   \includegraphics[width=0.77\textwidth]{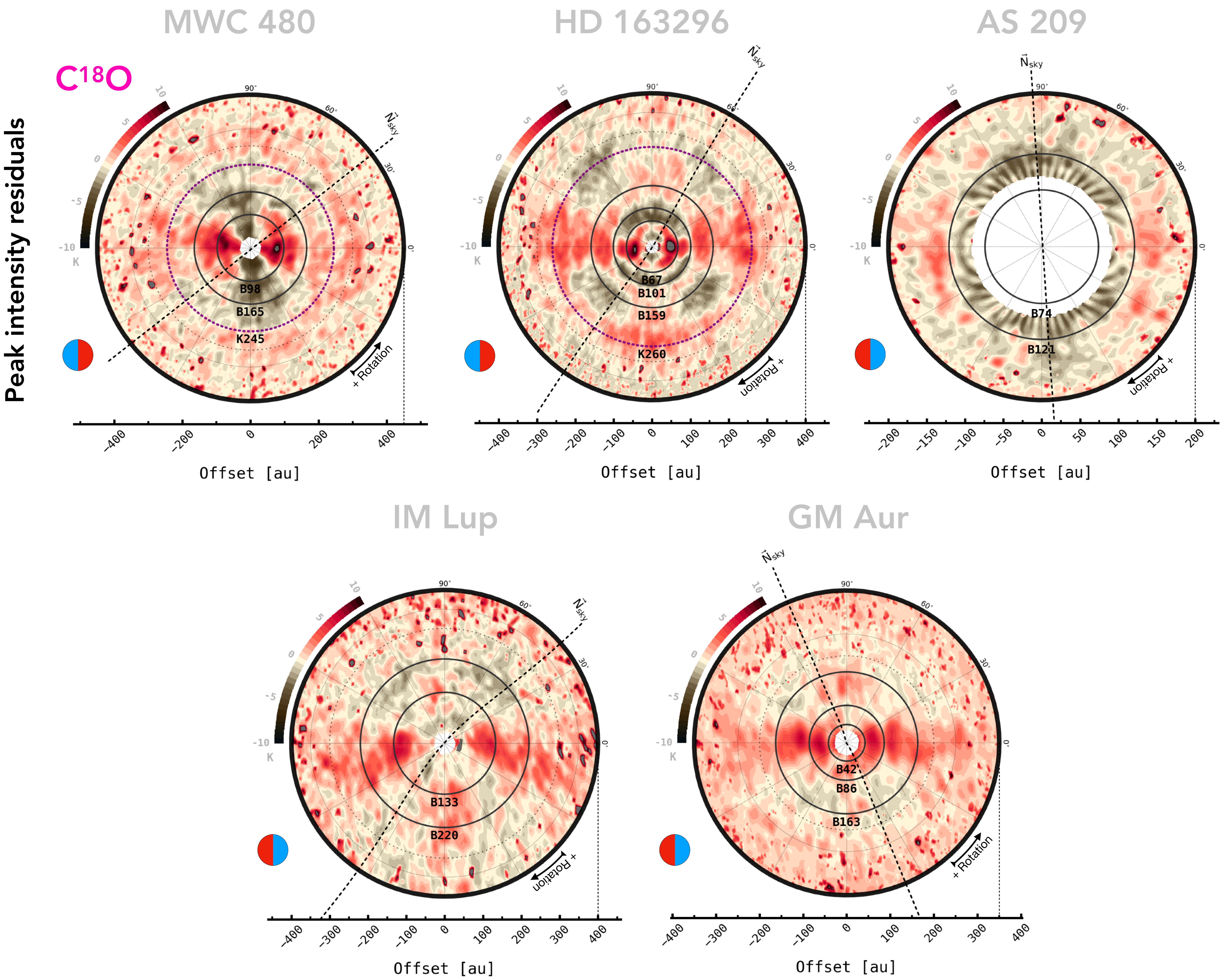}
    \caption{As Fig. \ref{fig:residuals_kinematics_12co} but for peak intensity residuals from \eiCO{}.
              }
         \label{fig:residuals_peakint_c18o}
   \end{figure*}

   \begin{figure*}
   \centering
   \includegraphics[width=1.0\textwidth]{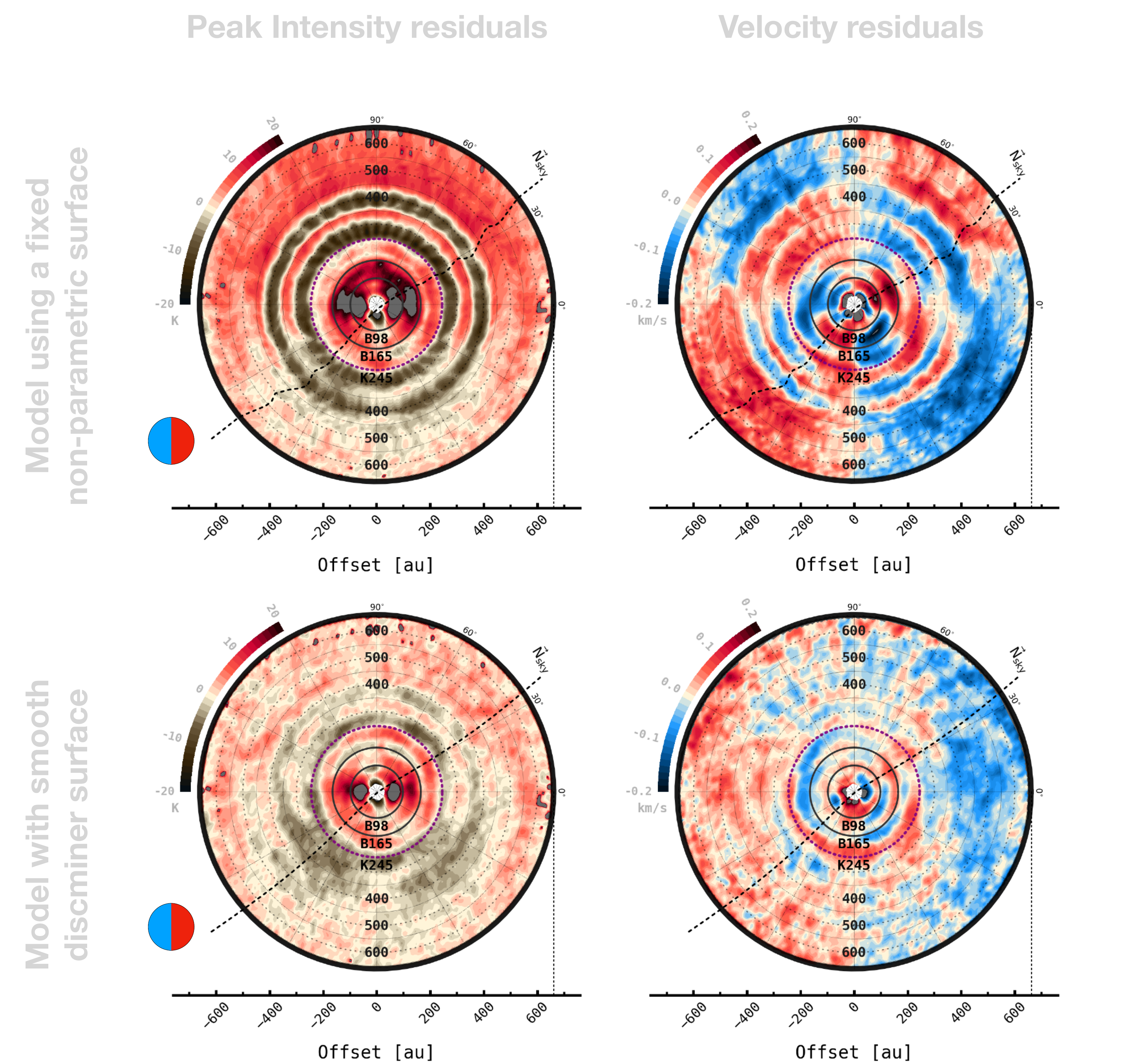}
    \caption{Illustrating the impact of using an irregular (top) and a smooth parametric (bottom) emission surface on the peak intensity and velocity residuals extracted for the \twCO{} disc of \mwc{}. Unphysical quadruple-like patterns observed at $R=300, 400$\,au and at $R>400$\,au in the velocity residuals resulting from the non-parametric surface suggest that a smooth (and less tapered) layer provides a better representation of the \twCO{} emission surface of this disc. 
              }
         \label{fig:surface_residuals_mwc480}
   \end{figure*}

   \begin{figure*}
   \centering
   \includegraphics[width=1.0\textwidth]{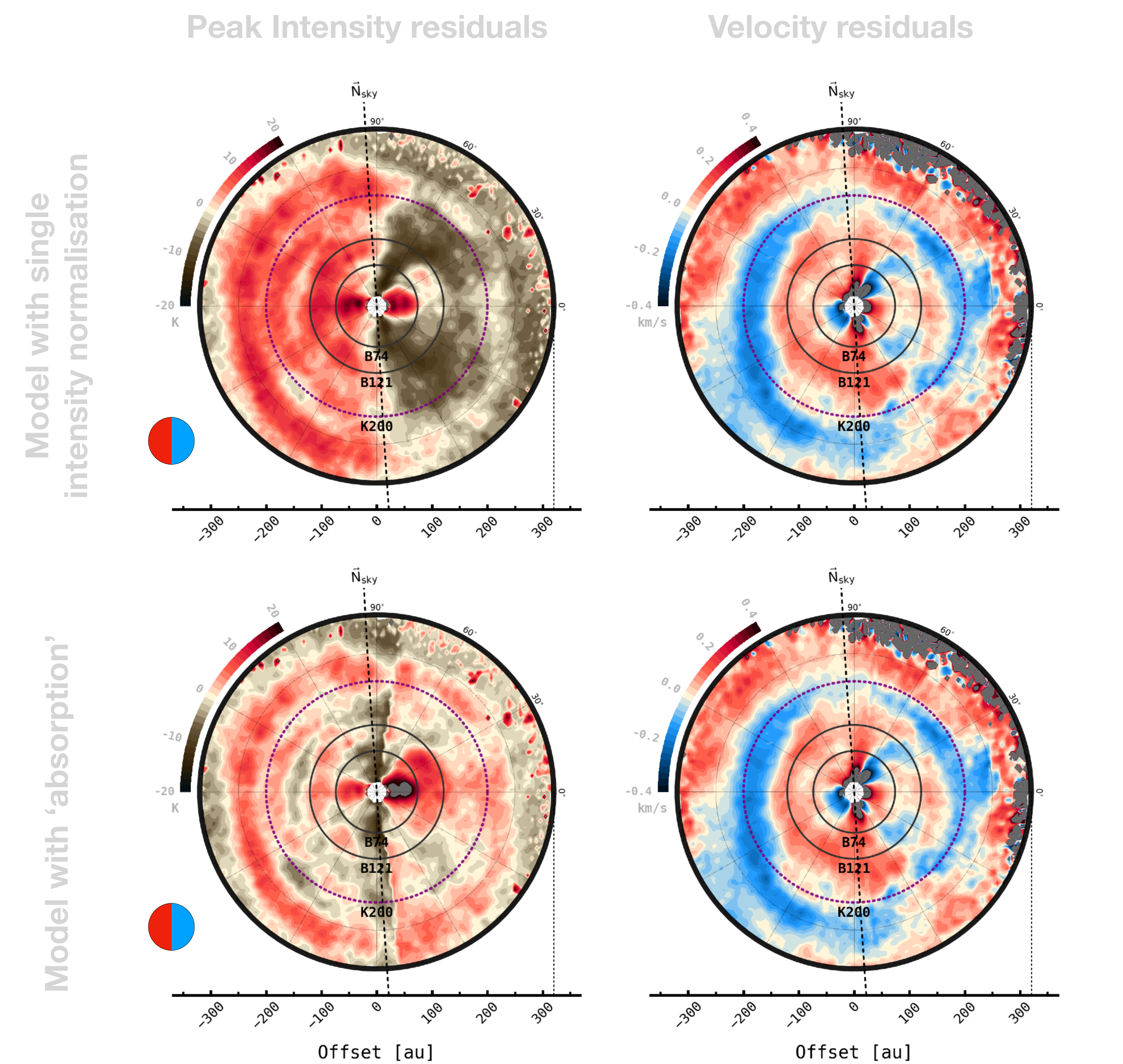}
    \caption{Illustrating the influence of having one (top) or two (bottom) model intensity normalisations on peak intensity and velocity residuals extracted for the \twCO{} disc of \as{}. Although it is irrelevant for the retrieved kinematic signatures, we adopt the two-normalisation approach which aims to emulate absorption by foreground material occurring on the blueshifted side of the disc, allowing us to better capture fluctuations in the intensity field.
              }
         \label{fig:absorption_as209}
   \end{figure*}

\begin{figure*}
   \centering  
      \includegraphics[width=0.98\textwidth]{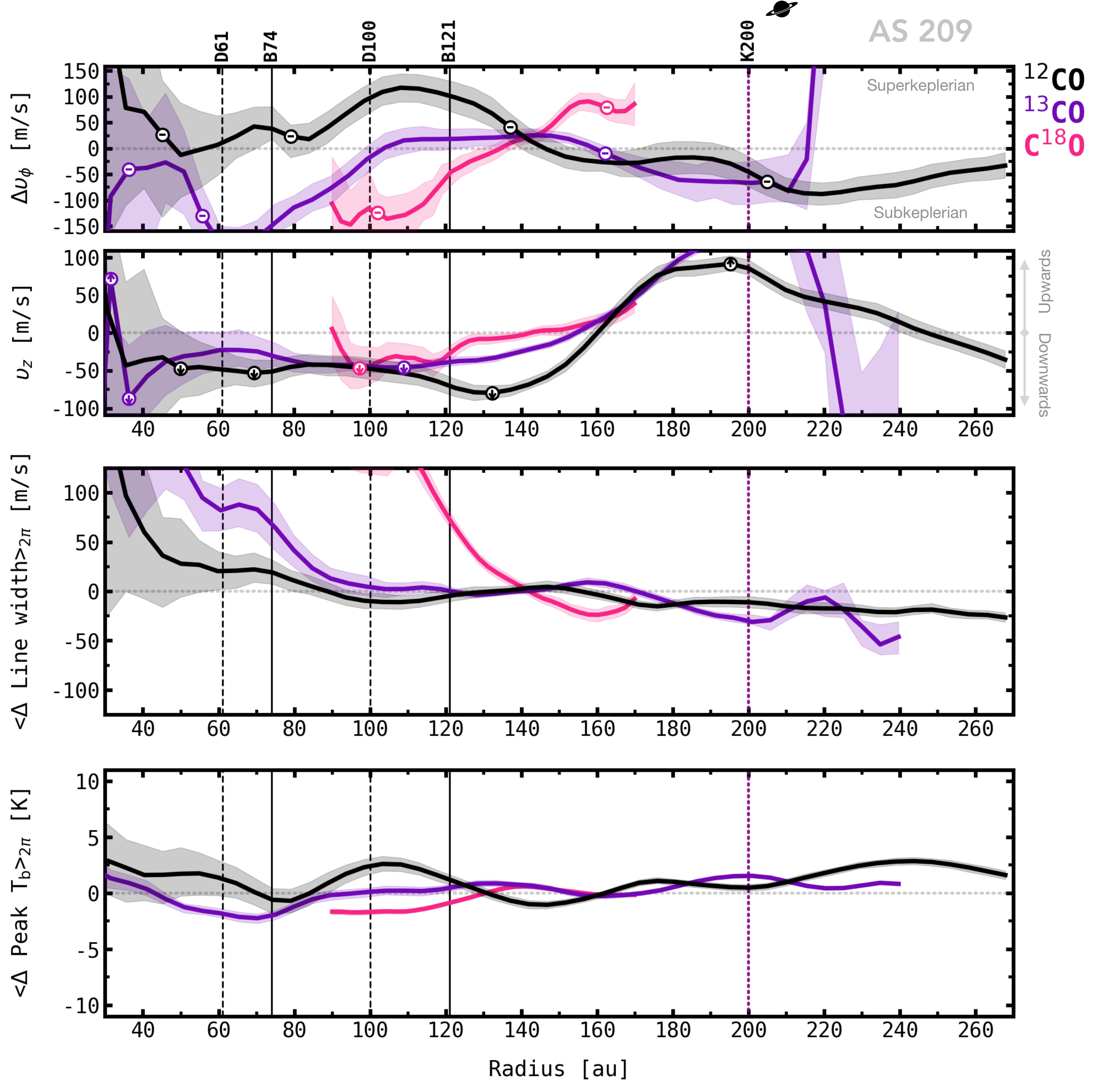}
      \caption{As Fig. \ref{fig:averaged_residuals_mwc480} but for the disc of \as{}. The planet marker indicates the orbital radius of the planet candidate associated with line width enhancements near the location of the CPD candidate proposed by \citet[][]{bae+2022} (see Table \ref{table:planet_locations}).
              }
         \label{fig:averaged_residuals_as209}
    
   \end{figure*}
   

\begin{figure*}
   \centering  
      \includegraphics[width=0.98\textwidth]{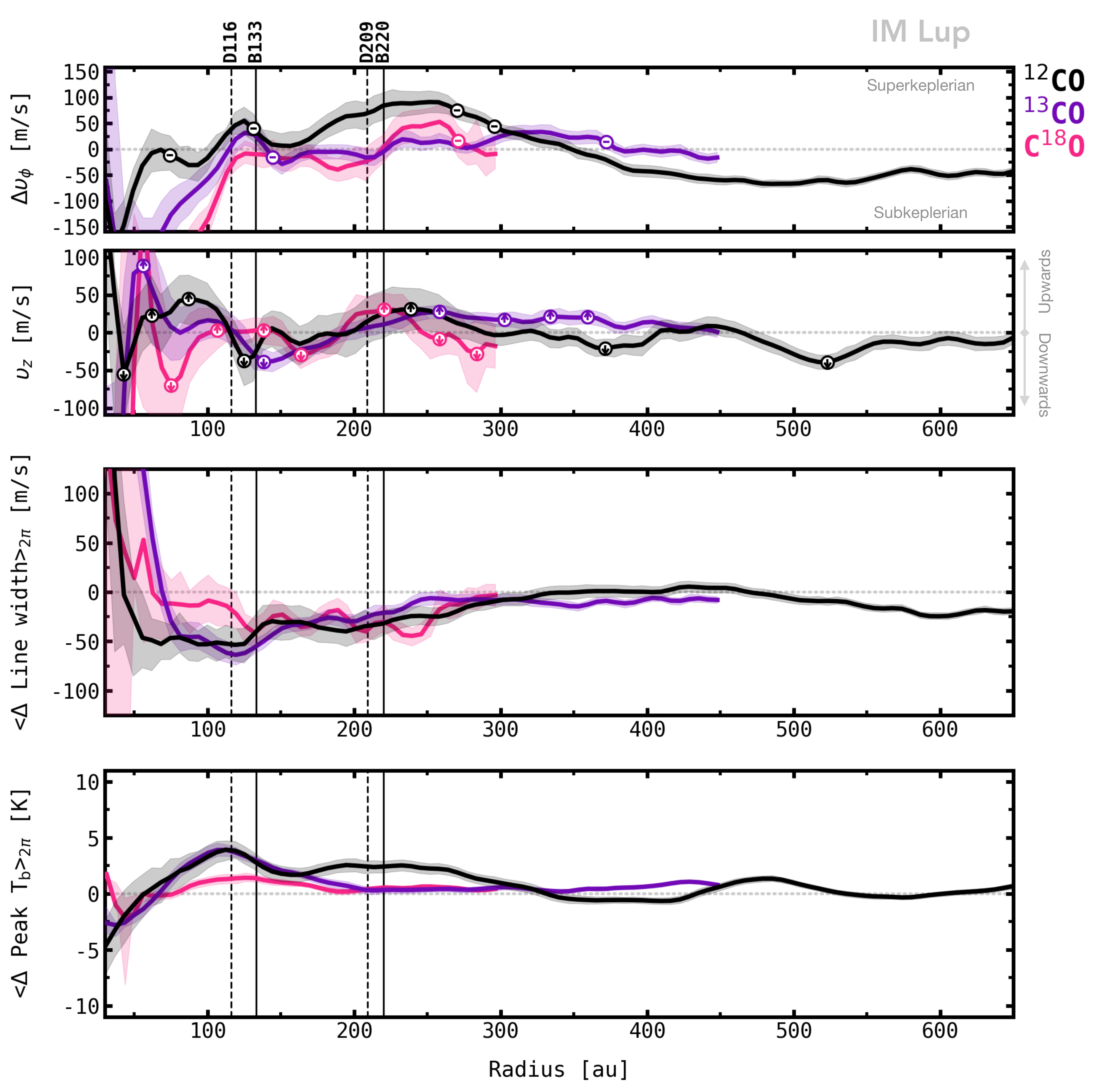}
      \caption{As Fig. \ref{fig:averaged_residuals_mwc480} but for the disc of \im{}. The calculation of the velocity profiles illustrated in this Figure excludes $\pm30^\circ$ wedges around the (north and south) disc minor axis to avoid contamination from possibly unphysical velocity residuals (see Sect. \ref{sec:extended_perturbations}, \textit{IM Lup}).
              }
         \label{fig:averaged_residuals_imlup}
    
   \end{figure*}

\begin{figure*}
   \centering  
      \includegraphics[width=0.98\textwidth]{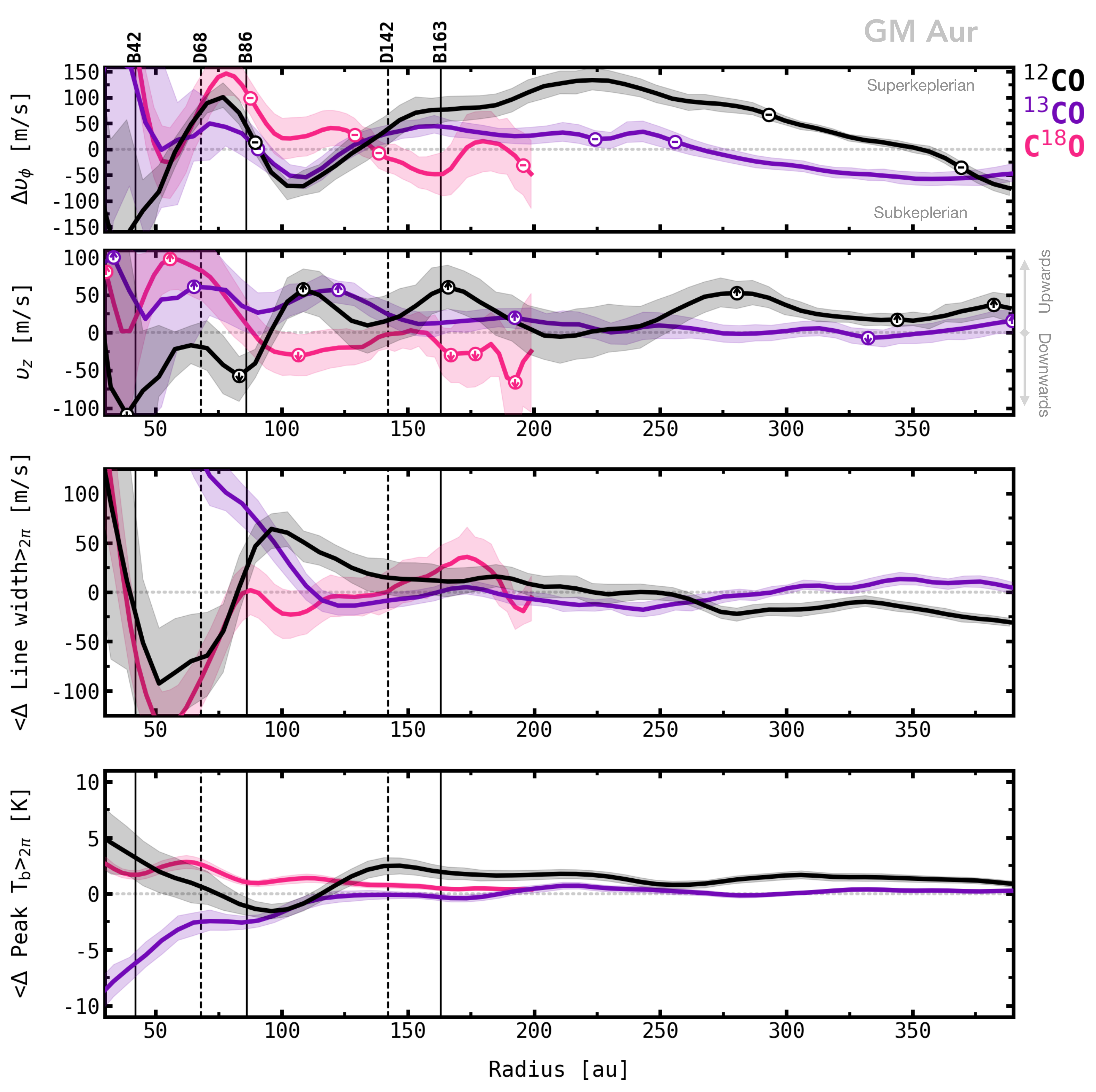}
      \caption{As Fig. \ref{fig:averaged_residuals_hd163296} but for the disc of \gm{}. The calculation of velocity profiles illustrated in this Figure excludes $\pm45^\circ$ wedges around the (north and south) disc minor axis to minimise contamination from infalling material (see Sect. \ref{sec:extended_perturbations}, \textit{GM Aur}).
              }
         \label{fig:averaged_residuals_gmaur}
    
   \end{figure*}

   \begin{figure*}
   \centering
   \includegraphics[width=0.9\textwidth]{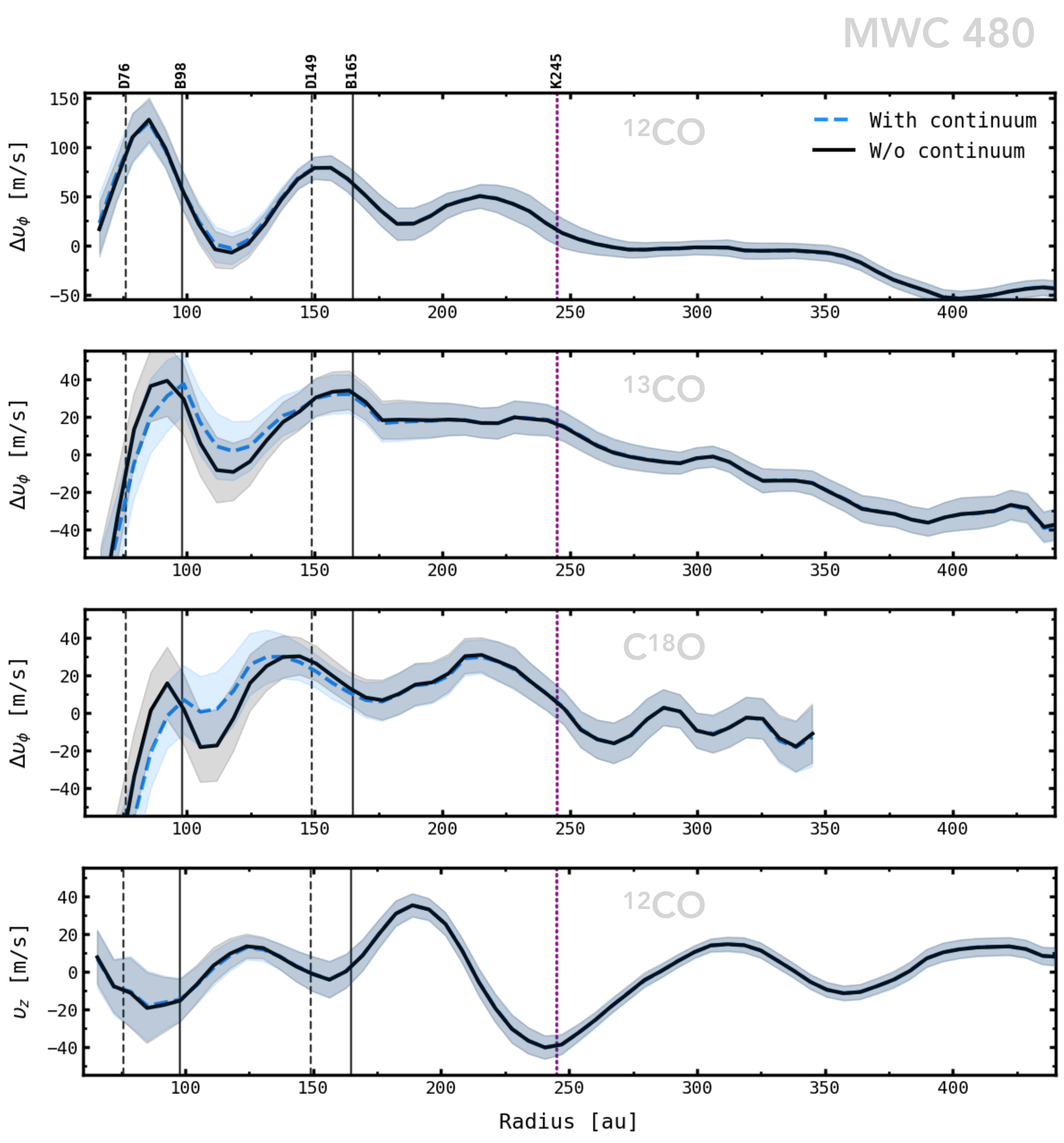}
    \caption{Comparison of azimuthal and vertical velocity profiles extracted with the decomposition method introduced in Sect. \ref{sec:azimuthal_meridional_velocities} for the disc of \mwc{} from cubes with and without continuum emission. 
              }
         \label{fig:wcont_vprofiles_mwc480}
   \end{figure*}

\section{Azimuthal average of absolute velocity residuals} \label{sec:appendix_absolute_residuals}

In this section, we demonstrate that a radial profile of the azimuthal component of the velocity field, $\upsilon_\phi$, or $\Delta\upsilon_\phi$ if referred to an axisymmetric model of the same component, can be obtained through the computation of azimuthal averages of absolute velocities, or absolute velocity residuals, respectively. The latter are defined as follows,

\begin{align} \label{eq:dv_abs}
\left|\upsilon_{0, d}'\right| - \left|\upsilon_{0, m}'\right| \equiv& \quad \left|\upsilon_{0, d} - \upsilon_{{\rm sys}, m}\right| - \left|\upsilon_{0, m} - \upsilon_{{\rm sys}, m}\right| \notag \\ 
=& \quad \left|\upsilon_{\phi, d}\cos{\phi}\sin{i} - \upsilon_{r, d}\sin{\phi}\sin{i} - \upsilon_{z, d}\cos{i}\right| \notag\\& - \left|\upsilon_{\phi, m}\cos{\phi}\sin{i}\right|.
\end{align}
Now, it is useful to note that under the reasonable assumption that the observed line-of-sight velocities are dominated by azimuthal velocities\footnote{Valid for intermediate and high inclinations where the sine is similar or much larger than the cosine of the disc inclination, and for almost all azimuthal angles $\phi$, owing to the fact that azimuthal velocities, $\upsilon_{\phi, d}$, are normally between one and two orders of magnitude larger than radial and vertical motions, $\upsilon_{r, d}$ and $\upsilon_{z, d}$.}, the absolute value of the data and model velocities can also be written as,
\begin{align} \label{eq:v_abs}
\left|\upsilon_{0}'\right| \equiv \begin{cases}
    + \upsilon_{0}', & \text{if $-\pi/2<\phi<\pi/2$}\\
    - \upsilon_{0}', & \text{if $\pi/2<\phi<3\pi/2$},
  \end{cases}
\end{align}
for clockwise rotation and positive inclination. Using eq. \ref{eq:v_abs}, an azimuthal average applied to eq. \ref{eq:dv_abs} can thus be split into different independent terms,
\begin{align} \label{eq:average_abs}
\left<\left|\upsilon_{0, d}'\right| - \left|\upsilon_{0, m}' \right|\right>_{2\pi} & \nonumber \\
     =& \quad \frac{1}{2\pi} \left[ \int_{-\pi/2}^{\pi/2} \upsilon_{\phi, d}\cos{\phi}\sin{i} \,d\phi \right. \nonumber \\ 
     & \quad - \cancelto{0}{\int_{-\pi/2}^{\pi/2} \upsilon_{r, d}\sin{\phi}\sin{i} \,d\phi} \nonumber \\ 
     & \quad - \bcancel{\int_{-\pi/2}^{\pi/2} \upsilon_{z, d}\cos{i} \,d\phi} \nonumber \\
     & \quad - \int_{\pi/2}^{3\pi/2} \upsilon_{\phi, d}\cos{\phi}\sin{i} \,d\phi \nonumber \\ 
     & \quad + \cancelto{0}{\int_{\pi/2}^{3\pi/2} \upsilon_{r, d}\sin{\phi}\sin{i} \,d\phi} \nonumber \\ 
     & \quad + \bcancel{\int_{\pi/2}^{3\pi/2} \upsilon_{z, d}\cos{i} \,d\phi} \nonumber \\
     & \quad - \int_{-\pi/2}^{\pi/2} \upsilon_{\phi, m}\cos{\phi}\sin{i} \,d\phi \nonumber \\
     & \quad + \left. \int_{\pi/2}^{3\pi/2} \upsilon_{\phi, m}\cos{\phi}\sin{i} \,d\phi \right] \nonumber \\
    \iff \left<\right>_{2\pi} =& \quad \frac{1}{2\pi} \sin{i}  \left(\upsilon_{\phi, d}-\upsilon_{\phi, m}\right)  \left[ \sin{\phi} \biggr|_{-\pi/2}^{\pi/2} - \sin{\phi} \biggr|_{\pi/2}^{3\pi/2} \right], \nonumber \\
    =& \quad \frac{2}{\pi} \sin{i}  \left(\upsilon_{\phi, d}-\upsilon_{\phi, m}\right), \nonumber
\end{align}
from which the azimuthal component of the velocity is followed immediately,
\begin{equation} \label{eq:vphi}
    \Delta\upsilon_\phi \equiv \upsilon_{\phi, d}-\upsilon_{\phi, m} = \frac{\pi}{2\sin{i}} \left<\left|\upsilon_{0, d} - \upsilon_{{\rm sys}, d}\right| - \left|\upsilon_{0, m} - \upsilon_{{\rm sys}, m}\right|\right>_{2\pi}
\end{equation}

Finally, it is possible to generalise the above expression if the azimuthal section considered for the azimuthal average, with a total extent of $\psi$ radians and not necessarily continuous, is symmetric around the disc major and minor axes,
\begin{equation} \label{eq:vphi}
    \Delta\upsilon_\phi = \frac{\psi}{4\sin{\frac{\psi}{4}}\sin{|i|}} \left<\left|\upsilon_{0, d} - \upsilon_{{\rm sys}, m}\right| - \left|\upsilon_{0, m} - \upsilon_{{\rm sys}, m}\right|\right>_{\psi}.
\end{equation}

\end{appendix}
\end{document}